\documentclass[twocolumn,prd,floatfix,preprintnumbers,a4paper,nofootinbib,superscriptaddress]{revtex4-1}

\usepackage{epsfig,graphics,graphicx}
\usepackage{amsmath,amssymb,mathtools,amsfonts}
\usepackage{bm,soul}
\usepackage[usenames,dvipsnames]{color}
\usepackage{amssymb,url,psfrag,times}
\usepackage[varg]{txfonts}
\usepackage[colorlinks, pdfborder={0 0 0}]{hyperref}
\usepackage{lineno,verbatim}
\usepackage[utf8]{inputenc}
\usepackage{ulem}
\usepackage{xcolor}
\usepackage{dsfont}
\usepackage[utf8]{inputenc}
\usepackage{enumerate}

\usepackage[english]{babel} 
\usepackage{blindtext}

\definecolor{LinkColor}{rgb}{0, 0, 0.75}
\definecolor{CiteColor}{rgb}{0, 0, 0.75}
\definecolor{UrlColor}{rgb}{0, 0, 0.75}
\hypersetup{linkcolor=LinkColor}
\hypersetup{citecolor=CiteColor}
\hypersetup{urlcolor=UrlColor}
\newcommand{\Amsterdam}{Institute for High-Energy Physics, University of Amsterdam, Science Park 904, 1098 XH Amsterdam, The Netherlands}
\newcommand{\KCL}{King's  College  London,  Strand,  London  WC2R  2LS,  United Kingdom}
\newcommand{\MITLIGO}{MIT Kavli Institute for Astrophysics and Space Research and LIGO Laboratory, 77 Massachusetts Avenue, Cambridge, MA 02139, USA}

\usepackage{inputenc}
\usepackage{pgfplots}
\usepackage{tikz}
\usetikzlibrary{shapes,arrows}
\usetikzlibrary{positioning}
\usetikzlibrary{backgrounds}

\newcommand{\braket}[2]{ {\langle {#1} \, | \, {#2} \rangle} }
\newcommand{\bra}[1]{ \langle {#1} |  }
\newcommand{\ket}[1]{ | {#1} \rangle }

\definecolor{lightblue}{rgb}{.82,.88,0.95}
\definecolor{lightred}{rgb}{0.95,.86,0.86}
\definecolor{yellow}{rgb}{0.95,0.95,0.86}
\definecolor{green}{rgb}{.90,1,0.95}
\definecolor{lightpurple}{rgb}{.95,0.85,0.95}

\tikzset{
    vertex/.style = {
        circle,
        fill            = black,
        outer sep = 2pt,
        inner sep = 1pt,
    }
}

\def\tk{Teukolsky}

\def\ee{Einstein's equations}

\def\gr#1{General Relativity#1
  (GR#1)\gdef\gr{GR}}

\def\gw#1{gravitational wave#1}

\def\grad#1{gravitational radiation#1}

\def\rm#1{\mathrm{#1}}

\def\Apx#1{Appendix~(\ref{#1})}
\def\apx#1{Appx.~(\ref{#1})}
\def\capx#1{Appx.~\ref{#1}}

\def\Sec#1{Section~\ref{#1}}
\def\sec#1{Sec.~\ref{#1}}
\def\csec#1{Sec.~\ref{#1}}

\def\tk#1{Teukolsky#1}

\def\Fig#1{Figure~\ref{#1}}
\def\fig#1{Fig.~\ref{#1}}

\newcommand{\Figsa}[2]{Figures~(\ref{#1}) and (\ref{#2})}
\def\Eqn#1{Equation~(\ref{#1})}
\def\eqn#1{Eq.~(\ref{#1})}
\def\ceqn#1{Eq.~\ref{#1}}
\newcommand{\Eqns}[2]{Equations~(\ref{#1}-\ref{#2})}
\newcommand{\Eqnsa}[2]{Equations~(\ref{#1}) and (\ref{#2})}
\newcommand{\eqns}[2]{Eqs.~(\ref{#1}-\ref{#2})}
\newcommand{\eqnsa}[2]{Eqs.~(\ref{#1}) and (\ref{#2})}
\newcommand{\ceqns}[2]{Eqs.~\ref{#1}-\ref{#2}}
\newcommand{\ceqnsa}[2]{Eqs.~\ref{#1} and \ref{#2}}

\def\sw{spin weighted}

\def\lal#1{LIGO Analysis Library#1
  (LAL#1)\gdef\lal{LAL}}
\def\nrda#1{\nr{} Data Analysis#1
  (NRDA#1)\gdef\nrda{NRDA}}
\def\tt#1{\textit{transverse--traceless}#1
  (TT#1)\gdef\tt{TT}}
\def\et#1{Einstein Telescope#1
  (ET#1)\gdef\et{ET}}
\def\ego#1{European Gravitational Observatory#1
  (EGO#1)\gdef\ego{EGO}}
\def\elisa#1{Evolved Laser Interferometer Space Antenna#1
  (eLISA#1)\gdef\elisa{eLISA}}
\def\ligo#1{Laser Interferometer Gravitational Wave Observatory#1
  (LIGO#1)\gdef\ligo{LIGO}}

\def\aligo#1{Advanced LIGO#1
  (Adv. LIGO#1)\gdef\aligo{Adv. LIGO}}
\def\snr#1{signal to noise ratio#1
  (SNR#1)\gdef\snr{SNR}}
\def\psd#1{power spectral density#1
  (PSD#1)\gdef\psd{PSD}}
\def\rom#1{reduced order model#1
  (ROM#1)\gdef\rom{ROM}}
\def\gatech#1{Georgia Institute of Technology#1
  (GaTech#1)\gdef\gatech{GaTech}}
\def\ffi#1{Fixed-Frequency Integration#1
  (FFI#1)\gdef\ffi{FFI}}
\def\sxs#1{Simulating Extreme Spacetimes#1
  (SXS#1)\gdef\sxs{SXS}}
\def\bam#1{Bifunctional Adaptive Mesh#1
  (BAM#1)\gdef\bam{BAM}}
\def\adm#1{Arnowitt-Deser-Misner
	(ADM#1)\gdef\adm{ADM}}
\def\frmse#1{Fractional Root-Mean Square Error
	(FRMSE)\gdef\frmse{FRMSE}}

\def\bh#1{black hole#1
 (BH#1)\gdef\bh{BH}}
\def\bbh#1{binary black hole#1
 (BBH#1)\gdef\bbh{BBH}}

\def\qnm#1{Quasi-Normal Mode#1
(QNM#1)\gdef\qnm{QNM}}
\def\eob#1{Effective One Body#1
  (EOB#1)\gdef\eob{EOB}}

\def\gw#1{gravitational wave#1}

\def\pn#1{Post-Newtonian#1
 (PN#1)\gdef\pn{PN}}
\def\pnl#1{post-Newtonian-like#1
  (PN-like#1)\gdef\pnl{PN-like}}

\def\nr{Numerical Relativity
 (NR)\gdef\nr{NR}}

\def\rd{ringdown}

\def\pca#1{principle component analysis#1
  (PCA#1)\gdef\pca{PCA}}
\def\svd#1{Singular Value Decomposition#1
  (SVD#1)\gdef\svd{SVD}}

\def\adj#1{{#1}^{\dagger}}

\def\A{{\L m}}

\def\mcl{\mathcal{L}}
\def\mct{{\mathcal{T}}}
\def\mcv{{\mathcal{V}}}

\def\mclo{{\mathcal{L}_o}}
\def\mcto{{\mathcal{T}_o}}
\def\mcvo{{\mathcal{V}_o}}

\def\tmcl{\tilde{\mcl}}
\def\mcp{\mathcal{P}}
\def\mcq{\mathcal{Q}}
\def\amcl{\adj{\mcl}}
\def\cmcl{{\mcl^*}}
\def\sjk{{\sigma_{\lp\ell}}}
\def\mcL{\mathcal{L}}

\def\Lo{ {\mathcal{K}} }
\def\I{{\mathbb{I}}}
\def\max{\mathrm{max}}

\def\lmn{{{\ell m n}}}

\def\lpmn{{{\ell' m n}}}
\def\lpmnp{{{\ell' m n'}}}
\def\lm{{{\ell m}}}
\def\lpm{{{\ell' m}}}
\def\l{{{\ell}}}
\def\n{{\bar{n}}}
\def\lmbn{{{\ell m \n}}}
\def\lpmbn{{{\ell' m \n}}}
\def\lp{{{\ell'}}}
\def\pp{p}

\def\k{{\lmn}}

\def\j{{\lpmn}}

\def\L{\bar{\ell}}

\def\Lop{\mathcal{L}_{\k}}
\newcommand{\brak}[2]{ \braket{#1}{#2} }
\newcommand{\as}{{adjoint-spheroidal}}
\newcommand{\ass}{{adjoint-spheroidals}}

\newcommand{\ashs}{{\as{} harmonics}}
\newcommand{\qnms}{\qnm{s}}

\newcommand{\cw}{\tilde{\omega}}

\def\gmvp#1{greedy-multivariate-polynomial#1
  (\texttt{GMVP}#1)\gdef\gmvp{\texttt{GMVP}}}
\def\gmvr#1{greedy-multivariate-rational#1
  (\texttt{GMVR}#1)\gdef\gmvr{\texttt{GMVR}}}

\begin{document} 


\title{ Bi-orthogonal harmonics for the decomposition of gravitational radiation I: angular modes, completeness, and the introduction of adjoint-spheroidal harmonics }

\author{Lionel T. London}
\email[]{lionel.london@kcl.ac.uk}
\affiliation{\KCL}
\affiliation{\Amsterdam}
\affiliation{\MITLIGO}

\begin{abstract}
	The estimation of radiative modes is a central problem in gravitational wave theory, with essential applications in signal modeling and data analysis. This problem is complicated by most astrophysically relevant systems' not having modes that are analytically tractable. A ubiquitous workaround is to use not modes, but multipole moments defined by spin weighted spherical harmonics. However, spherical multipole moments are only related to the modes of systems without angular momentum. As a result, they can obscure the underlying physics of astrophysically relevant systems, such as binary black hole merger and ringdown. In such cases, spacetime angular momentum means that radiative modes are not spherical, but spheroidal in nature.  Here, we work through various problems related to spheroidal harmonics.  We show for the first time that spheroidal harmonics are not only capable of representing arbitrary gravitational wave signals, but that they also possess a kind of orthogonality not used before in general relativity theory. Along the way we present a new class of spin weighted harmonic functions dubbed ``adjoint-spheroidal" harmonics.  These new functions may be used for the general estimation of spheroidal multipole moments via complete bi-orthogonal decomposition (in the angular domain). By construction, adjoint-spheroidal harmonics suppress mode-mixing effects known to plague spherical harmonic decomposition; as a result, they better approximate a system's true radiative modes. We discuss potential applications of these results. Lastly, we briefly comment on the challenges posed by the analogous problem with Teukolsky's radial functions.
\end{abstract}

\date{\today}

\maketitle

%
\section{Introduction}
\vspace{-5.1pt}
%
%
\par Central to \gw{} detection and the inference of source parameters is the representation of gravitational radiation in terms of multipole moments~\cite{LIGOScientific:2018mvr,LIGOScientific:2020stg}.
These functions of time or frequency allow the radiation's angular dependence to be given by spin weighted harmonic functions.
This leaves the radiation itself to be represented as a sum over harmonic functions, with each term weighted by a different multipole moment.
The choice of representation, namely the choice of which harmonic functions to use, is not unique.
Only the radiation's spin weight must be respected~\cite{Teukolsky:1973ha,NP66}. 
While there are multiple appropriate \sw{} functions, only one set of harmonic functions corresponds to the system's natural modes.
\begin{figure}[htb] 
	\begin{tabular}{c}
		\includegraphics[width=0.44\textwidth]{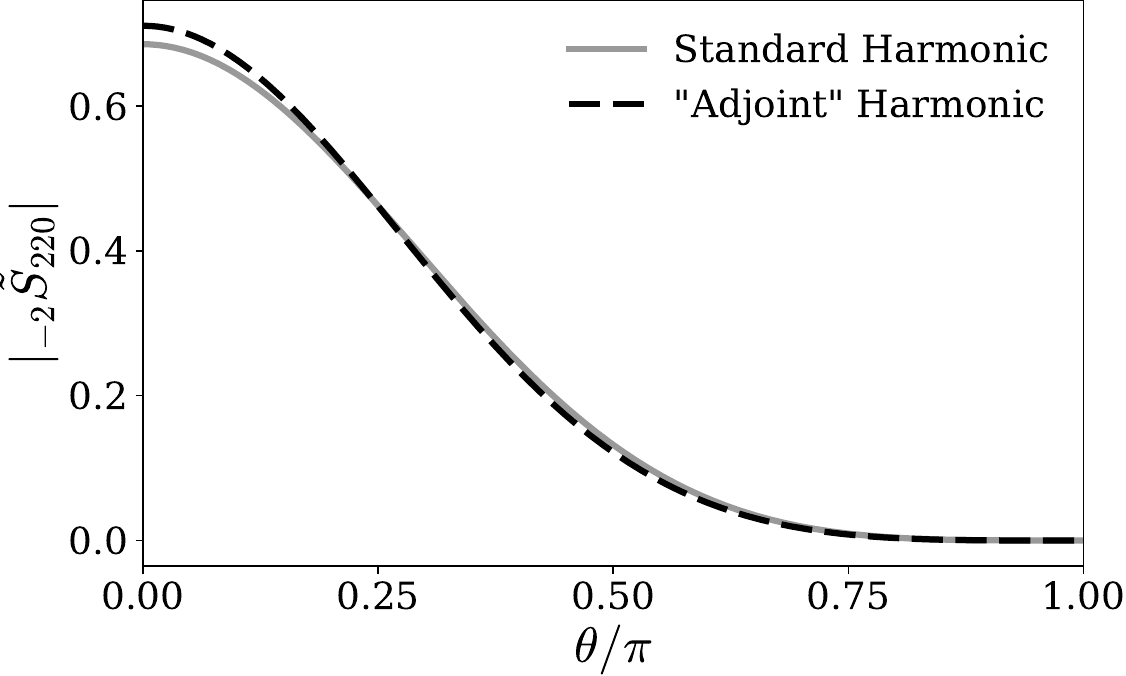} 
		\\
		\includegraphics[width=0.44\textwidth]{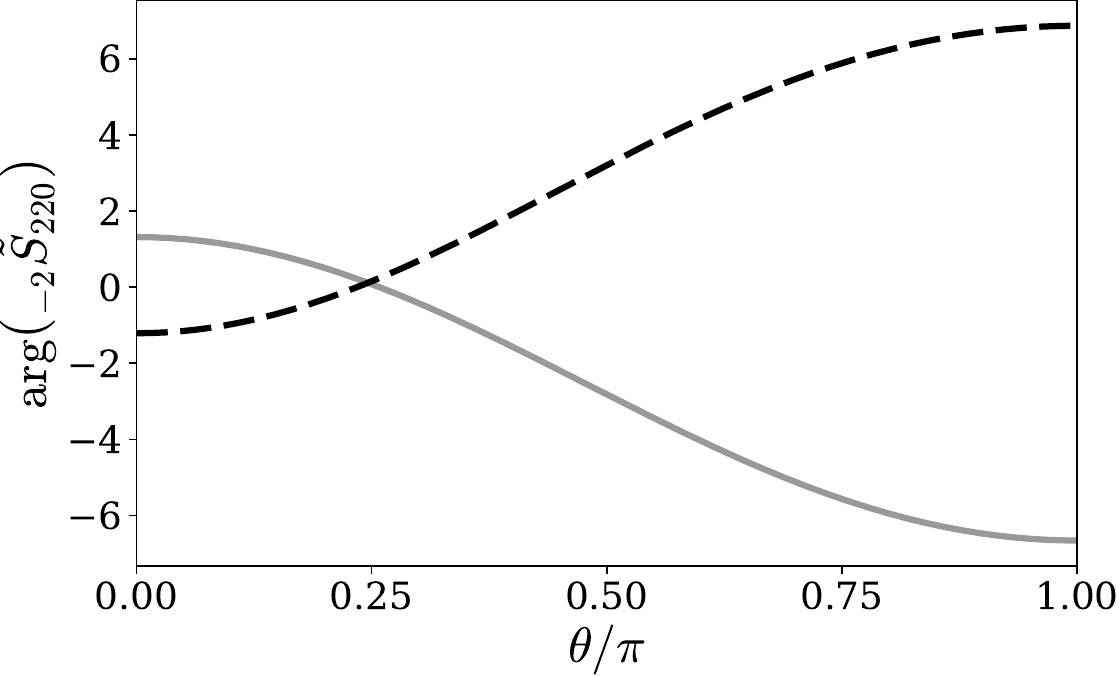}
	\end{tabular}
	\caption{Examples of this work's central result for spin weight $-2$, and Kerr spin parameter of $a=0.7$: 
	The new \ashs{}, ${_{-2}}\tilde{S}_{\lmn}$, differ from the spheroidal harmonics, ${_{-2}}{S}_{\lmn}$, in subtle but non-trivial ways. 
	\textit{Top}: a comparison of harmonic amplitudes for $(\l,m,n)=(2,2,0)$.
	\textit{Bottom}: a comparison of harmonic phases for $(\l,m,n)=(2,2,0)$. Note that, for ease of presentation, phases have been scaled by a factor of $100$.
	Here, $\arg(x+iy)=\tan^{-1}(y/x)$.
\vspace{-0.5cm}
	}
	\label{fig:new_harmonics}
\end{figure}
\par Spin-weighted spherical harmonics are perhaps the most commonly used functions for describing the angular behavior of \grad{}~\cite{Thorne:1980, Ruiz:2007yx}.
They are the simplest known functions fit for this purpose~\cite{Thorne:1980}. 
Their completeness and orthonormality make them straightforward to use~\cite{Breuer1977}.
Nevertheless, the spin weighted spherical harmonics are not always the most physically appropriate choice. 
\par This is readily seen in the study of single perturbed \bh{s}, where the analytic structure of gravitational radiation is understood in terms of the system's natural modes (eigenfunctions of \ee{})~\cite{Teukolsky:1973ha,Ruiz:2007yx,London:2014cma}.
On one hand, linear perturbations of spherically symmetric spacetimes (e.g.\ Schwarzschild \bh{s}) are known to yield radiative modes whose angular behavior is spherical harmonic~\cite{leaver85,Ruiz:2007yx,Thorne:1980}.
On the other hand, linear perturbations of spacetimes with angular momentum (e.g.\ Kerr \bh{s}) are known to yield radiative modes whose angular behavior is \textit{spheroidal} harmonic~\cite{Hughes:1999bq,leaver85,London:2014cma,London:2017bcn,Teukolsky:1973ha,Holzegel:2013kna,Berti:2005gp,Berti:2014fga}.
There, due to the complete and orthogonal nature of spherical harmonics, spherical harmonic multipole moments may indeed be used to represent gravitational radiation.  
However, doing so obscures the necessarily simpler information in the system's natural spheroidal modes~\cite{Kelly:2012nd,London:2014cma,Garcia-Quiros:2020qpx,Berti:2014fga}.
\par The use of spherical harmonics is known to complicate the morphology of \gw{} signal models, with downstream impact on data analysis
~\cite{London:2014cma,Garcia-Quiros:2020qpx,Blackman:2017pcm,Mehta:2019wxm,Carullo:2018sfu,Ghosh:2016qgn,Ota:2019bzl,Bhagwat:2019dtm}.
In particular, the artificial ``mixing'' of spheroidal modes is a potentially unnecessary complication, with direct impact on models \bbh{} merger and \rd{}~\cite{London:2014cma,Garcia-Quiros:2020qpx,Blackman:2017pcm,Berti:2014fga,Kelly:2012nd}. 
The need to overcome such complications drives ongoing interest in representing \gw{}s using harmonics that are as closely as possible related to the system's natural modes~\cite{Holzegel:2013kna,Garcia-Quiros:2020qpx,London:2017bcn,London:2014cma}.
\par In this context, spheroidal harmonics represent a logical alternative to spherical harmonics.
They are used extensively in black hole perturbation theory, and are integral to the calculation of \gw{}s from extreme mass-ratio inspirals~\cite{Hughes:1999bq,Mino:1997bx,OSullivan:2014ywd}. 
However, spheroidal harmonics have not been used more broadly in \pn{} theory or \nr{}, in part, for technical reasons.
\par In the late inspiral, merger and ringdown of extreme or comparable mass-ratio \bh{} coalescence, spheroidal harmonics are the complex valued, non-orthogonal eigenfunctions of \ee{}, which are themselves non-hermitian in these regimes~\cite{Hughes:1999bq,Blanchet:2013haa,Teukolsky:1973ha,leaver85}. 
Due in part to these features, it has thus far not been shown whether spheroidal harmonics possess the key properties that make spherical harmonics so useful: \textit{completeness} (the ability to exactly represent arbitrary \gw{} signals), and \textit{orthogonality} (the ability to decompose \gw{} signals into independent moments of information).  
\par Here, we will see how these technical hurdles can be overcome.
The primary result of this work is a class of new special functions that we will call the \textit{adjoint}-spheroidal harmonics.
They are related to the complex conjugates of the regular spheroidal harmonics, but differ from them in important ways. 
For example, \fig{fig:new_harmonics} shows that the absolute values of adjoint harmonics differ nontrivially from the traditional spheroidal harmonics.
In this work we lay the mathematical foundation for the \ashs{}. 
For the first time we show that they are complete when defined over the \qnms{}, and that they exhibit a kind of orthogonality in that setting.
These properties are naturally connected to the existence of the \ashs{}. 
\par In a companion paper (Paper II, Ref.~\cite{London:2021P2}), we present example applications of the \ashs{} to \gw{s} from extreme and comparable mass-ratio \bbh{s}, and we discuss potential applications of the \ashs{} in \gw{} theory.  
\par Although our presentation will focus on linear gravitational perturbations of Kerr, their \qnms, and thus their related spheroidal harmonics~\cite{leaver85}, we expect that the mathematical structure of our results applies (exactly or approximately) to any spin weighted harmonics related to the modes of axisymmetric spacetimes with angular momentum. 
\par For simplicity, we consider only spheroidal harmonics corresponding to pro- or retrograde \qnms{} (i.e.\ exclusively pro- or retrograde perturbations with respect to the \bh{} spin). 
The resulting \ashs{} can be used to calculate spheroidal harmonic multipole moments via \textit{bi}-orthogonal decomposition, i.e.\ orthogonality between two sets of functions rather than one~\cite{brauer1964,Christensen2003}.
Like the spin-weighted spherical harmonics, the spin-weighted spheroidal and \ashs{} allow the representation of general \gw{} signals.
Unlike the spherical harmonics, the spheroidal harmonics and their multipole moments are closely related to the natural modes of stationary spacetimes with angular momentum~\cite{leaver85,Teukolsky:1973ha}.
\par We will discuss \qnm{} orthogonality and completeness in the context of only the polar (i.e.\ $\theta$) dependence of each mode. 
In this sense, the presented work focuses on the solutions of Teukolsky's angular equation~\cite{leaver85,Teukolsky:1973ha}. 
One could alternatively focus on Teukolsky's master equation, which describes all spatial dependencies of perturbations, and separates into the radial and angular equations. 
In that setting, one would be interested in the master equation's self-adjointness, with respect to an appropriately constructed weight function. 
From that perspective, as well as what will be investigated here, the underlying premise is that the known uniqueness of \qnm{} eigenvalues (~see \csec{subsec:Subsets}) strongly implies the existence of a (bi-) orthogonal solution space. 
\par Here, we have chosen to investigate this topic by focusing on the angular equation because of its relative simplicity.
While this manuscript was being prepared, the author learned of ongoing and complementary work which investigates orthogonality of \qnms{} from the perspective of the master equation ~\cite{Green,Sberna:2021eui}. 
That work, as well as what we present here, are potential first steps towards a more general representation of \grad{} that is closely related to a spacetime's natural modes. 
%
\subsection{Resources for this work} 
%
\par The quantitative results of this work may be reproduced using routines from the openly available Python package, \href{https://github.com/llondon6/positive}{\texttt{positive}}~\cite{positive:2020}. 
Of principle use are the Kerr \qnm{} frequencies and the spheroidal harmonics. 
Both of these quantities may be determined using, for example, Leaver's analytic representation~\cite{leaver85}.
In \texttt{positive}, the \qnm{} frequencies may be accessed via the \texttt{positive.qnmobj} class, which automatically collects a \qnm{}'s frequency, spheroidal and radial harmonics.
The class contains convenient routines for calculating spherical-spheroidal inner-products, and can be made consistent with various popular \qnm{} conventions (see \texttt{positive.qnmobj.explain\_conventions}). 
Similarly, \texttt{positive} contains multiple inter-consistent routines for calculating the central objects of current interest, the spheroidal harmonic functions. 
These may be accessed via \texttt{positive.slm}, which uses Leaver's representation, and \texttt{positive.slmcg}, which uses a spherical harmonic representation.
This work's central result, namely the \ashs{}, may be accessed via \texttt{positive.calc\_adjoint\_slm\_subset}. 
%
\subsection{Notation \& Conventions}
%
\par We will at times adopt slightly different {notations} for convenience and brevity, and we will at times bypass mathematically rigorous definitions, language, and structure with the intent of expediting access to physical concepts. 
Proofs of various key ideas will be left to references~\cite{Christensen2003,brauer1964,Courant1954}.
We will work under geometrized units $G=c=1$ with $M=1$.
\par It will very often be useful to discuss different kinds of functions, e.g.\ different kinds of spheroidal harmonics. In each case, by ``kinds'', ``vector space'', or ``set'', we mean ordered sets of complex valued square-integrable functions which we may treat as abstract vectors.
\par   Outside of introductory sections we will drop the spin weight labels from the harmonics; for example, spheroidal harmonics ${_{s}S_{\lmn}}$ will be denoted $S_{\lmn}$. 
We will denote the spherical harmonics, ${_{s}}Y_\lm$, as $Y_\lm$.
While we will only be concerned with outgoing gravitational radiation (i.e.\ spin weight $-2$) most aspects of our discussion apply to all spin weights. 
In discussion where both spherical and spheroidal harmonics are relevant, we will denote spherical harmonic indices with an overbar.
We will be centrally concerned with the $\theta$ dependence of each harmonic; thus, $Y_{{\ell}{m}}$ and $S_{\ell m n}$ will refer to $Y_{{\ell} {m}}(\theta)$ and $S_{\ell m}(\theta;\gamma_{\ell m n})$, where $\gamma_{\lmn}$ is the \qnm{}'s \textit{oblateness}.
\par In some cases, we will consider the oblateness to not depend on $\ell$ and $n$. There, to emphasize the difference between the fixed-oblateness spheroidal harmonics and the physical ones, we will refer to the fixed-oblateness spheroidal harmonics as $Z_{\lm}(\theta;\gamma)$.
\par The reader should note that in both spherical and spheroidal settings, axisymmetry means that $\phi$ dependence of radiation is $e^{im\phi}$. 
Since $e^{im\phi}$ are orthogonal in $m$, any radiation may be decomposed into moments with like $m$.
Thus we will exclusively work in settings where where $m$ is fixed.
\par Regarding the spheroidal oblateness, we will let $a$ denote the \bh{} spin magnitude per unit mass, $a=J/M$, and $\cw_\lmn$ denote the complex valued \qnm{} frequency. This allows the oblateness to be defined as
\begin{align}
	\label{oblateness}
	\gamma_{\lmn} \; = \; a \, \cw_{\lmn} \; ,
\end{align}
where $n$ is an overtone label~\cite{Berti:2005ys,Nollert:1999ji,Andersson:1996cm}.
There will be special cases in which multiple overtones are irrelevant; in these cases the overtone label will be dropped, and the oblateness will simply be denoted as $\gamma$ (i.e.\ it need not be interpreted according to \eqn{oblateness}).
Related spheroidal harmonics will be written as $S_\lm$ or $S_\lm(\theta;\gamma)$.
\par Sums over indices will always be between some lower bound and infinity unless otherwise stated. For the spherical polar indices, we will use the usual bounds: $\max(|m|,|s|) \le \l < \infty$ and $-\l \le m \le \l$.
\par Bra-ket notation, $\braket{\cdot}{\cdot}$, will frequently be adopted as short-hand for the scalar product. 
We will use a standard polar inner-product, where the one dimensional integral is performed over $u=\cos(\theta)$, 
\begin{align}
	\label{prod}
	\braket{p}{q} \; = \; \int_{-1}^{1} \, p(u)^* \, q(u) \, \mathrm{d}u \; .
\end{align}
In \eqn{prod}, $p(u)$ and $q(u)$ are square-integrable functions of $u$, and $p(u)^*$ denotes the complex conjugate of $p(u)$.
A spheroidal harmonic ket e.g.\ $\ket{S_\lmn}$ is effectively short-hand for $S_\lm(\theta;\gamma_\lmn)$, except in the setting of the scalar product.
All harmonics are normalized with respect to \eqn{prod} unless otherwise stated. 
\par Lastly, we will only discuss sets of functions with like azimuthal index $m$, and spin weight $s$~\cite{NP66}. The identity operator, $\I$, will specifically refer to the space spanned by such functions. 
%
\subsection{Outline of the Problem}
\label{out}
%
General \gw{} signals can be represented in terms of spin-weight $-2$ spherical harmonic multipole moments, but here we wonder if another, perhaps more physical route is possible.
If we denote an arbitrary \gw{} signal (i.e.\ strain~\cite{Blanchet:2013haa}) as $h(r,t,\theta,\phi)$, then its spherical harmonic expansion is 
\begin{align}
	\label{hy0}
	h(r,t,\theta,\phi) \; = \; \frac{1}{r} \; \sum_{\l,m} \, h^Y_{\l m}(t) \, {_{-2}}Y_{\l m}(\theta) \, e^{i m \phi} \; . 
\end{align}
In \eqn{hy0}, $r$ is the radiation's luminosity distance, $\theta$ and $\phi$ are polar and azimuthal angles describing an observer's orientation with respect to a source centered frame, and $h^Y_{\l m}(t)$ is the signal's spherical harmonic multipole moment~\cite{Ruiz:2007yx,Blanchet:2013haa,London:2017bcn}.
\par Here we will interpret the natural starting point for \eqn{hy0} to be that the spherical harmonics are naturally related to the \qnm{s} of non-spinning (spherically symmetric) spacetimes~\cite{Ruiz:2007yx,leaver85}.  
In that context, $h^Y_{\l m}(t)$ is naturally a sum over possible overtone contributions~\cite{Andersson:1996cm,leaver85}. 
Each overtone \qnm{} ultimately originates from the radial part of the linearized \ee{}, and the physical situation's initial data determines how much each overtone is excited~\cite{leaver85,London:2014cma,Andersson:1996cm}.
For general \gw{} signals, $h^Y_{\l m}(t)$ may be understood to encode information about the structure and dynamics of the spacetime,  including the source~\cite{Blanchet:2013haa,Thorne:1980}. 
Despite only corresponding to the modes of spherically symmetric spacetimes, it is well known (e.g.\ from Sturm-Liouville theory) that the spin-weighted spherical harmonics are complete, orthogonal, and thereby readily applicable to general \gw{} signals~\cite{Courant1954}. 
\par Here, our primary goal is to understand whether the spheroidal harmonics are also applicable to general \gw{} signals, thereby justifying a spheroidal harmonic multipole moment expansion of the form
\begin{align}
	\label{hs0}
	h(r,t,\theta,\phi) \; = \; \frac{1}{r} \; \sum_{\ell,m} \, h^S_{\ell m}(t) \, {_{-2}}S_{\ell m}(\theta; \gamma_{\ell m } ) \, e^{i m \phi} \; . 
\end{align}
In \eqn{hs0}, $h^S_{\ell m}(t)$ are \textit{spheroidal} harmonic multipole moments, and $\gamma_{\ell m}$ are their closely related oblateness parameters.
Note that, like in the case of perturbed spherically symmetric spacetimes, for perturbed Kerr \bh{s}, each $h^S_{\ell m}(t)$ may contain information about multiple overtone modes.
\par In the present work we seek to understand whether \eqn{hs0} is physically well motivated and mathematically well defined. 
In Paper II we seek to understand whether the relationship $h^S_{\ell m}$ and spacetime modes (e.g.\ \qnm{s}) makes them useful tools for \gw{} astronomy~\cite{London:2021P2}.
\par Here, we will work through the following technical questions: 
\begin{enumerate}
	\item[\textit{i}.] It is well known that the spheroidal harmonics depend on an {oblateness} parameter~(\ceqn{oblateness}). In this way, each spheroidal harmonic with polar and azimuthal quantum numbers $\ell$ and $m$ also depends on \textit{additional} information: the spacetime angular momentum, and the mode frequency which encodes information about the spacetime's radial structure. What consequences does this additional information have for how we must think about the differential equations which define the \qnm{s'} spheroidal harmonics?
	\item[\textit{ii}.] Can the spheroidal harmonics be used to exactly represent arbitrary \gw{} signals, particularly during e.g.\ \bbh{} post-merger, but also during merger and inspiral? Equivalently, are the \qnm{s'} spheroidal harmonics \textit{complete}?
	\item[\textit{iii}.] Do the \qnm{s'} spheroidal harmonics possess a kind of orthogonality?
\end{enumerate}
In forthcoming sections our task is to answer each of these questions.
\par Along the way we will find that several ideas are intertwined.
Whether the spheroidal harmonics are complete is inseparable from how one defines \eqn{hs0}'s oblateness parameters, $\gamma_{\ell m}$.
Completeness of the spheroidal harmonics is sufficient to justify the existence of the \ashs{},
and the spheroidal harmonic multipole moments are well defined when the \ashs{} are themselves well defined.
This work concludes with a discussion of the adjoint-function's application in the representation of \gw{}s. \bh{} ringdown is chosen as a simple and concrete setting for this discussion.
The application of adjoint spheroidal functions to \grad{} from the inspiral, merger, and ringdown of extreme and comparable mass ratio \bbh{s} is the subject of Paper II~\cite{London:2021P2}. 
\par In \sec{sec:ManyOps} we address question (\textit{i}), for which a pedagogical review of the spheroidal harmonic differential equation is useful.
In short, the differential operator for which the spheroidal harmonics, or simply ``spheroidals'', are eigenfunctions can be shown to result from \ee{} linearized about the Kerr solution (i.e.\ \tk{'s} equations)~\cite{Teukolsky:1973ha,leaver85}. 
For the \qnm{s}, each spheroidal harmonic differential operator depends on the mode's oblateness, $\gamma_\lmn = a \, \cw_\lmn$, according to 
\begin{align}
	\label{LSa}
	\mcL_{\lmn} = V_S(u,\gamma_{\lmn}) + \partial_{u}(1-u^2)\partial_{u} \; ,
\end{align}
where, $u = \cos{\theta}$, and the operator's potential is 
\begin{align}
	\label{LSb}
	V_S(u,\gamma_{\lmn}) \; = \; s+ u \gamma_{\lmn} (u\gamma_{\lmn}-2s)-\frac{(m'+su)^2}{1-u^2} \;.
\end{align}
In \eqn{LSb}, $s$ is the field's spin weight, and $m'$ is the equation's analog of the associated Legendre index.
\par Since each \qnm{} corresponds to a different oblateness, each physical spheroidal harmonic is the eigenfunction of a \textit{different} differential operator.
In turn, each operator is of the associated Legendre type, with a potential given by \eqn{LSb}. 
Therefore each operator has its own set of eigenfunctions which we may label in $\lp$ and $m'$.
Our primary interest will be in the solutions for which $\lp=\ell$ and $m'=m$.
These are the \textit{physical spheroidal harmonics} relevant to \gw{} theory and experiment~\cite{leaver85,London:2014cma,Bhagwat:2017tkm,Berti:2005ys}. 
We will at times simply refer to these harmonics as the ``physical spheroidals''.
As there is an infinity of such harmonics, we are ostensibly faced with an infinity of related different differential operators.
\par This technical aspect of the \qnm{s} does not appear to have been investigated previously, thus in \sec{sec:ManyOps} we introduce conceptual tools (ideas and notation) that will help us navigate this ``issue of many operators'' and its related situations.
These tools draw upon ideas from functional analysis and quantum mechanics~\cite{Christensen2003,Mostafazadeh:2001jk}. 
\par In \sec{sec:CB}, we use these tools to address questions (\textit{ii}) and (\textit{iii}).
We will show that subsets of the physical spheroidal harmonics with fixed overtone label can support bi-orthogonality with the \ashs{}, and that related subsets can be complete. 
If we denote the \ass{} as $\tilde{S}_{\ell m}(\theta;\gamma_\lmn)$, then when we refer to them as the bi-orthogonal dual of the spheroidal harmonics, we simply mean that
\begin{align}
	\label{SOa}
	\int_{0}^{\pi}\, \tilde{S}_{\ell m}(\theta;\gamma_{\lmn}) \, S^*_{\lpm}(\theta;\gamma_{\lpmn}) \, \sin(\theta) \, \mathrm{d}\theta \; \propto \; \delta_{\ell \ell'} \; .
\end{align}
These conclusions are supported by two key ideas from functional analysis. 
The first is that physical spheroidal harmonics with fixed overtone label form a \textit{minimal} set, meaning that any one spheroidal harmonic cannot be exactly represented by an \textit{infinite} sum over the others~\cite{Christensen2003}.
The second key idea is that a set of functions can have a bi-orthogonal dual if and only if it is minimal~\cite{Christensen2003}.
The goal of \sec{sec:CB} is to to discuss each of these ideas for the  spheroidal harmonics of Kerr \qnm{s}.
\par \Sec{sec:CB} is the most technical section of this work. It is organized into two subsections. 
\Sec{subsec:Subsets} combines ideas presented in \sec{sec:ManyOps} with old and new perturbation theory results to show that the overtone solutions of Kerr are linearly independent, but \textit{not} minimal.
As will be described in \sec{subsec:Subsets}, this means that from the perspective of the angular harmonics, \qnm{s} with the same values of $\ell$ and $m$, but different overtone labels, $n$, do not encode distinct mode information. 
This is exactly the situation that one should expect from the Schwarzschild \qnm{s}~\cite{leaver85}.
The key corollary of this conclusion is that fixed overtone subsets \textit{are} minimal. They are therefore a simple and useful way to organize mode information.
\par Given that the physical spheroidal harmonics on a fixed overtone subset are minimal, the existence of the related \ashs{} is assured~\cite{Christensen2003}.  
While proof of this fact may be found in Ref.~\cite{Christensen2003}, the end of \sec{subsec:Subsets} provides a brief conceptual overview. 
The reader should note that the full proof does not immediately facilitate calculation of the \ass, but it does allow us to draw conclusions from their existence. 
\par In particular, existence of the \ashs{} allows the construction of a unique linear map (an isomorphism) between spherical and spheroidal harmonics. 
Once again drawing from results in functional analysis (e.g.~\cite{brauer1964}), \sec{subsec:YSMaps} shows that the uniqueness of this ``spherical-to-spheroidal'' map means that the spheroidal harmonics are complete~\cite{Christensen2003}.
Like the \ashs{}, the existence of a spherical-to-spheroidal map is assured by the properties of the spheroidal harmonics, but not in a way that immediately lends to calculation. 
However, for concreteness, \sec{subsec:YSMaps} illustrates a way to explicitly define spherical-to-spheroidal maps using the standard spherical harmonic expansion along with related raising and lowering operators defined in Ref.~\cite{Shah:2015sva}.
\par \Sec{sec:Calc} goes a step further by showing that spherical-to-spheroidal maps may be expressed as infinite dimensional matrices of inner-products. 
The truncation of these matrices enables practical non-perturbative calculation of the \ashs{}. 
\Sec{sec:Calc} presents an algorithm to this end. That algorithm is the central result of this work.
Numerical examples are provided for the dominant $m=2$ Kerr spheroidal harmonics.
\par \Sec{sec:Decomp} provides a pedagogical discussion of what spheroidal harmonic decomposition might look like in practice. 
This section completes our discussion of the potential role of overtones within spheroidal harmonic decomposition~(\ceqn{hs0}). 
In particular, a quantitative comparison of mode-mixing between spherical and spheroidal representations closes our discussion. 
\par \Sec{sec:Discuss} summarizes our work thus far and points the way to future development.
\par \Apx{apx:mixing} provides a perturbation theory derivation of spherical-spheroidal mixing coefficients that provides leading order estimates at arbitrary perturbative orders.
\Apx{apx:many} revisits the issue of many operators by introducing two new operators of relevance to the physical spheroidal harmonics.
The first operator is one for which all physical spheroidal harmonics are eigenfunctions.
The second is an operator for which all \ashs{} are eigenfunctions. 
While this second operator is simply the adjoint of the first, its introduction helps illuminate the role of isospectrality and operator inter-winding in the \ashs{}.  
%
\section{The Issue of Many Operators}
\label{sec:ManyOps}
%
Standard arguments for orthogonality and completeness assume that a \textit{single} operator defines the space of interest.
This is true e.g.\ for the spherical harmonics.
But this is not true for the physical spheroidal harmonics, with their oblateness values that depend on $\ell$.
Here we briefly work through standard arguments for orthogonality and completeness in the context of a simple kind of spheroidal harmonic wherein all oblateness values are the same for different values of $\ell$ and $m$ (i.e.\ not the physical spheroidal harmonics).
This \textit{special case} will allow us to briefly review the concepts of orthogonality and bi-orthogonality.
We then show why the standard arguments underpinning these concepts do not trivially generalize to the physical spheroidal harmonics.
We conclude with a summary of the ideas necessary to generalize the standard arguments to the physical spheroidals.
\par We will use bra-ket notation to facilitate various manipulations.
We will also use different symbols to distinguish the spheroidal harmonics with fixed oblateness from the physical spheroidal harmonics.
Thusly we will denote the spheroidal harmonics with fixed oblateness as $Z_{\ell m}(\theta;\gamma)$, and the associated kets as $\ket{Z_\lm}$.
Similarly, we will denote the physical spheroidal harmonics as $S_{\ell m}(\theta;\gamma_\lmn)$, and the associated kets as $\ket{S_\lmn}$.
The reader should note that, as discussed in the context of \eqn{LSb}, the fixed-oblateness spheroidals are related to the physical spheroidals when the oblateness varies with $\ell$ (and/or overtone label $n$) according to
\begin{align}
	\label{ZS0}
	{S_\lmn}(\theta,\gamma_\lmn) \; = \; {Z_\lm}(\theta,\gamma)|_{\gamma=\gamma_\lmn} \; .
\end{align}
\Eqn{ZS0} communicates that while each $Z_\lm$ and $S_\lmn$ are closely related, their key difference is whether they are members of a sequence of harmonics in which the oblateness parameter varies between different harmonics in the sequence.
\par Our current aim is to describe select properties of the fixed-oblateness spheroidals, and by doing so highlight key aspects of physical spheroidals.
In this setting, the spheroidal harmonic differential operator is 
\begin{align}
	\label{LSc}
	\mclo \; = \; \left( s+u\gamma(u\gamma-2s)-\frac{(m+su)^2}{1-u^2} \right) + \partial_{u}(1-u^2)\partial_{u} \; .
\end{align}
The fixed-oblateness spheroidal harmonic kets, $\ket{Z_\lm}$, are then eigenvectors of $\mclo$ with eigenvalues $-A_\lm$,
\begin{align}
	\label{LSd}
	\mclo \, \ket{Z_\lm} \; = \; - A_\lm \,  \ket{Z_\lm} \; .
\end{align}
In the context of perturbed Kerr \bh{}s, $A_\lm$ is the separation constant for \tk{}'s master equation~\cite{Teukolsky:1973ha}.
\par We are now prepared to demonstrate how the properties of $\mclo$ provide information about the completeness and orthogonality (or as we shall see bi-orthogonality) of the fixed-oblateness spheroidal harmonics.
Pedagogical arguments to this end for e.g.\ the spherical harmonics begin by determining the adjoint of their differential operator, and then analyzing the matrix elements of that operator in a spherical harmonic basis. 
Here we may proceed in the same manner, but we must take extra care, as $\gamma$ can be complex valued ~(e.g.\ \ceqn{LSa}).
As a result, the operator adjoint, $\adj{\mclo}$, as defined by 
\begin{align}
	\label{adjoint-condition}
	\brak{ p }{ \mclo \, | \, q } \; = \; \brak{ p }{ \mclo \, q } \; = \; \brak{ \, \adj{\mclo} p }{ q } \; ,
\end{align}
can be shown (via integration by parts) to simply be 
\begin{align}
	\label{ajdL}
	\adj{\mclo} \; = \; \mclo ^* \; .
\end{align}
The first equality of \eqn{adjoint-condition} simply communicates that $\mclo$ acting on an arbitrary ket $\ket{q}$ results in a new ket, $\ket{\mclo \, q}$.
\Eqn{ajdL} is a slight departure from Sturm-Liouville theory which, if $\gamma$ is real, simply yields that $\adj{\mclo} = \mclo \; $ (i.e.\ if $\gamma$ is real, then $\mclo$ is self-adjoint)~\cite{Courant1954}.
Equations (\ref{ajdL}) and (\ref{LSd}) can be used to show that 
\begin{align} 
	\label{LSe}
	\adj\mclo \, \ket{Z^*_\lm} \; = \; - A^*_\lm \,  \ket{Z^*_\lm} \; ,
\end{align}
meaning that the eigenvectors of $\adj{\mclo}$ are simply conjugates of the spheroidal harmonics.
\par We are now prepared to write matrix elements of $\mclo$ in a vector space for which rows are spanned by eigenvectors of $\adj\mclo$, and columns are spanned by eigenvectors of $\mclo$.
Doing so yields two equivalent expressions:
\begin{align}
	\label{elementLa}
	\brak{ \, Z^*_{\lpm} }{ \mclo \, | \, Z_{\lm} } &= \brak{ \, Z^*_{\lpm} }{ \mclo \,  Z_{\lm} }  =  -A_\lm \, \brak{ \, Z^*_{\lpm} }{ Z_{\lm} }
\end{align}
and
\begin{align}
	\label{elementLb}
	\brak{ \, Z^*_{\lpm} }{ \mclo \, | \, Z_{\lm} } &=  \brak{ \adj\mclo \, Z^*_{\lpm} }{   Z_{\lm} }  =  -A_\lpm \, \brak{ \, Z^*_{\lpm} }{ Z_{\lm} } \; .
\end{align}
\Eqn{elementLa} uses the eigenvalue relation for $\mclo$, while \eqn{elementLb} uses the definition of the adjoint~(\ceqn{adjoint-condition}) and the eigenvalue relation for $\adj\mclo$.  
Since \eqns{elementLa}{elementLb} represent the same quantity in two different ways, their difference must be zero.
Subtracting \eqn{elementLa} from \eqn{elementLb} yields 
\begin{align}
	\label{elementLc}
	( \, A_\lpm-A_\lm \, ) \;  \brak{ \, Z^*_{\lpm} }{ Z_{\lm} } \; = \; 0 \; .
\end{align}
For $\l' \neq \l$, \eqn{elementLc} communicates that $\brak{ \, Z^*_{\lpm} }{ Z_{\lm} } \; = \; 0 \;$ or, equivalently,
\begin{align}
	\label{elementLd}
	\brak{ \, Z^*_{\lpm} }{ Z_{\lm} } \; \propto \; \delta_{\l' \l} \; .
\end{align}
\par There are many lessons to be learned from this simple example. 
Broadly, these lessons can help us distinguish between orthogonality, bi-orthogonality, and the relevance of many operators for the physical spheroidal harmonics. 
These lessons are prerequisites for our ultimately understanding the physical spheroidal harmonics' bi-orthogonality and completeness.
\par One lesson pertains to the case of zero oblateness. 
There, $\gamma=0$, and \eqn{LSc} can be used to show that $\mclo$ reduces to the spherical harmonic differential operator. 
In that setting \eqn{elementLd} reduces to the known fact that the spin-weighted spherical harmonics are orthogonal in $\l$~\cite{NP66}.
%
%
\par Another lesson pertains to cases where the oblateness is complex valued. 
In that case, \eqn{LSc} means that the spheroidal harmonics with fixed oblateness are not orthogonal with themselves, but rather with their complex conjugates. 
Because two sets of functions are needed, \eqn{elementLd} is a statement of \textit{bi}orthogonality~\cite{brauer1964,Christensen2003,Courant1954}.
Thus it is said that the \textit{conjugate spheroidal harmonics}, $Z^*_\lm$, are bi-orthogonal duals of $Z_\lm$. In particular, we note that \eqn{elementLd} differs from the usual statement of orthogonality due to the presence of $Z^*_\lm$ rather than $Z_\lm$ in its ket. 
In \sec{subsec:Subsets}, we will begin to generalize this simple kind of bi-orthogonality to the physical spheroidal harmonics.
While \eqn{elementLd} is a simple result that has been known to functional analysis for some time~(e.g.\ Refs.~\cite{brauer1964} and \cite{Courant1954}), this appears to be the first time it has been pointed out in the context of spheroidal harmonics relevant to \bh{s}. 
\smallskip
\par We now turn to the key lesson of \eqns{elementLa}{elementLd}, and the motivating issue of this section.
Our previous conclusions of orthogonality or bi-orthogonality hinge upon the fact that $\mclo$ does not depend explicitly on $\l$.
It is this point that allows us to calculate matrix elements in \ceqnsa{elementLa}{elementLd}.
To examine this point, we may begin to consider the case of the physical spheroidal harmonics and their operators, $\mcL_\lmn$~(i.e.\ \ceqns{LSa}{LSb}).
\par Following \eqns{elementLa}{elementLd}, we may attempt to write down the matrix elements of $\mcL_\lmn$ using the physical spheroidal harmonics and their conjugates.
As with the fixed-oblateness spheroidals, this yields two equivalent expressions:
\begin{align}
	\label{elementLe}
	\brak{ \, S^*_{\lpmn} }{ \mcL_\lmn \, | \, S_{\lmn} } &= \brak{ \, S^*_{\lpmn} }{ \mcL_\lmn \,  S_{\lmn} }  
	\\
	\label{elementLf}
	&=  -A_\lmn \, \brak{ \, S^*_{\lpmn} }{ S_{\lmn} }
\end{align}
and
\begin{align}
	\label{elementLg}
	\brak{ \, S^*_{\lpmn} }{ \mcL_\lmn \, | \, S_{\lmn} } &=  \brak{ \adj\mcL_\lmn \, S^*_{\lpmn} }{   S_{\lmn} }  
	\\
	\label{elementLh}
	&\neq  -A_\lpmn \, \brak{ \, S^*_{\lpmn} }{ S_{\lmn} } \; .
\end{align}
Since $\mcL_\lmn$ and $\ket{S_\lmn}$ have the same oblateness parameter, namely $\gamma_\lmn$, \eqnsa{elementLe}{elementLf} simply communicate that $\ket{S_\lmn}$ is an eigenket of $\mcL_\lmn$.
Thus far, this mirrors the result of our special case. 
\par However, since $\mcL_\lmn$ and $\ket{S_\lpmn}$ have different oblateness parameters, namely $\gamma_\lmn$ and $\gamma_\lpmn$, $\ket{S_\lpmn}$ is {not} an eigenvector of $\mcL_\lmn$.
Thus \eqn{elementLg} is not equal to \eqn{elementLh}.
This turn of events means that standard arguments for orthogonality and bi-orthogonality do not apply to the physical spheroidal harmonics.
\par One could also investigate the completeness properties of the spheroidal harmonics with fixed oblateness, $Z_\lm(\theta,\gamma)$.
Since their operator, $\mclo$, is of Sturm-Liouville form, there is a standard argument in functional analysis to show that $Z_\lm(\theta;\gamma)$ form a complete set:
The Sturm-Liouville form of $\mclo$ means that there exists an invertible operator, say $\mcto$, that transforms spherical harmonics into spheroidal harmonics. 
As maths go, one may prove that $\mcto$ exists without showing explicitly \textit{how} to calculate it~\cite{brauer1964,Christensen2003}.
Nevertheless, the existence of $\mcto$ can then be used to prove that $Z_\lm(\theta,\gamma)$ form a complete set.
Crucially, as with $\mclo$, a key assumption is that $\mcto$ is independent of $\ell$.
Thus this standard argument also falls short of applying to the physical spheroidal harmonics.
(This line of reasoning will be revisited in \sec{subsec:YSMaps}.)
\par With the breakdown of standard arguments comes questions. 
Given that we know the \qnm{} frequencies $\cw_\lmn$, and therefore the oblatenesses $\gamma_\lmn=a\cw_\lmn$, we therefore know all of the operators for which the physical spheroidal harmonics are eigenfunctions.
Does this not imply that we should be working within a framework that explicitly accounts for the spheroidal harmonics' multiple operators?
In such a framework, what form should the collection of spheroidal harmonic operators take?
Is it possible to use this framework to determine whether the physical spheroidals form a complete set?
Does this framework shed light on whether there exist functions that, along with the physical spheroidals, form a bi-orthogonal set?
%
%
%
\par Hints to each of these question may be found in \bh{} perturbation theory, quantum mechanics and functional analysis literature.
\par Single \bh{} perturbation theory provides algorithms for computing the physical spheroidal harmonics~\cite{leaver85,Cook:2014cta}.
In effect, these algorithms provide a computational definition of a single operator, $\mcl$, for which all physical spheroidals are eigenfunctions, i.e.
\begin{align}
	\label{mcl}
	\mcl \, \ket{S_\lmn} \; = \; -A_\lmn \, \ket{S_\lmn} \; \text{ for all }(\l,m,n).
\end{align}
But it appears that this knowledge has never been articulated mathematically rather than algorithmically. 
In particular, although we may algorithmically understand the action of $\mcl$, we cannot begin to determine whether it has an adjoint, $\adj{\mcl}$, without a better understanding of its precise mathematical form.
We revisit the form of $\mcl$ in \apx{apx:many}.
\par From quantum mechanics, Ref.~\cite{Mostafazadeh:2001jk} studies the properties of operators (Hamiltonians) whose eigenfunctions are bi-orthogonal. 
There it is useful to work with a vector space representation of operators (i.e.\ as is typically done in quantum mechanics with the identity operator).
\par Lastly, in Refs.~\cite{brauer1964,Christensen2003} (and many others), the existence of a bi-orthogonal dual and the completeness of the related vector space are discussed in the language of functional analysis.
In that setting it is known that a sequence of vectors has a bi-orthogonal dual if it is not only linearly independent, but also minimal (in the sense described in \sec{out})~\cite{Christensen2003}. 
\par Our current task is to apply these hints to questions about the physical spheroidal harmonics, and their many operators $\mcL_\lmn$.
%
\section{Spheroidal harmonic bi-orthogonality and completeness }
\label{sec:CB}
%
\par Here we work through two problems regarding the bi-orthogonality and completeness of the physical spheroidal harmonics. 
For concreteness, our discussion will center about the spheroidal harmonics of Kerr \qnms{}.
The first problem to be addressed has to do with \textit{how} we should conceptualize the \qnm{s'} spheroidal harmonics.
The result of this discussion is in effect an existence proof of the \ashs{}. 
The second problem has to do with the completeness of the physical spheroidal harmonics. 
Given the existence of the \ashs{}, as well as the existence of an operator that maps spherical to spheroidal harmonics, this section concludes that the spheroidal harmonics are complete, and and therefore may be used to represent arbitrary \grad{}. 
\par Regarding how the spheroidal harmonics are conceptualized, it is well known that the \qnm{s} of Schwarzschild \bh{s} are naturally organized into spherical harmonic moments, where each may be a sum of overtones,
\begin{align}
	\label{hQNMs}
	h^{\mathrm{QNM}}_{\mathrm{Schwarzschild}}  =  \frac{1}{r}  \sum_{\l,m}  \left(  \sum_{n=0}^{\infty}  b_{\lmn} \, e^{- i \cw_\lmn t}  \right)  {_{-2}Y_{\lm}}(\theta) \, e^{i m \phi} \; .
\end{align} 
In \eqn{hQNMs}, $b_{\lmn}$ is a complex valued \qnm{} amplitude, and the net expression describes the \qnm{} part of \bh{} \rd{}~\cite{leaver85}. 
It may be seen in \eqn{hQNMs} that for Schwarzschild \bh{}s, every overtone mode labeled with $\ell$ and $m$ has exactly the \textit{same} angular ``shape'' given by ${_{-2}Y_{\lm}}(\theta)$.
\par The matter would seem to be very different for Kerr \qnm{s}.
The oblateness parameter's appearance in the spheroidal harmonic differential operator~(\ceqns{LSa}{LSb}) means that every overtone labeled with $\ell$ and $m$ is associated with a \textit{different} spheroidal harmonic, ${_{-2}S_\lm}(\theta;\gamma_\lmn)$.
Restricting our consideration to (exclusively) only pro- or retrograde Kerr \qnm{s} \cite{London:2018nxs,MaganaZertuche:2021syq} 
\begin{align}
	\label{hQNMk}
	h^{\mathrm{QNM}}_{\mathrm{Kerr}}  =  \frac{1}{r}  \sum_{\l,m}  \left(  \sum_{n=0}^{\infty}  b_{\lmn} \, e^{- i \cw_\lmn t}    {_{-2}S_{\lm}}(\theta;\gamma_\lmn) \right) \, e^{i m \phi} \; ,
\end{align} 
where we use the convention that
\begin{align}
	\label{cwconvention}
	\cw_{\l m n} \; = \; -\cw_{\l\,-m\,n}^* \; .
\end{align}
The simplified perspective of \eqn{hQNMk} is relevant to e.g.\ the post-mergers of non-precessing \bbh{s} (e.g.~\cite{London:2014cma,Kamaretsos:2011um,Husa:2015iqa,Khan:2015jqa}), and it has been shown to apply to a large variety of precessing systems~\cite{Hamilton:2021pkf}.
We will discuss the implications of that simplification in \sec{sec:Discuss}. 
For now, with \eqns{hQNMs}{hQNMk} in mind, the key matter of concern is the extent to which it is meaningful to think of different Kerr overtones as having \textit{different} angular shapes given by  ${_{-2}S_\lm}(\theta;\gamma_\lmn)$. 
\par In \sec{subsec:Subsets} we will show that the multiple overtones in \eqn{hQNMk} provide \textit{redundant} angular information.
This will be accomplished by an investigation of the harmonics' large-$\l$ behavior in three limits: zero oblateness, linear in oblateness, and general oblateness.
We will show that, to linear order in $\gamma_\lmn$, the spheroidal harmonics become \textit{identical} to the spherical harmonics as $\l \rightarrow \infty$, regardless of overtone number.
We will also see that this conclusion generalizes in a simple way to arbitrary values of $\gamma_\lmn$. 
In this sense we will see that the conceptual structure of \eqn{hQNMs} prevails -- it is not robust to think of the different overtones' spheroidal harmonics as being distinct from one another.  
\par It is then fair to wonder whether there exists an alternative representation of \eqn{hQNMk} that explicitly accounts for redundancy in the overtones' spheroidal harmonics.
In other words, is there a systematic way to organize the physical spheroidal harmonics into minimal (i.e.\ non-redundant) subsets?
While there are surely many mathematical answers to this question, \sec{subsec:Subsets} adopts an approach that is physically motivated:
The fundamental (i.e.\ $n=0$) \qnms{} are known to be the most excited, and persist in time domain signals for the longest duration~\cite{London:2014cma,leaver85,Berti:2016lat,Khan:2015jqa,Kamaretsos:2012bs}.
In physical scenarios, such as the nonlinear \bbh{} merger, where it may be possible for higher overtones to play a significant role, it is currently unclear whether \qnm{s} apply at all, given that the background spacetime is changing at its fastest rate in the entire coalescence, strongly implying that it is not stationary and thus not Kerr~\cite{Keitel:2016krm}. 
This reasoning motivates our consideration of what we will call fixed \textit{overtone subsets} of the spheroidal harmonics. 
In particular, numerical examples in \sec{subsec:Subsets} provide evidence that the set of $n=0$ spheroidal harmonics is minimal, and thus supports a bi-orthogonal dual.
\par The reader should note that when we refer to \ashs{} we specifically mean those defined on a single overtone subset, and that all numerical results pertain to the $n=0$ subset which is known to be both astrophysically relevant and spectrally stable~\cite{Jaramillo:2020tuu}.
\par \Sec{subsec:YSMaps} addresses the question of whether physical spheroidal harmonics may, in principle, be used to exactly represent general \gw{} signals. 
In this discussion we begin to address the issue of many operators by constructing a spherical to spheroidal map that is appropriate for the physical spheroidal harmonics. 
We apply this map to a standard argument for completeness, and show that the physical spheroidal harmonics with $n=0$ form a complete set.
\par While these discussions of bi-orthogonality and completeness rely on functions and operators that have been shown to simply exist without explicit definition, \sec{subsec:YSMaps} provides an approximate expression for the basic spherical to spheroidal map.  The reader may look to \sec{sec:Calc} for a non-perturbative definition of that map, and the \ashs{}.
%
\subsection{Minimal spheroidal harmonic subsets }
\label{subsec:Subsets}
%
\par We will now show that \textit{overtone subsets} are minimal, and therefore support the existence of the \ashs{}.
By overtone subsets, we mean sets of Kerr spheroidal harmonics where all members have the same overtone index $n$.
For example, all spheroidal harmonics with $n=0$ define the ``lowest'' or ``fundamental'' overtone subset. 
By minimal, we mean that 
\begin{align}
	\label{Ma}
	\ket{S_\lpmn} \neq \sum_{\l \neq \lp} \, c_{\l} \, \ket{S_\lmn},\; \text{for all possible }c_\l\text{ and }\lp \; .
\end{align}
\Eqn{Ma} expresses that we cannot equate any member of the overtone subset in terms of a linear combination of \textit{all} other members. 
While this may remind the reader of linear independence, it should be noted that linear independence strictly applies to sets of finite size, and so is not quite applicable here. 
In particular, since \eqn{Ma} sums over $\l$ through infinity, we must investigate the spheroidal harmonics in that limit.
Our goal is to determine wether a kind of linearly \textit{de}pendent behavior emerges asymptotically.
\par To proceed, it suffices to apply a standard argument for linear independence, and then consider the limit as $\l \rightarrow \infty$ in that context.
To show that a finite subset of harmonics is linearly independent, we may rely on a standard lesson from linear algebra: 
If the eigenvalues of an operator are unique, then that operator's eigenfunctions are linearly independent.
While there exists a standard proof for this statement (e.g.\ Ref.~\cite{axler2015linear}), the first half of this section provides a brief overview for transparency and convenience.
The latter half of this subsection provides a large-$\l$ analysis of the spheroidal harmonic eigenvalues, followed by a brief discussion of why minimal sets support bi-orthogonality.  
\par Towards the linear independence of a finite subset of harmonics, we begin in the spirit of contradiction:
we may assume that any two spheroidal harmonics, with labels $(\l,m,n)$ and $(\lp,m,n')$, are linearly \textit{dependent},
\begin{align}
	\label{LI1}
	c_\l \ket{S_\lmn} + c_{\lp} \ket{S_{\lpmnp}} = 0 \; .
\end{align}
We may then apply $\mcl$ from \eqn{mcl},
\begin{align}
	\label{LI2}
	A_\lmn \, c_\l \ket{S_\lmn} \;+\; A_{\lpmnp} \,c_{\lp} \ket{S_{\lpmnp}} = 0 \; .
\end{align}
We may also scale \eqn{LI1} by $A_{\lpmnp}$,
\begin{align}
	\label{LI3}
	A_{\lpmnp} \, c_\l \ket{S_\lmn} \;+\; A_{\lpmnp} \,c_{\lp} \ket{S_{\lpmnp}} = 0 \; .
\end{align}
Subtracting \eqn{LI2} from \eqn{LI3} gives
\begin{align}
	\label{LI4}
	(A_\lpmnp  - A_{\lmn}) \, c_\l \, \ket{S_\lmn} = 0 \;.
\end{align}
\par \Eqn{LI4} is a key pedagogical step towards our connecting linear dependence to eigenvalues.
The left-hand side of \eqn{LI4} can only be zero if $c_\l$ is zero, or $A_\lmn$ equals $A_{\lpmnp}$.
\Eqn{LI1} means that if $c_\l$ is zero, then $c_\lp$ must also be zero; i.e.\ only trivial linear dependence is possible if eigenvalues are distinct.
Thus, if $A_\lmn$ and $A_{\lpmnp}$ are distinct, then we must conclude that $c_\l$ and $c_\lp$ are zero, and so $\ket{S_\lmn}$ and $\ket{S_\lpmnp}$ are linearly \textit{independent}.
Whether applied to an arbitrary finite subset of spheroidal harmonics (e.g.\ one that includes multiple overtones), or specifically to an overtone subset, this argument extends to multiple spheroidals by induction.
\begin{figure*}[ht] 
	\begin{tabular}{cc}
		\includegraphics[width=0.48\textwidth]{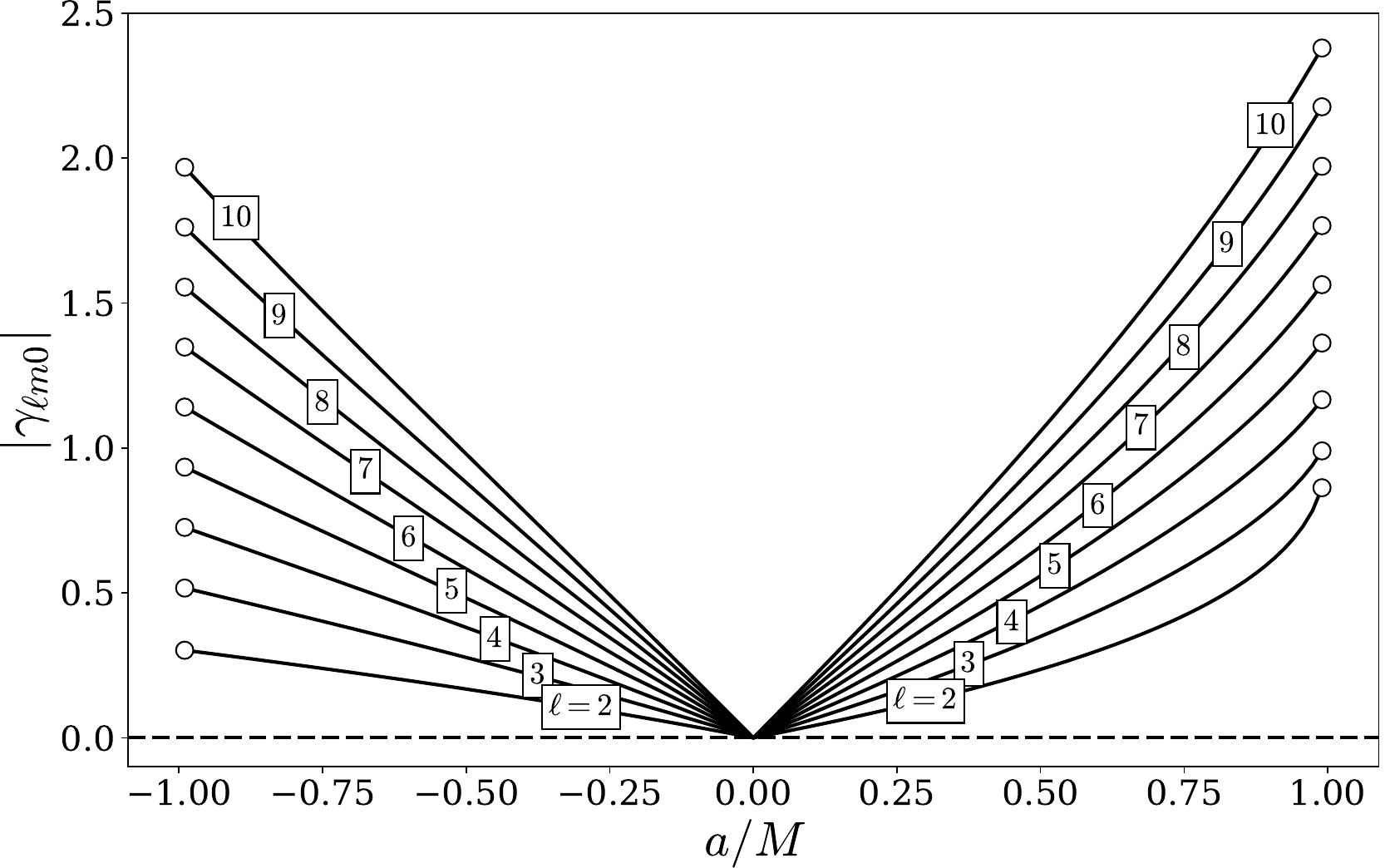} &
		\hspace{10pt}\includegraphics[width=0.48\textwidth]{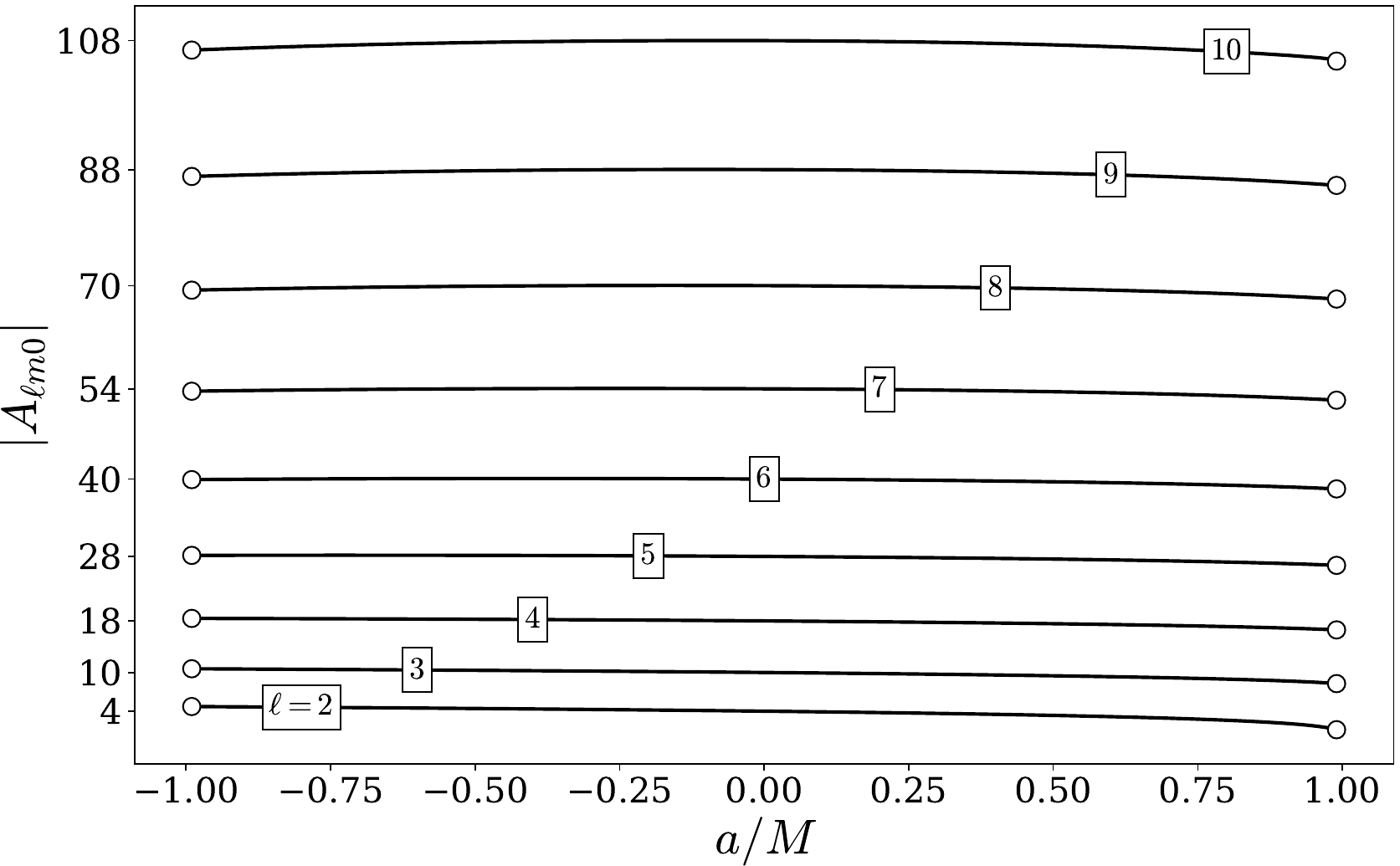}
	\end{tabular}
	\caption{
		Examples of how physical spheroidal harmonics have distinct oblatenesses for each $\ell$, and that the related eigenvalues are also distinct. \textit{Left:} The absolute value of oblatenesses for different Kerr \bh{} spins, $a = J/M$, for spin weight $s=-2$, azimuthal index $m=2$, and overtone number $n=0$, and $\ell$ from $2$ to $10$. The horizontal dashed line marks zero oblateness, where the spheroidal harmonics are equal to the sphericals. Open circles signify that extremal \bh{} spins are not shown. \textit{Right:} Eigenvalues for the left panel's physical spheroidal harmonics. The $y$-axis of this panel has tick-marks defined by the spherical harmonic eigenvalue, $\l (\l+1)-s (s+1)$. 
	}
	\label{scgamma} 
\end{figure*}
\smallskip
\par We now turn to the physical spheroidal harmonics' eigenvalues~( \ceqnsa{LSd}{mcl}).
Our aim is to apply the preceding argument of linear independence to spheroidal harmonics that differ in $\l$ and/or $n$.
If, for a fixed spin parameter $a$, each $A_\lmn$ is \textit{distinct} as $\l\rightarrow\infty$, then the related spheroidal harmonics constitute a set that is not only linearly independent, but also minimal.
\par The spheroidal harmonic eigenvalues may be calculated to perturbative orders in $\gamma_\lmn$ (e.g.\ ~\cite{Seidel:1988ue,Breuer1977,Press:1973zz,Berti:2005gp}), or numerically via e.g.\ Leaver's method~\cite{leaver85}.
We will first use a perturbative approximate to investigate general analytic properties, and then a numerical estimate particular to the $n=0$ spheroidals of Kerr \qnm{s}.
For the numerical check, Leaver's method will be used to compute the \qnm{} frequencies, $\cw_\lmn$, from the \bh{} spin parameter, $a$. 
From this, the oblateness values $\gamma_\lmn=a\cw_\lmn$ will be used to calculate $A_\lmn$ according to Leaver's algorithm~\cite{leaver85}. 
\par Using the results of Ref.~\cite{Seidel:1988ue} to expand $A_\lmn$ to second order in $\gamma_\lmn$ gives 
\begin{subequations}
\label{AlmnAll}
\begin{align}
	\label{Almn1}
	A_{\lmn} \; &= \; \l (\l+1)-s(s+1)  
	\\ 
	\label{Almn2}
	&\;- \; \gamma_\lmn  \, \frac{2 s^2 m}{\ell (\ell+1)} 
	\\ 
	\label{Almn3}
	&\; + \;  \gamma_\lmn^2 \, [ \;f(\l+1)-f(\l)-1 \;] \;+\;\mathcal{O}(\gamma_\lmn^3/\l^4)   \; ,
\end{align}
\end{subequations}
where 
\begin{align}
	\label{Almn4}
	f(\l) \; = \; \frac{\left(\l^2-\l_{\min }^2\right) \left(\l^2-s^2\right) \left(\l^2-\frac{m^2 s^2}{\l_{\min }^2}\right)}{2 \left(\l-\frac{1}{2}\right) \l^3 \left(\l+\frac{1}{2}\right)} \; ,
\end{align}
and $\l_{\min} = \max(|m|,|s|)$.  
\Eqn{Almn1} has been written to highlight the perturbative and large-$\l$ properties of $A_\lmn$.
The first line of \eqn{Almn1} is simply the spherical harmonic eigenvalue.
\Eqn{Almn2} shows the first order correction in $\gamma_\lmn$, and \eqn{Almn3} shows the 2nd order correction along with the order of the remainder. 
In \eqn{Almn3} we have taken care to note that the dominant part of the remainder is proportional to both $\gamma_\lmn^3$ and $\ell^{-4}$. 
Similarly, higher order corrections are inversely proportional to $\l$ at increasing powers~\cite{Seidel:1988ue}.
\par We may now use \eqns{Almn1}{Almn4} to inspect the large-$\l$ behavior $A_\lmn$. 
The {spherical} harmonic eigenvalues are distinct in $\l$ but degenerate in $n$, thus the same is true for the contribution shown in \eqn{Almn1}.
\Eqn{Almn2} and \eqn{Almn3}'s last term vanish as $\l\rightarrow\infty$, and an asymptotic expansion of \eqn{Almn3}'s $\gamma_\lmn^2$ term shows that its asymptote is $-\gamma_\lmn^2/2$.
In particular, it is not hard to show that as $\ell\rightarrow\infty$,
\begin{align}
    f(\ell + 1) - f(\ell) - 1 &\sim -\frac{1}{2} + \mathcal{O}(\ell^{-2})\;.
\end{align}
Together these points constrain the large-$\l$ (i.e.\ asymptotic) behavior of the eigenvalues,
%
\begin{align}
	\label{Almn5}
	A_\lmn \; \sim \; \l (\l+1)-s(s+1)  -  \gamma_\lmn^2/2 \; .
\end{align}
In \eqn{Almn5}, ``$\sim$'' denotes asymptotic equivalence.
\par As written, \eqn{Almn5} helps us inspect three limits: the zero-oblateness limit, the linear-in-oblateness limit, and the general oblateness limit where the $\l$ dependence of $\gamma_\lmn$ plays a key role. 
We will now briefly discuss the spheroidal eigenvalues in each of these contexts.
\par At $\gamma_\lmn=0$, \eqn{Almn5} communicates that the spheroidal eigenvalues are equal to the spherical ones. 
While this fact is also evident from \eqnsa{LSb}{AlmnAll}, we emphasize here that the overtone harmonics are neither linearly independent nor minimal for \textit{all physically relevant} oblatenesses.
\par We may also draw from \eqn{Almn5} that, at linear order in oblateness, the spheroidal harmonic eigenvalues are equal to the spherical harmonic ones (i.e.\ one takes $\gamma_\lmn^2$ to zero at linear order in $\gamma_\lmn$).
Thus it is not only that the spheroidal harmonics reduce to the spherical harmonics at zero oblateness, but also that in the small oblateness limit, large-$\l$ spheroidal harmonics have eigenvalues that are asymptotically equivalent to those of the spherical harmonics.
Since the spherical eigenvalues only depend on $l$ and $s$, the different overtones' spheroidal harmonic in $n$ are asymptotically equivalent.
\par Finally, for large but physical values of the oblateness parameter, \eqn{Almn5} helps us imagine scenarios where the $\l$ dependence of $\gamma_\lmn$ plays a central role.
For this we may draw from the fact that $\gamma_\lmn=a \, \cw_\lmn$, and that the \qnm{} frequencies, $\cw_\lmn$, are known (e.g.\ from Ref.~\cite{Yang:2012he}) to have the following large-$\l$ form
\begin{align}
	\label{LCW}
	\cw_\lmn \; \sim \; \l\, \omega_\mathrm{Orb} + i g_L (n+1/2) \; .
\end{align}
In \eqn{LCW}, $\omega_\mathrm{Orb}$ is the Keplerian orbital frequency for a circular photon orbit, and $g_L$ is related to the Lyapunov exponent of that orbit.
For simplicity, we have written \eqn{LCW} to only the dominant terms in $\l$.
It is important to note that both $\omega_\mathrm{Orb}$ and $g_L$ are geometric quantities, and are thus independent of $\l$.
It is also important to note that $n$ appears in \eqn{LCW} additively with respect to $\l$, meaning that for finite $n$, as $\l \rightarrow \infty$, $n$ becomes fractionally insignificant.
\par With \eqn{LCW} in hand, we may now consider the regime where $\l$ is large with respect to $n$ for $n \ge 0$, and the $\l$ dependence of $\gamma_\lmn$ dominates (i.e.\ where $\l \gg n$ ).
There, the $n$-dependent terms in \eqn{LCW} may be neglected, yielding
\begin{align}
	\label{LCW2}
	\cw_\lmn \; \sim \; \l\, \omega_\mathrm{Orb} \; .
\end{align}
When combined with \eqn{Almn5}, the asymptotic behavior \eqn{LCW2} allows us to further distill the large-$\l$ behavior of the spheroidal eigenvalues.
Keeping only the largest powers of $\l$, this yields
\begin{align}
	\label{Almn6}
	A_\lmn \; \sim \; \l^2 \, (1  -  a^2 \,  \omega_\mathrm{Orb}^2/2 \, ) \; .
\end{align}
Thus, in the large-$\l$ limit, and for large and physical values of oblateness, $A_\lmn$ asymptotically lose their dependence on the overtone index $n$.
In other words, the related overtones' spheroidal harmonics become (again)  asymptotically equivalent: 
for any two overtone harmonics with label $n$ and $n'$ (and like values of $s$ and $m$), there always exists some $\l \gg n$ and $\l \gg n'$, such that $1-|A_\lmn / A_{\lmn'}|$ is arbitrarily small.
\smallskip
\par From these three settings (zero-oblateness, linear-in-oblateness, and general oblateness), we may conclude that the standard argument for linear independence holds for different overtones if $\l$ is finite, but does not hold as $\l \rightarrow \infty$.
This is because different overtones with the same value of $\l$ have the same asymptotic (i.e. large-$\ell$) behavior. 
As a result, the full set of spheroidal harmonics, including all overtones, is \textit{not minimal} according to the infinite sum in \eqn{Ma}.
Conversely, and perhaps more importantly, we may also conclude that any subset of physical spheroidal harmonics for which every value of $\l$ uniquely labels one harmonic \textit{is minimal} according to \eqn{Ma}.
In what follows we take that the simplest and most physically relevant minimal subset is that with $n=0$ (i.e.\ the fundamental \qnm{}  subset\footnote{Some authors use $n=1$ to denote the fundamental overtones. We choose to not do this here.}).
\smallskip
\par \Fig{scgamma} graphically demonstrates that, for all allowed spin parameters $a$, the $n=0$ Kerr oblatenesses are distinct, and so are their related eigenvalues.
While it is known that values of $\ell$ between $2$ and $5$ are more than sufficient to accurately represent \grad{} for current ground based detectors~(e.g.\ \cite{London:2017bcn,Blackman:2017pcm,Blanchet:2013haa,Cotesta:2018fcv}), \fig{scgamma} shows up to $\l=10$ as might be relevant for future detectors.  
The left and right panels plot quantities with respect to the spin parameter, $a$.
We use the convention that $a<0$ corresponds to perturbations that are retrograde to the \bh{} spin direction ~\cite{Husa:2015iqa,London:2018nxs}.
By the convention used in e.g.\ Ref.~\cite{Berti:2014fga}, our $a<0$ corresponds to $m<0$.
In the left panel of \fig{scgamma}, the sharp feature about $a=0$ corresponds to $\gamma_\lmn$ passing through zero and being (approximately) negated.
The absolute value of $\gamma_\lmn$ is plotted; the sharp feature simply indicates reflection.
The underlying real and imaginary parts of $\gamma_\lmn$ are smooth functions of $a$~\cite{London:2018nxs,Husa:2015iqa}.
Outside of $|a|\approx 0$, it is clear that for the \qnm{s} shown, $|\gamma_\lmn|$ are distinct in $\l$.
\par The right panel of \fig{scgamma} shows the absolute value of spheroidal eigenvalues derived from $\gamma_\lmn$ in the right panel of \fig{scgamma} using Leaver's method~\cite{leaver85}.
The $y$-axis' tick marks are defined by the spherical harmonic eigenvalues.
By comparing the $y$-axis' tick-marks to the curves for each eigenvalue, it is clear that the spherical harmonic contribution to each eigenvalue dominates.
For each spin value shown, each $n=0$ eigenvalue is distinct, supporting the minimal nature of the fundamental \qnm{}  subset for all spins.
\smallskip
\par Having reviewed the minimal nature of overtone subsets, what remains is to understand the connection between a minimal subset and the existence of that subset's bi-orthogonal dual (i.e.\ the \ashs{}).
The claimed connection is that the physical spheroidal harmonics of overtone subsets, $\ket{S_\lmn}$ (where only $\l$ varies), have bi-orthogonal duals, $\ket{\tilde{S}_\lmn}$, \textit{if and only if} the overtone subset is minimal.  
Although we refer the reader to Ref.~\cite{Christensen2003} for the full proof of this statement, we conclude this section by outlining the proof's key ideas.
\par The claim has two assertions: (\textit{i}) that $\ket{\tilde{S}_\lmn}$ exist if $\ket{S_\lmn}$ are minimal, and (\textit{ii}) if $\ket{S_\lmn}$ are minimal, then $\ket{\tilde{S}_\lmn}$ exist. 
The proof in question must address each of these assertions separately. 
\par Assertion (\textit{i}) is perhaps the simplest to demonstrate, as the bi-orthogonality of $\ket{\tilde{S}_\lmn}$ and $\ket{S_\lmn}$ mean that 
\begin{align}
	\label{ORTH4}
	\brak{ S_\lmn }{ \tilde{S}_\lpmn } \; \propto \; \delta_{\l,\lp} \; .
\end{align}
This idea is then applied to the notion of whether any data that can be exactly represented in a linear combination of $\ket{S_\lmn}$ may also be exactly represented using $\ket{\tilde{S}_\lmn}$.
In particular, it can be shown that \eqn{ORTH4} means that $\ket{S_\lmn}$  cannot be represented as a linear combination of the remaining spheroidals with $\l \neq \lp$.
This statement is equivalent to \eqn{Ma}. 
Thus, if bi-orthogonal duals exist, then the related set is minimal.
\par The proof of assertion (\textit{ii}) is somewhat more technical. 
It can be shown that the right-hand-side of \eqn{Ma} is related to a projection operator, $\mathcal{U}_\l$, which projects a ket onto the space of all spheroidal harmonics with the exception of $\ket{S_\lmn}$.
Note that $\mathcal{U}_\l$ is labeled by the same value of $\ell$ as $\ket{S_\lmn}$.
By definition, $\mathcal{U}_\l$ is such that $\brak{S_\lpmn}{(\I-\mathcal{U}_\l) \, S_\lmn}\propto \delta_{\lp,\l}$, meaning that $\ket{ \tilde{S}_\lmn } \propto \ket{(\I-\mathcal{U}_\l) \, S_\lmn}$.
Thus if $\ket{S_\lmn}$ are minimal, then $\ket{ \tilde{S}_\lmn }$ exist.
%
\subsection{Maps between spherical and spheroidal harmonics}
\label{subsec:YSMaps}
%
\par The existence of the \ashs{} means that we may, in principle, decompose arbitrary \gw{} signals into spheroidal harmonics moments. 
In this, the right-hand-side of \eqn{hs0} is justified, and \sec{sec:Calc} will provide a method to calculate the adjoint harmonics.
For now, a key remaining issue is whether we may generally \textit{equate} any square-integrable \gw{} signal with its spheroidal harmonic decomposition.
In other words, the current topic of discussion is whether the physical spheroidal harmonics, and their adjoint-functions, are complete.
\par It is useful to frame this topic with a few pedagogical ideas:
The spin-weighted spherical harmonics are known to be complete because they are closely related to the trigonometric functions, and the trigonometric functions themselves are known to be complete in a rudimentary way (e.g.\ the Fourier series)~\cite{Courant1954,brauer1964}.
Completeness of the spherical harmonics means that any square-integrable \gw{} signal (ket), say $\ket{h}$, may be equated with,
\begin{align}
	\label{hs1}
	\ket{h} \; = \; \sum_{\l} \, \ket{Y_{\l m} }\brak{ Y_{\l m} }{h} \; .
\end{align} 
Equivalently, we may define an identity operator in terms of the spherical harmonics, 
\begin{align}
	\label{iy}
	\I \; = \; \sum_\l \, \, \ket{Y_{\l m}}\bra{ Y_{\l m} } \; ,
\end{align}
such that \eqn{hs1} can be compactly written as $\ket{h} = \I \, \ket{h}$.
In this language, our present task is to determine whether the identity operator in \eqn{iy} may be alternatively represented in terms of the physical spheroidal harmonics and their adjoint functions,
\begin{align}
	\label{M0}
	\I \; = \; \sum_{\l} \, \ket{ S_\lmn } \bra{ \tilde{S}_\lmn } \; .
\end{align}
In \eqn{M0}, the reader should note that $\I$ is specifically the identity operator for all functions (e.g.\ parts of \gw{} signals) corresponding to a fixed $m$ and $s$.
\par To proceed, we will first rely on a standard argument from functional analysis applied to the fixed-oblateness harmonics, $\ket{Z_{\l m}}$. 
This argument essentially says that the mean difference between the spherical and spheroidal harmonics is proportional to $1/\l$, and so the two harmonics are ``close'' and as so have related properties.
These related properties are defined by a linear operator that \textit{maps} between spherical and spheroidal harmonics.
In this section we will focus on how the existence of such a map means that the physical spheroidal harmonics inherit completeness from the sphericals.
The details of supporting arguments are left to the appendix.
\par In particular, that the spherical and spheroidal harmonics are close is shown in \apx{apx:mixing}.
There, perturbative and recursive methods are used to show that any spheroidal harmonic may be written as a sum of spherical harmonics,
\begin{align}
	\label{M1}
	\ket{Z_\lm} \; = \; c_{\l}\,\ket{Y_\lm} \; + \; c_{\l}\,\sum_{\lp \neq \l} \, \sigma_{\lp \l} \, \ket{Y_{\lp m}} \; , 
\end{align}
where $c_{\l}$ is a normalization constant, and
\begin{align}
	\label{M2}
	\sigma_{\lp \l} \; &= \; \brak{ Y_{\lp m} }{ Z_\lm } 
	\\
	\label{M3}
	&\approx \; \frac{1}{|\lp-\l|!} \, \left( \frac{-\gamma s}{ 2 \l }  \right)^{ |\lp-\l| } \; .
\end{align}
\Eqn{M1} is simply a spherical harmonic expansion.
\Eqn{M3} results from recognizing that perturbation theory has an inherently recursive structure which, in the case of the spherical-spheroidal inner-products, has an exact solution at leading order. 
The form of \eqn{M3} is key as it relates to the general convergence of \eqn{M1}, and (see e.g.\ Ref.~\cite{brauer1964}) the existence of an invertible operator, $\mcto$, that maps spherical to spherical harmonics,
\begin{align}
	\label{M4}
	\mcto \, \ket{ Y_{\l m} } \; = \; \ket{Z_{\l m}} \; .
\end{align}
It is the existence and invertibility of $\mcto$ that can be used to show that the spheroidal harmonics with fixed oblateness, $\gamma$, inherit the completeness of the spherical harmonics~\cite{brauer1964,Christensen2003}.
\par We are presently concerned with generalizing the arguments of e.g.\ Ref.~\cite{brauer1964} to the $n=0$ physical spheroidal harmonics to show that they too are complete, and can therefore exactly represent arbitrary \gw{} signals.
This will be done by defining an operator for the physical spheroidals, $\mct$, that is the generalization of $\mcto$.
Together $\mcto$, $\mct$, and their inverses, $\mcvo$ and $\mcv$, will play central roles in our showing that the spheroidals are complete, 
\begin{align}
	\label{is}
	\I \;= \; \sum_{\l} \, \ket{S_\lmn}\bra{\tilde{S}_\lmn} \, \; .
\end{align}
Per discussion in \sec{subsec:Subsets}, \eqn{is} is defined for fixed $n$. 
As implied by \eqn{is}, a key component of our discussion will be the fact that \ashs{} may exist on overtone subsets.
\smallskip
\par Let us begin. 
Given the closeness of the spherical and spheroidal harmonics (e.g.\ \ceqn{M3}), we may be assured that there exists a linear operator $\mcto$ that transforms spherical harmonics into spheroidals~\cite{Christensen2003,brauer1964}.
For now we need only know that such an operator exists.
Nevertheless, it may be worthwhile to briefly discuss its construction.
For example, it must be the case that $\mcto$ acts on $\ket{Y_\lm}$ to exactly the effect of \eqn{M1}'s right hand side. 
It is also true that spherical harmonics with different values of $\l$ are related by linear differential ``raising'' and ``lowering'' operators~\cite{Shah:2015sva}.
Combining these ideas allows us to think of $\mcto$ as a kind of differential operator.
If we let $\mathcal{P}_\l$ and $\mathcal{Q}_\l$ be the $\l$ raising and lowering operators (Eqns. 29 and 30 of ~Ref.\cite{Shah:2015sva}) then,
\begin{align}
	\label{M5}
	\mcto \; \approx \; c_\l \;(1\; + \; \sigma_{\l-1,\l} \,\mathcal{Q}_\l \; + \; \sigma_{\l+1,\l} \,\mathcal{P}_\l \,)\; ,
\end{align}
with
\begin{align}
	\label{M5.1}
	\mathcal{Q}_\l \, Y_\lm \; = \; Y_{\l-1,m} \; \text{  and } \;\;
	\mathcal{P}_\l \, Y_\lm \; = \; Y_{\l+1,m}\; .
\end{align} 
In \eqn{M5.1} the raising and lowering operators are of the form $c_0(\theta) + c_1(\theta)\,\partial_\theta$ \cite{Shah:2015sva}.
In \eqn{M5} we have only kept the adjacent spherical harmonic contributions for simplicity. 
We will see in \sec{sec:Calc} that it is much more useful to think of $\mcto$ as an infinite dimensional matrix whose elements are simply related to spherical-spheroidal inner products $\sigma_{\lp \l}$.  
\par {Given the existence of $\mcto$, our first task is to determine a generalization for the physical spheroidal harmonics: $\mcto$ transforms a spherical harmonic into a spheroidal harmonic with fixed oblateness (\ceqn{M5}).
Its generalization should transform a spherical harmonic into a physical spheroidal harmonic.
The development of this generalization begins} with the our recalling that each physical oblateness, $\gamma_\lmn$, defines a set of fixed oblateness harmonics.
Therefore there exists a sequence spherical-to-spheroidal maps that are essentially $\mcto$ but parameterized by the different physical oblateness, $\gamma_\lmn$. 
We will refer to each of these maps as $\mct_\lmn$, where
\begin{align}
	\label{M6}
	\mct_\lmn \, \ket{Y_\lm} \; = \; \ket{S_\lmn} \; .
\end{align}
Similarly, we will refer to related inverse maps as $\mcv_\lmn$, where
\begin{align}
	\label{M7}
	\mcv_\lmn \, \ket{S_\lmn} \; = \; \ket{Y_\lm} \; .
\end{align}
We are now tasked with determining whether there exists some generalization, say $\mct$ and $\mcv$, such that 
\begin{align}
	\label{M8}
	\mct \, \ket{Y_\lm} \;\, \; &= \; \ket{S_\lmn} \text{ for {all} } \l \; ,
	\\
	\label{M8b}
	\mcv \, \ket{S_\lmn} \; &= \; \ket{Y_\lm} \; \,\text{ for {all} } \l \; .
\end{align}
We may consider the following vector space representations
\begin{align}
	\label{M9}
	\mct \; &= \; \sum_{\lp} \,  \mct_\lpmn \, \ket{Y_\lpm} \bra{Y_\lpm} \; 
	\\
	\label{M10}
	&= \; \sum_{\lp} \, \ket{S_\lpmn} \bra{Y_\lpm} \; ,
\end{align}
and
\begin{align}
	\label{M9b}
	\mcv \; &= \; \sum_{\lp} \,  \mcv_\lpmn \, \ket{S_\lpmn} \bra{\tilde{S}_\lpmn} \; 
	\\
	\label{M10b}
	&= \; \sum_{\lp} \, \ket{Y_\lpm} \bra{\tilde{S}_\lpmn} \; .
\end{align}
\par In \eqn{M9}, we have explicitly written $\mct$ in terms of the objects we wish to generalize, $\mct_\lpmn$.
In \eqn{M10}, we apply $\mct_\lpmn$ to $\ket{Y_\lpm}$.
As represented in \eqnsa{M9}{M10}, $\mct$ may be easily shown to map spherical harmonics to physical spheroidal harmonics, as expected.
\par In \eqn{M9b}, we have also written $\mcv$ in terms of the objects we wish to generalize, $\mcv_\lpmn$.
In \eqn{M10b}, we apply the action of $\mcv_\lpmn$ on $\ket{S_\lpmn}$.
As with $\mcl$, $\mcv$ may be easily shown to have the correct behavior (i.e.\ \ceqn{M8}) when acting on spheroidal harmonics.
In this case, the correct behavior is assured by the existence and bi-orthogonality of the \ashs{}, $\tilde{S}_\lmn$.
\par What remains to be shown is whether $\mcv$ is a unique left and right inverse of $\mct$.
To this end, it suffices to evaluate $\mct \mcv$ and $\mcv \mct$.
From \eqnsa{M10}{M10b}, we have that
\begin{align}
	\label{M11}
	\mcv \mct \; &= \; \sum_{\lp,\l} \, \ket{Y_\lpm} \brak{\tilde{S}_\lpmn}{S_\lmn} \bra{Y_\lm} \; 
	\\
	\label{M12}
	&= \; \sum_{\l} \, \ket{Y_\lm} \bra{Y_\lm} \; = \; \I \; ,
\end{align}
and
\begin{align}
	\label{M11b}
	\mct \mcv \; &= \; \sum_{\lp,\l} \, \ket{S_\lmn} \brak{Y_\lm}{Y_\lpm} \bra{\tilde{S}_\lpmn} \; 
	\\
	\label{M12b}
	&= \; \sum_{\l} \, \ket{S_\lmn} \bra{\tilde{S}_\lmn} \; .
\end{align}
\par In \eqnsa{M11}{M11b} we have used the bi-orthogonality and orthogonality of the spheroidal and spherical harmonics. 
In \eqn{M12}, we simply find that $\mcv \mct$ is the identity operator represented in terms of spherical harmonic bras and kets.
However, it may not be immediately clear that \eqn{M12b} is this same identity operator represented with physical spheroidal harmonics.
To clarify the matter, it may help to consider that $\I^2=\I$, and so 
\begin{align}
	\label{M13}
	\I \; &= \; \mcv \mct \;  \mcv \mct \;  
	\\
	\label{M14}
	&= \; \mcv \left( \mct \mcv \right) \mct \; .
\end{align}
In \eqn{M13}, we have simply equated the identity operator with its square.
In \eqn{M14}, we have used the associative property of linear operators to group $\mct $ with $ \mcv$.
If $\mct \mcv $ is not $\I$, then we might use $\I$ to construct a contradiction:  $\mcv \, \I \, \mct \neq \mcv \left( \mct \mcv \right) \mct = \I$, which reduces to $\I \neq \I$.
Clearly, $\I=\I$, so we must conclude that $\mct \mcv = \I$, and equivalently
\begin{align}
	\label{M15}
	\I \; = \; \sum_{\l} \, \ket{S_\lmn} \bra{\tilde{S}_\lmn} \; .
\end{align}
An alternative but ultimately equivalent argument is that since each physical spheroidal harmonic within an overtone subset is uniquely associated with a single spherical harmonic (via $\mct_\lmn$), $\mct \mcv$ must transform to the same space as $\mcv \mct$. 
Thus if $\mcv \mct = \I$, then so must $\mct \mcv$~\cite{Christensen2003}.
\par In \eqn{M15}, we have essentially found that overtone subsets of the physical spheroidal harmonics are complete.
The ideas and arguments leading to this conclusion have a number of relevant reductions and alternative framings. 
For example, if we consider the fixed-oblateness spheroidals instead of the physical spheroidals, then it may be shown that \eqn{M15} reduces to 
\begin{align}
	\label{M16}
	\I \; = \; \sum_{\l} \, \ket{Z_\lm} \bra{{Z}^*_\lm} \; .
\end{align}
Further, since the identity operator is self-adjoint (i.e.\ $\I=\adj\I$), we may also conclude that (reversing the location of $\tilde{S}_\lmn$ and ${S}_\lmn$ yields)
\begin{align}
	\label{M17}
	\I \; = \; \sum_{\l} \, \ket{\tilde{S}_\lmn} \bra{S_\lmn} \; .
\end{align}
Similarly, one might redevelop \eqns{M9}{M15} with the adjoint operators, $\adj{\mct}$ and $\adj{\mcv}$.
These may be found by simply adjugating \eqnsa{M10}{M10b},
\begin{align}
	\label{M18}
	\adj{\mct} \; = \; \sum_{\lp} \, \ket{Y_\lpm} \bra{S_\lpmn} \; ,
	\\
	\label{M19}
	\adj{\mcv} \; = \; \sum_{\lp} \, \ket{\tilde{S}_\lpmn} \bra{Y_\lpm} \; .
\end{align}
From \eqns{M18}{M19} it is straightforward to use bi-orthogonality and orthogonality to show that  
\begin{align}
	\label{M20}
	\adj{\mct} \, \ket{\tilde{S}_\lmn} \; &= \; \ket{Y_\lm} \; \, \text{ for all } \l \; ,
	\\
	\label{M21}
	\adj{\mcv}  \, \ket{Y_\lm} \,\;\; &= \; \ket{\tilde{S}_\lmn} \text{ for all } \l \; .
\end{align}
\par With \eqns{M15}{M21} we have an abundance of conceptual tools to help us work with and think about the physical spheroidal harmonics.
Our next task is to use these tools to non-perturbatively calculate the adjoint spheroidal harmonics.
\begin{figure}[t] 
	\begin{tabular}{c}
		\includegraphics[width=0.44\textwidth]{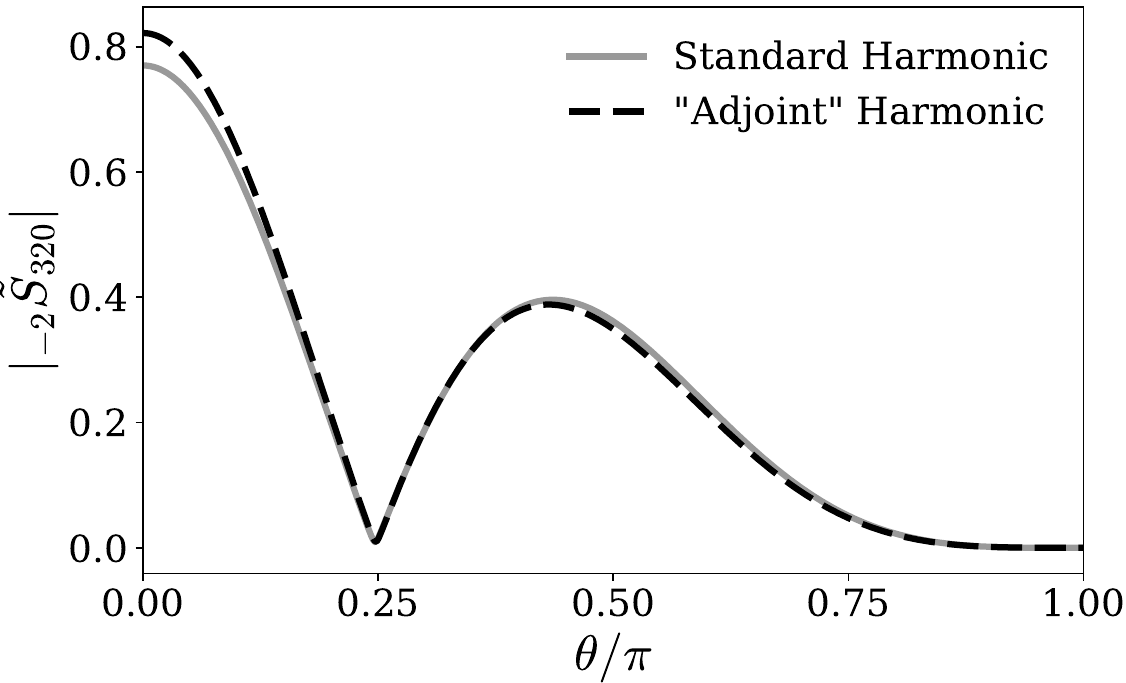} 
		\\
		\includegraphics[width=0.44\textwidth]{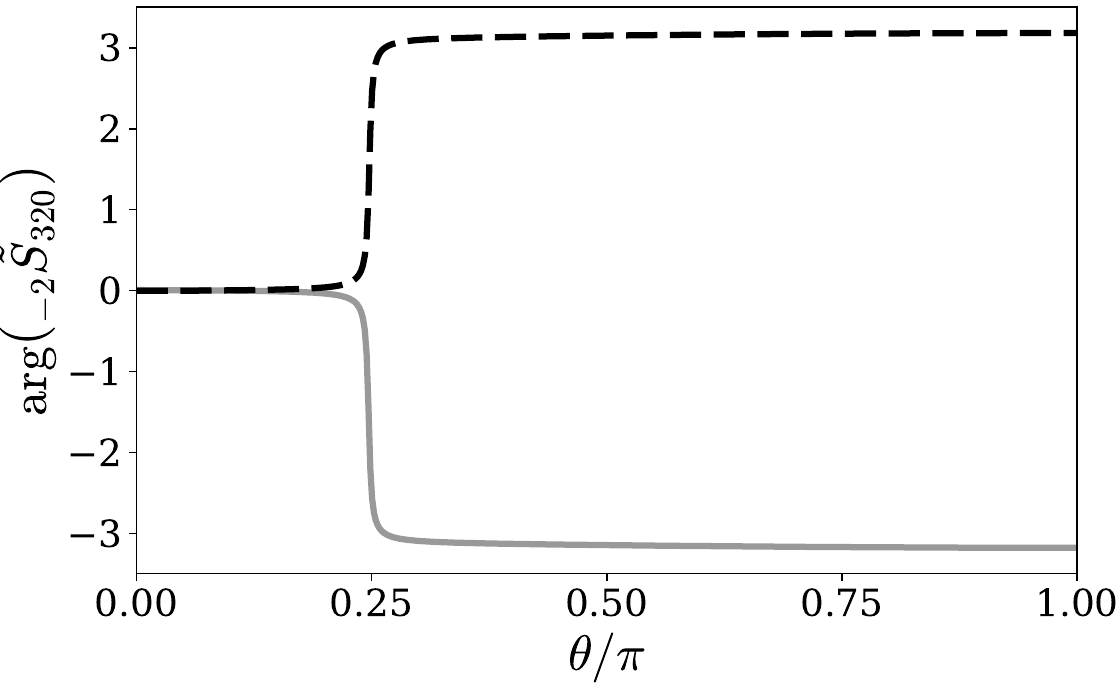}
	\end{tabular}
	\caption{
		The same as \fig{fig:new_harmonics} but for $(\l,m,n)=(3,2,0)$.
		Examples of this work's central result for spin weight $-2$, and Kerr spin parameter of $a=0.7$: 
	\textit{Top}: a comparison of harmonic amplitudes for $(\l,m,n)=(3,2,0)$.
	\textit{Bottom}: a comparison of harmonic phases for $(\l,m,n)=(3,2,0)$.
	Here, $\arg(x+iy)=\tan^{-1}(y/x)$.
	}
	\label{fig:new_harmonics_2}
\end{figure}
%
%
%
%
%
\section{Calculation of the \ashs{}}
\label{sec:Calc}
%
\par Here we present a non-perturbative algorithm for calculating the physical \ashs{}.
The starting point of our discussion is the completeness of the physical spheroidal harmonics on fixed overtone subsets.
This result, and related spherical-spheroidal maps, will be used to show that the \ashs{} may be calculated using a simple spherical harmonic expansion,
\begin{align}
	\label{C0}
	\ket{\tilde{S}_\lmn} \; = \; \sum_{\lp} \, \ket{Y_\lpm} \brak{Y_\lpm}{ \tilde{S}_\lmn } \; .
\end{align}
In this, our core task is to determine the inner-product values between spherical and \ashs{}, $\brak{Y_\lpm}{ \tilde{S}_\lmn }$.
\par We begin by recalling that \eqn{M18} provides us with $\adj\mct$, which transforms \ashs{} into spherical ones.
We may write $\adj\mct$ as an infinite dimensional matrix by expanding its spheroidal harmonic bras in spherical harmonics,
\begin{align}
	\label{C1}
	\adj\mct \; = \; \sum_{\l,\lp} \, \ket{Y_\lpm} \brak{S_\lpmn}{Y_\lm} \bra{Y_\lm} \; . 
\end{align}
\Eqn{C1} simply communicates that, in the spherical harmonic basis, $\adj\mct$ is simply a matrix of spherical-spheroidal inner-products.
For practical numerical calculations, we may consider the $N \times N$ dimensional truncation of $\adj\mct$,
\begin{align}
	\label{C2}
	\adj\mct_{(N)} \; = \; \sum_{\l,\lp}^{N} \, \ket{Y_\lpm} \brak{S_\lpmn}{Y_\lm} \bra{Y_\lm} \; . 
\end{align}
In \eqn{C2}, $\l$ and $\lp$ are between and $\max(|s|,|m|)$ and $N$.
\par Noting that $\adj\mcv$ is the inverse of $\adj\mct$, we may use matrix inversion to numerically estimate the matrix elements of $\adj\mcv_{(N)}$, the $N \times N$ truncation of $\adj\mcv$,
\begin{align}
	\label{C3}
	\adj\mcv_{(N)} \; &\approx \; {\adj\mct}^{-1}_{(N)} \; 
	\\
	\label{C4}
	 \; &\approx \; \sum_{\l,\lp}^{N} \, \ket{Y_\lpm} \brak{Y_\lpm}{\tilde{S}_\lmn} \bra{Y_\lm} \; . 
\end{align}
In \eqn{C3} we denote $\adj\mcv_{(N)}$ as being approximately equal to ${\adj\mct}^{-1}_{(N)}$, up to truncation error.
In \eqn{C4}, we have expanded \eqn{M19}'s \ashs{} in sphericals to highlight that the matrix elements of $\adj\mcv_{(N)}$ are the inner-products of interest. 
In particular, if we denote the matrix elements of $\adj{\mcv}_{(N)}$ as 
\begin{align}
	\label{C5}
	\tilde{\sigma}_{(N)\, \lp \l} \; &= \; \bra{Y_\lpm} \, \adj{\mcv}_{(N)}\, \ket{Y_\lm} \; ,
	\\
	\label{C6}
	&\approx \; \brak{Y_\lpm}{\tilde{S}_\lmn} \; .
\end{align}
then the adjoint spheroidal harmonics may be numerically estimated as
\begin{align}
	\label{C7}
	\ket{\tilde{S}_\lmn} \; &\approx \; \sum_{\lp}^{N} \; \tilde{\sigma}_{(N)\, \lp \l} \, \ket{Y_{\lm}} \; .
\end{align}
Equivalently, if we write the \ass{} as functions (in full notation) rather than vectors, then
\begin{align}
	\label{C8}
	{_{-2}}{\tilde{S}_\lm(\theta;\gamma_\lmn)} \; &\approx \; \sum_{\lp}^{N}  \, {{_{-2}}Y_{\lm}(\theta)}\; \tilde{\sigma}_{(N)\, \lp \l} \; .
\end{align}
\par \Eqnsa{C7}{C8} encapsulate the key result of this section.
\Eqn{C7} is a way of non-perturbatively calculating the \ashs{}, given the spherical-spheroidal inner-products. 
The approximately diagonal nature of $\adj\mct$ means that values along the $N^{\mathrm{th}}$ row and column of $\adj\mcv_{(N)}$ are least accurate.
In practice, it is found that $N^{\mathrm{th}}$ row and column elements of $\adj\mct$ rapidly converge with increasing $N$, with $N=\ell+6$ being sufficient to estimate the harmonics to machine precision.
An implementation of \eqns{C2}{C7} is included in \texttt{positive.aslmcg}~\cite{positive:2020}.
\smallskip
\par \Figsa{fig:new_harmonics}{fig:new_harmonics_2} show example evaluations of the $(\l,m,n)=(2,2,0)$ and $(3,2,0)$ \ashs{}.
There each harmonic is normalized when integrated over the solid angle.
Prograde \qnm{s} are used (i.e.\ those with positive \qnm{} frequencies at $m=2$).
Each adjoint-harmonic is derived from the $n=0$ overtone subset according to \eqn{C7}, with $|\ell-\ell'|\leq 8$.
This choice corresponds to the $|\ell-\ell'|=8$ terms' contributing less than $0.01\%$ in amplitude for each case.
Related spheroidal harmonics are calculated from Leaver's method (i.e.\ Ref.~\cite{leaver85}), and \texttt{positive.physics.qnmobj} class has been used to reference \qnm{} frequencies and related spheroidal harmonics with consistent conventions~\cite{positive:2020}.
%
\section{Spheroidal harmonic decomposition}
\label{sec:Decomp}
%
\par We have now developed an understanding of why the \ashs{} exist, and why an arbitrary \gw{} signal may be equated with its spheroidal harmonic expansion.
While there are still questions of theory within immediate reach, such as whether there exists an operator for which $\ket{\tilde{S}_\lmn}$ are eigenvectors~(See \capx{apx:many}), for now, we may begin to focus on somewhat more practical matters.
In this section we will be concerned with how one might apply the \ashs{} to physical problems. 
\par For simplicity and concreteness we will consider the application of Kerr $n=0$ spheroidal harmonics to arbitrary \gw{} signals.
We will then discuss the specific case of Kerr \rd{} (i.e.\ a sum of \qnms{}).
Lastly, we will discuss the conditions for which a signal's spheroidal multipole moments may be exactly equated with the physical system's modes. 
\par Let's begin by considering the basic situation of \gw{} theory wherein we wish to represent a \gw{} signal, $h(r,t,\theta,\phi)$, in terms of radiative multipole moments. 
In the case of e.g.\ \pn{} theory, one might want to analytically relate the radiative multipole moments of $h$ to the source's multipole moments~\cite{Blanchet:2013haa,Thorne:1980}.
In the case of \nr{}, including the numerics of particle perturbation theory, one might be provided with numerical radiation, and then want to decompose that data into multipole moments that are useful for e.g.\ the development of signal models~\cite{Hughes:2019zmt,London:2018gaq,Cotesta:2018fcv,Blackman:2017pcm,Garcia-Quiros:2020qpx,London:2017bcn}.
At the intersection of perturbative and non-perturbative \gw{} theory, one might want to represent the information perturbing an isolated \bh{} in a way that is closely aligned with the \bh{}'s intrinsic modes~\cite{LeTiec:2009yg,London:2018gaq,Kelly:2012nd,Garcia-Quiros:2020qpx}. 
In all of these settings, a spheroidal harmonic representation is of potential use.
For each, a choice of oblateness must be made prior to pursuing a spheroidal harmonic decomposition.  
\par In principle, the oblatenesses may be developed to suit the specific physical problem.
For example, the Kerr $n=0$ spheroidal harmonics have oblatenesses determined by the \bh{} spin parameter $a$, and the pro- or retrograde \qnm{} frequencies $\cw_{\lmn}$. 
In that setting oblateness values are ordered by $\l$, and relate to two physical quantities: the spacetime angular momentum, and its linear mode frequencies.
The mode frequencies themselves are largely determined by the problem's radial structure~\cite{Yang:2012he,leaver85}.
One might also imagine a physical settings and related mathematical frameworks wherein a (fixed background + adiabatic foreground) Kerr \bh{}'s geometry changes adiabatically (e.g. Post-Newtonian and particle perturbation theory \cite{Blanchet:2013haa,OSullivan:2014ywd,Sberna:2021eui}). 
In that case it might be natural to also consider oblateness values that evolve in time (or frequency)~\cite{Nollert:1999ji}.
Such a framework may be the topic of future work.
\par For now, it is illustrative to consider oblateness values given by Kerr overtone subsets.
This course allows us to concretely discuss applications while maintaining the basic structure of general spheroidal systems (i.e.\ oblateness values that depend on $\l$).
While the $n=0$ subset is of primary interest, we will proceed by referring to subset oblateness values as $\gamma_\lmbn$, where it should be understood that $\n$ is fixed.
In cases where the overtone index is not associated with a fixed overtone subset, $n$ will be used.
\par Given oblateness values, $\gamma_\lmbn$, one might apply the general form of a spheroidal harmonic expansion,
\begin{align}
	\label{hs0a}
	h(r,t,\theta,\phi) \; = \; \frac{1}{r} \; \sum_{\ell,m} \, h^S_{\ell m \n}(t) \, {_{-2}}S_{\ell m}(\theta; \gamma_{\lmbn } ) \, e^{i m \phi} \; , 
\end{align}
where, the spheroidal harmonic multipole moment, $h^S_{\ell m \n}$, is 
\begin{align}
	\label{hs0b}
	h^S_{\ell m \n}   =  \int_{0}^{2\pi} \int_{0}^{\pi} \, &{_{-2}}\tilde{S}^*_\lm(\theta;\gamma_\lmbn) \, e^{-im\phi} 
	\\ \nonumber
	&\times \, h(r,t,\theta,\phi) \, \sin(\theta)\, \mathrm{d}\theta \, \mathrm{d}\phi \, .
\end{align}
In \eqn{hs0a}, ${_{-2}}{S}_\lm(\theta;\gamma_\lmbn)$ are the physical spheroidal harmonics as may be calculated e.g.\ by Leaver's method~\cite{leaver85}.
In \eqn{hs0b}, ${_{-2}}\tilde{S}_\lm(\theta;\gamma_\lmbn)$ are the \ass{} defined by \eqn{C8}.
It may be easily verified that the bi-orthogonality of the physical spheroidals makes \eqnsa{hs0a}{hs0b} inter-consistent.
For this, one would begin by substituting the right-hand-side of \eqn{hs0a} into \eqn{hs0b}.
One would then apply the following bi-orthogonality relationship,
\begin{align}
	\label{SOb}
	\int_{0}^{\pi}\, \tilde{S}^*_{\ell m}(\theta;\gamma_{\lmbn}) \, S_{\lpm}(\theta;\gamma_{\lpmbn}) \, \sin(\theta) \, \mathrm{d}\theta \;  = \; \frac{\delta_{\ell \ell'}}{2 \pi} \; .
\end{align}
For consistency with \eqnsa{hs0a}{hs0b}, in \eqn{SOb} we define the spheroidal harmonics and their adjoint functions to be normalized when integrated over the solid angle, not just over the polar dimension.  This introduces the factor $1/2\pi$, which accounts for the fact that $\int e^{i(m-m')\phi}d\phi = 2\pi\delta_{mm'}$.  
\par With \eqns{hs0a}{SOb} we have at our disposal the ability to calculate the spheroidal harmonic expansion of arbitrary \gw{} signals.
If, rather than $h$, many spherical harmonic moments are provided, a standard change-of-basis approach may be preferable to the direct integration of \eqn{hs0b}.
However, the accuracy of that method is inherently limited by the number of available spherical harmonic moments.
Direct integration and change-of-basis are equivalent if the latter method is applied with enough spherical moments to reproduce $h$ up to the desired numerical precision. 
For both methods, completeness of the spheroidal harmonics allows them to encode \gw{s} from arbitrary physical scenarios.
\par While this is also true of the spherical harmonics, the potential benefit of the spheroidal harmonics is their proximity to the underlying modes of axisymmetric systems. 
To illustrate this point let us consider \bh{} \rd{}, where the underlying spheroidal mode structure is provided by analytic relativity~\cite{leaver85,Berti:2005ys,Teukolsky:1973ha}.
If we denote the \grad{} from \bh{} \rd{} as $h^{\mathrm{RD}}(r,t,\theta,\phi)$, then linear \bh{} perturbation theory has that 
\begin{align}
	\label{hRD1}
	r \, h^{\mathrm{RD}} \;=\;  \sum_{\l,m}  \, e^{i m \phi} \,  \sum_{n=0}^{\infty} \;\; h_\lmn^{\mathrm{Pro}}\;&{_{-2}S_{\lm}}(\theta;\gamma_\lmn)
	\\ \nonumber
	&\;\;+\;h_\lmn^{\mathrm{Ret}} \;{_{-2}S_{\lm}}(\theta;\gamma'_\lmn)   \; .
\end{align}
where 
\begin{align}
	\label{hRD2}
	h^{\mathrm{Pro}}_\lmn  \;=\; b_{\lmn} \, e^{- i \cw_\lmn t}  \; ,
	\\
	\label{hRD3}
	h^{\mathrm{Ret}}_\lmn  \;=\; b'_{\lmn} \, e^{- i \cw'_\lmn t}  \; .
\end{align}
In the left-hand-side of \eqn{hRD1} we have written $r h^{\rm {RD}}$ to simplify our consideration of the right-hand-side's terms which are all independent of $r$. We recall that $r$ is the source's luminosity distance~(\ceqn{hy0}).  
We have written \eqn{hRD1} to emphasize that all spheroidal moments have the same azimuthal dependence, $e^{im\phi}$.
We have also written \eqn{hRD1} to emphasize that \bh{} perturbation theory predicts the existence of radiative modes corresponding to perturbations prograde and/or retrograde with respect to the \bh{} angular momentum direction.
In \eqns{hRD1}{hRD3}, $h^{\mathrm{Pro}}_\lmn$ and $h^{\mathrm{Ret}}_\lmn$ respectively correspond to pro- and retrograde \qnms{}.
Similarly, in \eqns{hRD1}{hRD3} we denote prograde \qnm{} frequencies with $\cw_\lmn$, and retrograde ones with $\cw'_\lmn$.
The related oblatenesses are $\gamma_\lmn=a\cw_\lmn$ and $\gamma'_\lmn=a\cw'_\lmn$. 
\par \Eqn{hRD1} is the fully general form of \eqn{hQNMk}, and as was done there for $\cw_\lmn$, we use the convention that $\cw'_{\lmn}=-\cw'^*_{\l\,-m\,n}$.
Under this convention, the pro- and retrograde frequencies are related by 
\begin{align}
	\label{cwconvention2}
	\cw_{\lmn}(a) \; = \; \cw'_\lmn(-a) \; .
\end{align}
In \eqn{cwconvention2}, we note that the \qnm{} frequencies may be parameterized by the \bh{} spin, just as was done in \fig{scgamma}.
It should also be noted that $a \rightarrow -a$ has the principal effect of inverting the \bh{}'s spin axis.
In this sense, \eqn{hRD1} communicates that \bh{} ringdown may generally correspond to concurrent pro- and retrograde excitations.
\par We now wish to decompose \eqn{hRD1}'s $rh^{\rm{RD}}$ into spheroidal harmonic moments, as defined by the $n=\n$ overtone subset.
Our aim is to better understand the relationship between the \qnm{s} of perturbation theory, and the spheroidal multipole moments, $h^{\rm S}_{\lmbn}$.
To proceed we will focus on sets of like $m$ by defining
\begin{align}
	\label{hm1}
	h^{\rm{RD}}_m \; &= \; r\, \int_{0}^{2\pi} \, e^{-im\phi} \, h^{\rm{RD}} \, \rm{d}\phi\;
	\\
	\label{hm2}
	\; &= \; 2 \pi \, \sum_{\l ,n} \, h_\lmn^{\mathrm{Pro}}\;{_{-2}S_{\lm}}(\theta;\gamma_\lmn) 
	\\ \nonumber 
	& \quad\quad\quad\quad\quad\quad\quad + \; h_\lmn^{\mathrm{Ret}}\;{_{-2}S_{\lm}}(\theta;\gamma'_\lmn) \; .
\end{align}
In \eqns{hm1}{hm2} we define $h_m^{\rm{RD}}$ by simply applying the orthogonality of the complex exponentials to \eqn{hRD1}.
It is convenient to rewrite \eqn{hm2} using the more compact bra-ket notation, 
\begin{align}
	\label{hm3}
	\ket{h_m^{\rm{RD}}} \; = \; 2\pi \, \sum_{\l, n}\, \left( \, h_\lmn^{\mathrm{Pro}}\, \ket{S_\lmn} + \, h_\lmn^{\mathrm{Ret}} \,\ket{S'_\lmn} \, \right) \; .
\end{align}
In \eqn{hm3}, $\ket{S'_\lmn}$ correspond to the retrograde spheroidal harmonics, ${_{-2}S_{\lm}}(\theta;\gamma'_\lmn)$.
\par With \eqn{hm3}, the spheroidal moments of $h_m^{\rm{RD}}$ are determined according to
\begin{align}
	\label{hm4}
	h^\rm{S}_{\lmbn}  &=  \brak{\tilde{S}_\lmbn}{ h_m^{\rm{RD}} } \; 
	\\
	\label{hm5}
	&=  2\pi \, \sum_{\lp, n} \, h_\lpmn^{\mathrm{Pro}} \brak{\tilde{S}_\lmbn}{S_\lpmn} + \, h_\lpmn^{\mathrm{Ret}} \brak{\tilde{S}_\lmbn}{S'_\lpmn}  .
\end{align}
In \eqn{hm4}, the inner-product (i.e.\ \ceqn{prod}) simply corresponds to the $\theta$ integral of \eqn{hs0b}.
In \eqn{hm5}, we introduce $\lp$ to sum over polar indices. 
\par We may find in \eqn{hm5} a starting point for many practical insights. 
In particular, \eqn{hm5} may be used to consider two basic cases: one, where only pro- or retrograde modes are present, and another where both are present.
The first case is well known to be relevant to \bbh{} merger remnants from nonprecessing to moderately-precessing progenitors~\cite{Hamilton:2021pkf,Ossokine:2020kjp,London:2018gaq,Khan:2015jqa,Hughes:2019zmt}.
The second case is known to be most relevant to \bbh{s} which undergo significant precession just prior to merger~\cite{Hamilton:2021pkf,Hughes:2019zmt}. 
\par For the first and simplest case, we may hold that only prograde modes are excited, leaving 
\begin{align}
	\label{hm6}
	{h^\rm{S}_{\lmbn}} \; &= \; {2 \pi}\,\sum_{\lp, n} \, h_\lpmn^{\mathrm{Pro}} \brak{\tilde{S}_\lmbn}{S_\lpmn} 
	\\
	&= \; h_\lmbn^{\mathrm{Pro}} \, 
	\label{hm6a}
	\\
	& \;\;\; + \; {2 \pi}\,\sum_{\lp \neq \l} \, h_\lmn^{\mathrm{Pro}} \brak{\tilde{S}_\lmn}{S_\lmn} 
	\label{hm6b}
	\\
	& \;\;\; + \; {2 \pi}\,\sum_{ n\neq \n} \, h_\lpmn^{\mathrm{Pro}} \brak{\tilde{S}_\lmbn}{S_\lpmn} \;
	\label{hm6c}
	\\
	& \;\;\; + \; {2 \pi}\,\sum_{\lp\neq \l, \, n\neq \n} \, h_\lpmn^{\mathrm{Pro}} \brak{\tilde{S}_\lmbn}{S_\lpmn} \; . 
	\label{hm6d}
\end{align}
In \eqn{hm6} we have simply written a prograde-only \rd{}.
In \eqns{hm6a}{hm6d} we have organized the right-hand-side of \eqn{hm6} into four parts.
\par The first part is \eqn{hm6a}.
This is simply the term for which $\ell'=\ell$ and $n=\n$.  
There, $2\pi\braket{\tilde{S}_{\lmbn}}{S_{\ell m \n}}=1$, making the term likely to dominate.
For all other terms, \eqns{hm6b}{hm6d}, the inner-product is necessarily smaller than one,
\begin{align}
	2\pi\,\braket{\tilde{S}_{\lmbn}}{S_{\ell' m n}}<1 \;\;\; \text{for all } \l \neq \l' \text{ and } n \neq \n \; .
\end{align}
\par We might next consider the remaining terms for which $n=\bar{n}$ and $\l \neq \lp$, \eqn{hm6b}.
In the case of spherical harmonic decomposition (e.g. replacing $\tilde{S}_{\lm}$ with $Y_{\lm}$), these terms would be the next largest, and are known to be the cause of non-physical mode-mixing effects~\cite{Garcia-Quiros:2020qpx,Cotesta:2018fcv,London:2014cma}.
Here, due to bi-orthogonality of the \ass ~(\ceqn{SOb}), these terms are zero, meaning that the use of the adjoint spheroidal harmonics completely suppressed the primary cause of mode mixing, 
\begin{align}
	\label{hm7}
	{2 \pi}\,\sum_{\lp \neq \l} \, h_\lmn^{\mathrm{Pro}} \brak{\tilde{S}_\lmn}{S_\lmn} \; = \; 0 \; .
\end{align}
\par We might next consider \eqn{hm6c}, which collects terms for which $\ell'=\ell$ and $n \neq \n$. 
In the limit of spherical symmetry (i.e.\ zero oblateness), these terms lose their dependence on $\braket{\tilde{S}_{\lmbn}}{S_{\ell' m n}}$, and reduce to a sum over overtone contributions, exactly as one would expect from \eqn{hQNMs}.
In this sense, these terms are inherent to the physical situation, and do not result from our choice of spheroidal basis.
\par Lastly, we are left with \eqn{hm6d} which collects terms for which $\ell' \neq \ell$ and $n \neq \n$. 
In the limit of spherical symmetry, these terms become exactly zero. 
These terms exist in axisymmetry because of our choice of basis.
However, the asymptotic equivalence of different overtone harmonics means that these terms' inner-products are generally small relative to unity.
Thus these terms are likely to contribute the least. 
\par So far, the ideas applied to \eqn{hm6} apply to any choice of $\n$ (i.e.\ any overtone subset), and our conclusions would not change if we were to consider only retrograde ringdown with \ashs{} derived in that setting.
We will now briefly consider cases where pro- and retrograde modes are excited such that both are needed to accurately describe the \grad{}.
In this setting, many of the ideas discussed thus far apply. 
If, as in \eqn{hm5}, we wish to decompose the net signal into prograde spheroidal moments, then we will still be left with the four parts seen in \eqns{hm6a}{hm6d}.
However, due to the presence of retrograde modes, we will have four additional parts: analogs of \eqn{hm6}'s four parts corresponding to mixing between pro- and retrograde modes.
Using the ideas of \csec{subsec:Subsets}, it may be shown that the pro- and retrograde harmonics represent redundant spacial information (e.g.\ they are and exactly equivalent in the zero-oblatenesses limit, and are each minimal).
Thus, like overtones, pro- and retrograde modes cannot be separated by decomposition into only angular harmonics.
\par This situation is not dissimilar from what one would encounter during spherical harmonic decomposition.
However, the key difference is that terms for which $\l\neq \lp$ and $n=\n$ are either nullified (as in the case of the prograde sector, \eqn{hm7}) or lessened (as in the case of mixing between pro- and retrograde modes).
Nevertheless, just as in spherical harmonic decomposition, it is clear that multipole moments from spheroidal decomposition alone are \textit{not generally} modes.
\par With that in mind, we conclude this section with a brief discussion of exactly when spheroidal multipole moments, $h^{\rm S}_{\lmbn}$, may be exactly identified with the modes of physical systems.
We will limit this discussion to \rd{}'s spheroidal decomposition.
We expect aspects of that context transfer to other settings in which \ashs{} may be developed.
\par For \rd{}'s spheroidal decomposition, $h^{\rm S}_{\lmbn}$ will only correspond to a mode when the radiation is dominated by perturbations that are linear, either pro- or retrograde, and excite only one overtone subset.
While ostensibly narrow, these cases are known to include \bbh{} ringdown from systems with weak or no precession~\cite{Hamilton:2021pkf,Ossokine:2020kjp,London:2018gaq,Khan:2015jqa,Hughes:2019zmt}. 
However, even in such astrophysically relevant scenarios, there are limitations. 
Within the \qnm{s}, there is currently uncertainty regarding the importance of overtones~\cite{Jaramillo:2020tuu,Giesler:2019uxc}.
Furthermore, linearly perturbed black holes are known to, in principle, generate various kinds of \grad{}\cite{Nollert:1999ji}. 
The \qnm{s} are known to be by far the most dominant, but other types include power-law tails, and direct emission~\cite{Andersson:1996cm}.
Like overtones, neither power-law tails nor direct emission are amenable to decomposition with angular harmonics. 
For these reasons the spheroidal harmonic decomposition discussed here represents a tool for estimating, but not exactly extracting, information about spheroidal modes.
However, relative to spherical harmonics decomposition, the explicit lack of mode-mixing on the chosen overtone subset is spheroidal decomposition's primary advantage.
%
\section{Concluding Remarks}
\label{sec:Discuss}
%
\par When seeking to represent \grad{} in terms of multipole moments, there has been a tension. 
The spherical harmonics are the typical choice for defining radiative multipole moments~\cite{Thorne:1980,Ruiz:2007yx,Blanchet:2013haa}.
However, they are most appropriate for systems with zero angular momentum, of which, in nature, we may expect none~\cite{Ruiz:2007yx,leaver85,LIGOScientific:2020ibl,LIGOScientific:2018mvr}.
In this context, we have investigated the inclusion of angular momentum in how we represent \gw{s}.
\par By considering the spheroidal harmonics, and their angular momentum dependent oblateness parameters, we have adopted the simplest known physically motivated alternative to spherical harmonics~\cite{Teukolsky:1973ha,leaver85}.
In doing so, we have encountered multiple challenges.
\par In \sec{sec:ManyOps} we have illustrated a previously uninvestigated aspect of the Kerr spheroidal harmonics:
each spheroidal harmonic differential operator depends on an oblateness parameter, $\gamma_\lmn$, and each Kerr \qnm{} posses a \textit{different} oblateness.
In this sense, each \qnm{'s} spheroidal harmonic is an eigenfunction a \textit{different} differential operators.
From this ``issue of many operators'' follows many non-standard properties of the physical spheroidal harmonics. 
While these properties are not standard in \gw{} physics, they are familiar to other fields.
\par In \sec{sec:CB}, we have drawn from functional analysis to show that (\textit{i}) physical spheroidal harmonics posses a kind of orthogonality, and (\textit{ii}) that the physical spheroidal harmonics are complete.
Most importantly, notion (\textit{ii}) means that spheroidal harmonics of e.g.\ Kerr \bh{s} may be used to exactly represent arbitrary \gw{} signals.
In \sec{subsec:Subsets} we have shown that only subsets of spheroidal harmonics with the same overtone index, $n$, encode \textit{unique} angular information, and we have used results from functional analysis to conclude that these ``fixed overtone subsets'' possess a kind of \textit{bi-orthogonality}.
In this sense, \sec{subsec:Subsets} concludes that there exist angular harmonics, $\tilde{S}_\lmn$, that are orthogonal to the physical spheroidal harmonics: $\brak{\tilde{S}_\lpmn}{{S}_\lmn} \propto \delta_{\lp\l}$.
We have named these new angular functions the ``\ashs{}''.
\par In \sec{subsec:YSMaps} we have drawn inspiration from quantum mechanics and functional analysis literature to show that the spheroidal harmonics are complete~\cite{Mostafazadeh:2001jk,ROSASORTIZ201826,Breuer1977,Christensen2003}.
We have shown that the spheroidal harmonics may be related to the spherical harmonics by an invertible operator, $\mct$.
In adopting a vector space construction of $\mct$,  we took our first step towards overcoming the issue of many operators.
This, in turn, allowed us to show that the spherical harmonics and their adjoint functions support a kind of decomposition.
This ``spheroidal harmonic decomposition'' shares many features with spherical harmonic decomposition, but accomplishes our goal of including spacetime angular momentum directly in the definition of the radiative moments. 
\par In \sec{sec:Calc} we place the \ashs{} on concrete footing by showing how they may be calculated. 
In deeming it essential to first provide a non-perturbative method to calculate the adjoint-harmonics, we have left a thorough analytic treatment for future work. 
We provide example evaluations of the new harmonics in \Figsa{fig:new_harmonics}{fig:new_harmonics_2}.
\par In \sec{sec:Decomp} we have outlined what spheroidal harmonic decomposition looks like in practice. 
There we have addressed general applications, as well specific cases, such as the spheroidal harmonic decomposition of radiation from \gw{} \rd{}.
\Sec{sec:Decomp} concluded with a general discussion of when spheroidal decomposition allows for the exact extraction of a system's intrinsic modes, rather than simply multipole moments which may only approximate modes. 
There we illustrated that the suppression of mode-mixing is spheroidal decomposition's primary advantage.
\par Many aspects of the presented work may be refined and expanded upon.
For example, we have only briefly discussed the potential applications of spheroidal decomposition.
Multifaceted investigations are needed to better determine the potential use of the \ass. 
Paper II, with its focus on applications to extreme and comparable \bbh{s}, is one such investigation~\cite{London:2021P2}.
It may also be possible to better understand spacetime oblatenesses, particularly for spacetimes that, unlike Kerr, are not stationary.
This direction may also be followed in future work. 
\par Each of these potential investigations brings new and potentially useful questions.
Does the analytic structure of \ashs{} inform the broader non-hermitian nature of \ee{}? 
Can any of the techniques used here also be applied to solutions to Teukolsky's radial equation?
How should the oblateness parameter be defined in systems where mass and spin are radiated non-adiabatically?
And can the answer to these questions inform yet unprobed aspects of \bbh{} merger?
%
%
\section*{Acknowledgments}
%
Work on this problem was supported at Massachusetts Institute of Technology~(MIT) by National Science Foundation Grant No. PHY-1707549 as well as support from MIT’s School of Science and Department of Physics. This work was supported at the University of Amsterdam by the GRAPPA Prize Fellowship. Lastly, this work was supported at King's College London by the Royal Society University Research Fellowship, Grant No. URF{\textbackslash}R1{\textbackslash}11451. Many thanks are extended to Scott Hughes for his invaluable support as well as his granting access to his calculations for the Kerr spheroidal harmonics for the occasional sanity check. Additional thanks are extended to Richard Price, Richard Melrose, Halston Lim, Mark Hannam, Stephen Fairhurst, Vitor Cardoso and Paolo Pani for their helpful questions and input. Additional thanks are extended to Alessandra Buonanno and Ajit K. Mehta for their feedback.
%
\appendix
%
\section{Perturbation theory approximation of the spherical-spheroidal mixing coefficients}
\label{apx:mixing}
%
Perturbation theory arguments may be used to estimate the spherical-spheroidal mixing coefficients.
The preamble to these arguments is largely insensitive to the details of the problem at hand; however, they are useful for the efficient clarification of the matter. 
In this section, we will use beyond linear order perturbation theory to derive \eqn{M3},
\begin{align}
	\label{apx-sigma-scale}
	\sigma_{\ell \pm p,\ell} \; \approx \; \frac{1 }{p!} \; \left( \frac{ -\gamma s}{2 \ell} \right)^{\,p} \; .
\end{align}
Relative to \eqn{M3}, in \eqn{apx-sigma-scale} we have labeled the oblateness as $\gamma$ rather than $\gamma_\lmn$ as the statement hold regardless of whether the spheroidal oblateness is fixed with respect to physical indices. 
Here, we have chosen to define $p=|\lp-\l|$, making \eqn{M3}'s $\lp=\l \pm p$.
Without loss of generality we will continue to consider both $m$ and $s$ fixed.
We will at times find it useful to use alternative notation for the polar index $\l$.
For example, $\ket{Y_\A}$ represents a spherical harmonic where $\l=\L$.
\par We begin by framing the general perturbative problem as a kind of recursion relation. We then use the specific nature of the spheroidal potential to show that each perturbative order depends on the absolute difference $|\lp-\ell|$. 
\par Let the zero oblateness spheroidal operator be $\Lo$~(i.e.\ \ceqn{LSc} with $\gamma=0$), and the perturbing potential be 
\begin{align}
	\label{eq:AV}
	V^{(\mathrm{S})} \; \approx \;  - 2 s u  \; . 
\end{align}
In \eqn{eq:AV} we deliberately neglect the full potential's $\gamma u^2$ term as, at every perturbative order, it introduces higher order terms which are peripheral to our final approximation (i.e.\ including the $\gamma u^2$ term produces contributions that decay in $p$ at least as fast as \ceqn{apx-sigma-scale}).
From this perspective, the spheroidal harmonics are eigenfunctions of the operator
\begin{align}
	\label{eq:A00}
	\mclo \; = \; \Lo + \gamma V^{(\mathrm{S})} \; ,
\end{align}
and the spherical harmonics of eigenfunctions of $\Lo$.
Using kets to represent the harmonics, the eigen relationships are
\begin{align}
	\label{eq:A0}
	\mclo \, \ket{Z_\lm} \; &= \; -A_{\lm} \, \ket{S_\lm} \; ,
	\\
	\label{eq:A1}
	\Lo \, \ket{Y_{\lpm}} \; &= \; -E_{\lpm} \, \ket{Y_{\lpm}} \; .
\end{align}
Towards \eqn{apx-sigma-scale}, our first choice in representing $\ket{S_\lm}$ is a non-perturbative one. 
The completeness of the spherical harmonics as well as the natural reduction of the spheroidals to the sphericals when $\gamma=0$ mean that a good ansatz for $\ket{S_\lm}$ is 
\begin{align}
	\ket{Z_\lm} \; = \; \sum_{\lp} \, \sjk \, \ket{Y_{\lpm}} \; ,
\end{align}
where $\sjk$ is the spherical spheroidal mixing coefficient of interest, $\sjk=\brak{Y_{\lpm}}{S_\lm}$.
Using \eqn{eq:A0} to apply this ansatz to \eqn{eq:A1} gives 
\begin{align}
	\label{eq:A2}
	(\Lo  +  \gamma \, V^{(\mathrm{S})} )\sum_{\lp} \, \sjk \, \ket{Y_{\lpm}} = -A_{\lm} \, \sum_{{\lp}} \, \sjk \, \ket{Y_{\lpm}} \; .
\end{align}
In \eqn{eq:A2} we can see that the quantity $\mclo\,\ket{ S_\lm}$ can be written in terms of only spherical harmonics. 
With this in mind, acting on \eqn{eq:A2} with $\bra{Y_\A}$, and then applying the spherical harmonics eigenvalue relation (\ceqn{eq:A1}) yields
\begin{align}
	\label{eq:A3}
	\sum_{\lp} \, \gamma \, \sjk \brak{Y_\A}{V^{(\mathrm{S})} \,| Y_{\lpm}} \; = \; (E_\A - A_{\lm}) \, \sigma_{\L \ell}.
\end{align}
It is well known that $\brak{Y_\A}{V^{(\mathrm{S})} \,| Y_{\lpm}}$ is only non-zero when $|\L-\lp| \leq 1$; thus, \eqn{eq:A3} is in effect a 3-term recursion relation.
While one may be tempted to investigate its solutions via the roots of its characteristic polynomial, here
we will look for approximate solutions using standard perturbation theory ansatzes: 
\begin{align}
	\label{eq:A4}
	\sjk \; = \; \sum_{j=0} \, \sjk^{(j)} \, \gamma^j \; ,
	\\
	\label{eq:A4b}
	A_{\lm} \; = \; \sum_{q=0} \, A_{\lm}^{(q)} \, \gamma^q \; .
\end{align}
Applying \eqn{eq:A4} to \eqn{eq:A3}, and for brevity defining $V^{(\mathrm{S})}_{\L\lp}=\brak{Y_\A}{V \,| Y_{\lpm}}$ yield
\begin{align}
	\label{eq:A5}
	\sum_{j, q} \gamma^{j+q} \, A_{\lm}^{(q)} \, \sigma_{\L \ell}^{(j)}  =  E_\A \sum_{j} \sigma_{\L \ell}^{(j)} \gamma^{j} \, - \,\sum_{{\lp}, j} \sjk^{(j)} \gamma^{j+1} V^{(\mathrm{S})}_{\L\lp} \,.
\end{align}
Having applied our perturbative ansatz, our aim is to enforce that \eqn{eq:A5} holds for each power of $\gamma$.
To this end, we are free to rewrite sums such that coincident powers of $\gamma$ appear in each.
This may be accomplished in the left-hand side of \eqn{eq:A5} by letting $j+q=v$, and on the right-hand side of \eqn{eq:A5} by letting $j+1=z$ with $z>0$. 
These changes along with relabeling back to $p$ yields 
\begin{align}
	\label{eq:A6}
	\sum_{j=0}\,\gamma^j \left(\sum_{v=0}^{j}A_{\lm}^{(j-v)}\sigma_{\L \ell}^{(v)}\right) = E_\A \sum_{j=0}\sigma_{\L \ell}^{(j)} \gamma^{j}-\sum_{\lp,j=1} \sjk^{(j-1)} \gamma^j V^{(\mathrm{S})}_{\L\lp}\,.
\end{align}
For clarity, all summation lower bounds are written in \eqn{eq:A6}.
Enforcing that the summed coefficients of $\gamma^j$ amount to zero gives 
\begin{align}
	\label{eq:A7}
	\sum_{{\lp}} \sjk^{(j-1)} V^{(\mathrm{S})}_{\L\lp} \; = \; E_\A \sigma_{\L \ell}^{(j)} - \sum_{v=0}^{j} \, A_{\lm}^{(j-v)} \sigma_{\L \ell}^{(v)} \; ,
\end{align}
where if $j=0$, then 
\begin{align}
	\label{eq:A8}
	\sigma_{\L \ell}^{(0)} \;( E_\A - A_{\lm}^{(0)} ) \; = \; 0.
\end{align}
\Eqn{eq:A8} communicates that either $\sigma_{\L \ell}^{(0)}=0$ or $E_\A - A_{\lm}^{(0)}=0$.
The necessary coincidence between the $0th$ order approximant and $\gamma=0$ requires that 
\begin{align}
	\label{eq:A9}
	\sigma_{\L \ell}^{(0)} \;=\; \delta_{\L \ell} \; ,
	\\
	\label{eq:A9b0}
	A_{\lm}^{(0)}\;=\; E_\lm \; .
\end{align}
Using \eqn{eq:A9}, the $v=j$ term may be extracted from the sum in \eqn{eq:A7}'s right-hand side, allowing its dependence on $\sigma_{\L \ell}^{(j)}$ to be clarified. Thus, for $j>0$,
\begin{align}
	\label{eq:A7b}
	\sum_{{\L}} \sjk^{(j-1)} V^{(\mathrm{S})}_{\L\lp} \; = \; (E_\A-E_\lm) \, \sigma_{\L \ell}^{(j)} \; - \; \sum_{v=0}^{j-1} \, A_{\lm}^{(j-v)} \sigma_{\L \ell}^{(v)} \; .
\end{align}
\Eqn{eq:A7b} is useful: evaluating it for perturbative orders $j=1$ and greater allows the determination of $\sigma_{{\L}\ell}^{(j)}$.
\par For $j>0$, \eqn{eq:A7} represents a kind of variable order recursion relation. An analog of \eqn{eq:A7} may be derived for all perturbative expansions.
\Eqn{eq:A9} is the $j=0$ boundary condition.
\par For the linear in $\gamma$ approximant, we need only consider \eqns{eq:A7}{eq:A9} with $j=1$.
In this, it may be straightforwardly shown that the standard perturbation theory results follow:
\begin{align}
	\label{Aloa}
	A_{\lm}^{(1)} \; = \; -V^{(\mathrm{S})}_{\ell\ell}
\end{align}
and if $\L \neq \ell$, then 
\begin{align}
	\label{eq:A9b}
	\sigma_{\L \ell}^{(1)} \; = \; \frac{ V^{(\mathrm{S})}_{\L \ell} }{E_\A-E_\lm} \; ,
\end{align}
where if $\L = \ell$, then
\begin{align}
	\sigma_{\ell \ell}^{(1)}=0 \; .
\end{align}
Thus, to linear order in $\gamma$, we have the spherical-spheroidal mixing coefficients are 
\begin{align}
	\label{eq:A10}
	\sigma_{\L \l} = \left\{\begin{array}{lr}
        1, & \text{for } {\L}=\ell\\
        \gamma \,  \frac{ V^{(\rm S)}_{{\L} \ell} }{E_{\lpm}-E_\lm}, & \text{for } {\L} \neq \ell
        \end{array}\right\} \;.
\end{align}
\par \Eqn{eq:A10} marks the end of our case insensitive preamble.
To make progress, we must apply problem specific knowledge about $V^{(\mathrm{S})}_{{\L}\ell}$.
According to our approximate $V^{(\mathrm{S})}=-2us$, it follows that its spherical harmonic averages, $V^{(\mathrm{S})}_{{\L}\ell}$, involve $\brak{Y_{\L m}}{u\, |Y_\lm} $, which are well known in terms of Clebsh-Gordan coefficients~\cite{Teukolsky:1973ha,Mino:1997bx,OSullivan:2014ywd},
%
%
\begin{align}
	\label{eq:A12a}
	V^{(\mathrm{S})}_{{\L}\ell} \; &= \; -2s \brak{Y_{\L m}}{u\, |Y_\lm} 
	\\
	\label{eq:A12b}
	&= \; -2s \left\{\begin{array}{lr}
        c_{\pm 1}(\ell), & \text{for } \L = \ell \pm 1\\
        0, & \text{otherwise}
        \end{array}\right\} \; ,
\end{align}
where 
\begin{align}
	\label{eq:A13}
	c_{-1}{(\ell)} = \frac{1}{\ell}\sqrt{\frac{(\ell-m) (\ell+m) (\ell-s) (\ell+s)}{(2 \ell-1) (2 \ell+1)}}
\end{align}
and
\begin{align}
	\label{eq:A14}
	c_{+1}(\ell) = -c_{-1}(\ell+1)\;.
\end{align}
\par In the zeroth and linear order approximants, we begin to see a pattern emerge.
\Eqn{eq:A9} communicates that orthogonality of the spherical harmonics means that at zeroth order in $\gamma$, $\sigma_{\L \l}$ is only non-zero when $\L=\ell$.
\Eqns{eq:A12a}{eq:A14} communicate that the structure of $V^{(\mathrm{S})}$ results in a linear in $\gamma$ approximant for $\sigma_{\L \l}$ that is non-zero only if $\L \in \{\ell-1,\ell+1\}$.
At second order in $\gamma$, evaluating \eqn{eq:A7b} with $p=2$ yields that, 
\begin{align}
	\label{eq:A15}
	\sigma_{{\L} \ell}^{(2)} \; = \; (E_{{\A}}-E_\lm)^{-1} \, \left( A_{\lm}^{(1)} \sigma_{\L \ell}^{(1)} + \, \sum_{\L} \sigma_{\L \ell}^{(1)} V^{(\mathrm{S})}_{\L \L}  \right) \; .
\end{align}
In this, the pattern extends at second order by activating non-zero contributions when $\L \in \{\ell-2,\ell-1,\ell+1,\ell+2\}$.
Owing to the nature of $V^{(\mathrm{S})}$~(\ceqns{eq:A12a}{eq:A12b}) leading order contributions for $\sigma_{\ell \pm 2, \ell}^{(2)}$ are necessarily the simplest. 
\begin{figure}[tb]
    \includegraphics[width=0.44\textwidth]{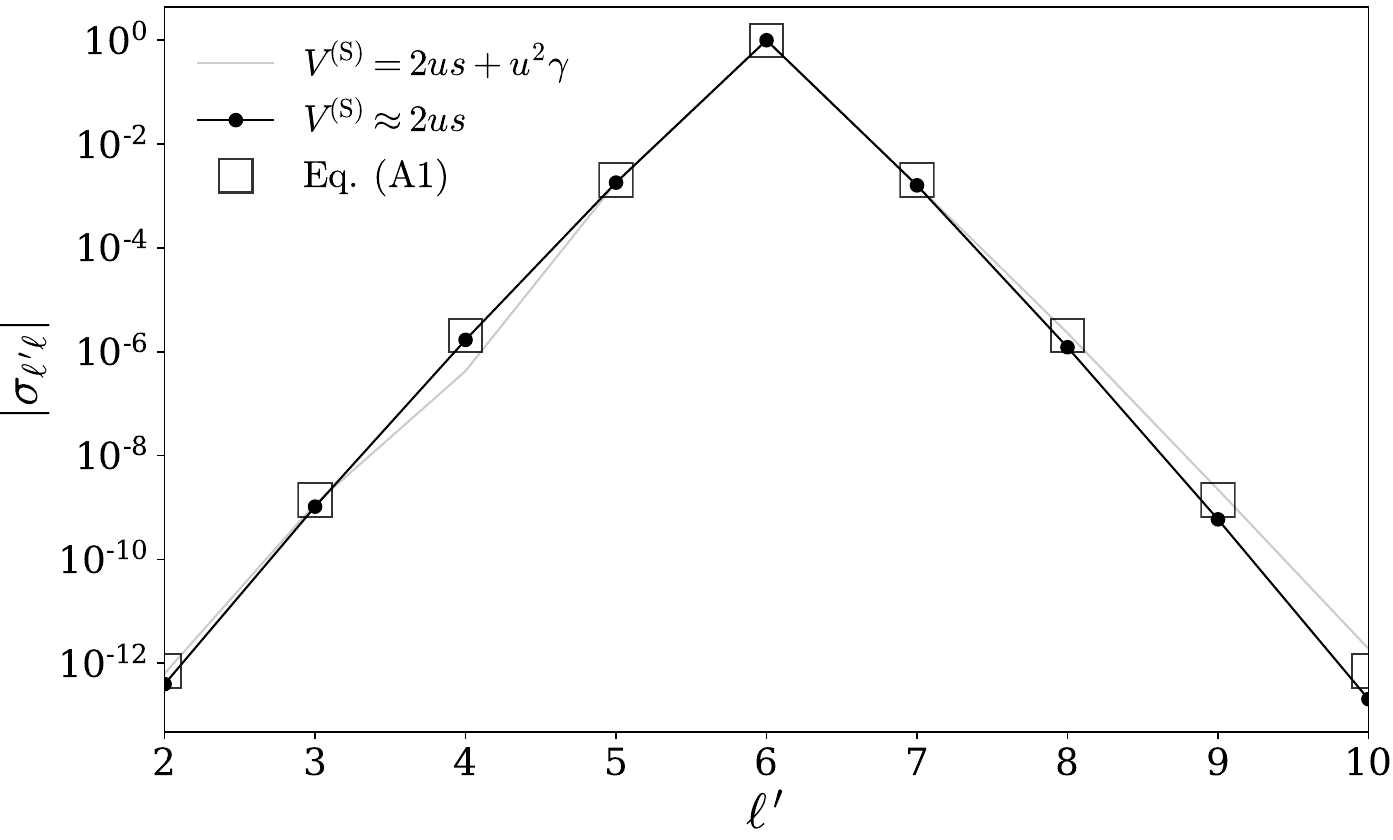} 	
	\caption{Example of approximate for spherical-spheroidal inner-products $\sigma_{\ell'\ell}$~(\ceqn{apx-sigma-scale}) for Kerr with dimensionless spin $a=0.01$ and $\ell=6$. 
	Curves compare exact results for the full  and approximate spheroidal potentials, $V^{(\mathrm{S})}=-2su+u^2\gamma$ and $V^{(\mathrm{S})}\approx -2su$ respectively.
	Note that the spheroidal differential equation depends on $\gamma V^{(\mathrm{S})}$~(See \textit{e.g.} \ceqn{eq:A00}).
	}
	\label{fig:sigma}
\end{figure}
They emerge from the last term of \eqn{eq:A15}, when $\L=\ell \pm 1$ and $\A = \ell \pm 2$,
\begin{align}
	\label{eq:A16}
	\sigma^{(2)}_{\ell \pm 2,\ell} \; = \; \sigma^{(1)}_{\ell \pm 1,\ell} \, \frac{V^{(\mathrm{S})}_{\ell \pm 2, \ell \pm 1}}{E_{\ell \pm 2}-E_{\ell}} \; . 
\end{align}
Subsequent orders follow this pattern, with the leading order behavior of each $\sigma_{\ell \pm \pp,\ell}$ inner product obeying the straight-forward generalization of \eqn{eq:A16},
\begin{align}
	\label{eq:A16b1}
	\sigma^{(\pp)}_{\ell \pm \pp,\ell} \; = \; \sigma^{(\pp-1)}_{\ell \pm (\pp-1),\ell} \, \frac{V^{(\mathrm{S})}_{\ell \pm \pp, \ell \pm (\pp-1)}}{E_{\ell \pm \pp}-E_{\ell}} \; . 
\end{align}
In \eqn{eq:A16b1}, we note the appearance of the absolute difference between $\ell$ and $\L$, namely,
\begin{align}
	\label{eq:A16b}
	\pp \; = \; | \L-\ell | \; .
\end{align}
With \eqn{eq:A16b1}, we have arrived at a recursive formula that is almost ready to lend qualitative insight into the behavior of the spherical-spheroidal inner-products, $\sigma_{\L \l}$.
\par To go further, we may consider the large $\pp$ behavior of $V_{\ell \pm n, \ell \pm (\pp-1)}$. 
It is also useful to recall that the spherical harmonic eigenvalue is
\begin{align}
	\label{eq:A17}
	E_{\ell m} \; = \; (\ell-s)(\ell+s+1) \; .
\end{align}
Thus, the Clebsh-Gordan coefficients (\ceqn{eq:A13}) along with \eqn{eq:A17} communicate that
\begin{align}
	\label{eq:A18}
	V^{(\mathrm{S})}_{\ell \pm \pp, \ell \pm (\pp-1)} \; &\sim \; -2 \gamma s \, \left( \frac{1}{2} + \mathcal{O}({1}/{\pp^2}) \, \right) \; ,
	\\
	\label{eq:A19}
	E_{\ell \pm \pp,m}-E_{\ell m} \; &= \; \pp\, ( 1 + 2\ell+\pp ) \; .
\end{align}
Applying \eqns{eq:A18}{eq:A19} to the recursion relation presented in \eqn{eq:A16} yields 
\begin{align}
	\label{eq:A20}
	\sigma^{(\pp)}_{\ell \pm \pp,\ell} \; \approx \; \frac{ -\gamma s }{\pp\, ( 1 + 2\ell+\pp )} \, \sigma^{(\pp-1)}_{\ell \pm (\pp-1),\ell} \; .
\end{align}
\Eqn{eq:A16} is the recursive formula for a series whose boundary condition is given by the zeroth order correction, $\sigma^{(0)}_{\ell\ell}=1$.
In \eqn{eq:A20} we have used \eqn{eq:A18} as it is qualitatively accurate for $p>1$.
Recursive evaluation of \eqn{eq:A20} yields the rather factorial heavy
\begin{align}
	\label{eq:A21}
	\sigma^{(\pp)}_{\ell \pm \pp,\ell} \; \approx \; \left( -\gamma s \right)^{\,\pp} \; \frac{(2\ell+1)! }{\pp! (\pp+2\ell+1)!} \; .
\end{align}
A somewhat simpler but more approximate picture emerges for large $\ell$, 
\begin{align}
	\label{bigl}
	\ell \gg \pp \gg 1
\end{align}
whence we may think of \eqn{eq:A20}'s denominator as
\begin{align}
	\label{eq:A19b}
	E_{\ell \pm \pp,m}-E_{\ell m} \; &\approx \; 2 \ell \pp  \; .
\end{align}
From this perspective the iterative solution to \eqn{eq:A20} becomes 
\begin{align}
	\label{eq:A21b}
	\sigma^{(\pp)}_{\ell \pm \pp,\ell} \; \approx \; \frac{1 }{\pp!} \; \left( \frac{ -\gamma s}{2 \ell} \right)^{\,\pp} \; .
\end{align}
In \eqn{eq:A21b}, we have nearly arrived at our destination. 
Although it was derived under a large $\ell$ assumption, it is qualitatively accurate for $\ell\ge |m|$.
\par To finish our proof, we need only note that \eqn{eq:A21} pertains to the leading order perturbative contribution, $\sigma^{(\pp)}_{\ell \pm \pp,\ell}$, rather than the full quantity $\sigma_{\ell \pm \pp,\ell}$ only as a matter of asymptotics. 
The full quantity, $\sigma_{\ell \pm \pp,\ell}$, is equal to $\sigma^{(\pp)}_{\ell \pm \pp,\ell}$ plus subdominant higher order terms. 
But as we are only interested in the leading order approximation, \eqn{eq:A21} is equivalent to our prompt, \eqn{apx-sigma-scale}.
\par \Fig{fig:sigma} presents an example of \eqn{apx-sigma-scale} applied to Kerr with $a=0.01$.
There, the assumption that $V^{(\mathrm{S})}\approx -2su$ and the final approximation~(\ceqn{apx-sigma-scale}) are compared to the exact numerical calculation. 
Good agreement is shown. 
%
%
\section{Revisiting the issue of many operators: Physical Spheroidal Harmonics as Eigenfunctions of a Single Operator}
\label{apx:many}
%
The structure of the spherical-spheroidal map $\mct$ and the existence of the physical adjoint spheroidal harmonics imply and intriguing possibility: we should now be able to construct a single operator for which all of the physical spheroidal harmonics are eigenfunctions.
In turn, this raises the possibility that there may be an operator of which the physical adjoint spheroidal harmonics are eigenfunctions.
We explore these possibilities in this section.
Along the way we encounter a so-called ``inter-winding'' relationship indicative of two operators which share eigenvalues~\cite{Shah:2015sva,ROSASORTIZ201826,Andrianov:1984er}.
%
\subsubsection{A unified operator for the physical spheroidal harmonics}
\label{l:1}
%
\par In the presence of spacetime angular momentum, a spacetime's natural modes are spheroidal in nature.
The physical spheroidal harmonics naturally emerge in this context. 
Unlike the fixed oblateness harmonics discussed previously, the physical spheroidal harmonics must be solved simultaneously with a spheroidal radial equation, and as a result have oblatenesses proportional to the polar index, $\ell$~\cite{Yang:2012he}.
Although the physical spheroidal harmonics are considered to be a single set of functions, each of these functions has typically been considered to be the eigenfunction of a distinct spheroidal harmonics operator, $\Lop$. 
This is the operator presented in \eqn{LSc}, and duplicated below for convenience,
\begin{align}
	\label{Lop3}
	\Lop = \left( s(1-s)+(u\gamma_{\k}-s)^2-\frac{(m+su)^2}{1-u^2} \right) + \partial_{u}(1-u^2)\partial_{u} \; .
\end{align}
\par Here, we are motivated by the possibility that bi-orthogonality in such systems is consistent with the existence of a single operator for which all physical spheroidal harmonics are eigenfunctions.
Further, we are motivated by the possibility that this implies the existence of a single such operator for the physical adjoint harmonics, as well as individual operators $\tilde{\mcl}_\k$ for which each $\ket{\tilde{S}_\k}$ is an eigenfunction. 
\par To this end, we start by noting that the physical spherical-spheroidal map $\mct$~(\ceqn{M9}) already has the key properties of such an operator: when acting on select functions with label $\k$, it has the effect of a $\k$-specific operator.
With this in mind, we seek an operator $\mcl$, such that 
\begin{align}
	\label{LopAction}
	\mcl \, \ket{S_\k} \; = \; \Lop \, \ket{S_k} \; = \; -A_\k \, \ket{S_\k} \; \text{for all} \; \k .
\end{align}
%
%
With this in mind, the structure of $\mct$ and the existence of the physical adjoint-harmonics allow for $\mcl$ of the form,
\begin{align}
	\label{Lop2}
	\mcl \; &= \; \sum_{\lp}^\infty \, \mcl_\j \, \ket{S_\j}\bra{\tilde{S}_\j} \; 
	\\
	\label{Lop2b}
	&= \; \sum_{\lp}^{\infty} \, -A_\j \, \ket{S_\j}\bra{\tilde{S}_\j} \; .
\end{align}
\Eqn{Lop2} is required for \eqn{LopAction} to hold, and \eqn{Lop2b} is simply the matrix representation of $\mcl$ in the bases of spheroidal harmonics (rows) and adjoint spheroidals (columns).
In this way the existence of the adjoint spheroidal harmonics enables the physical spheroidals to be unified under a single operator, $\mcl$.
%
\subsubsection{An operator for the physical adjoint spheroidal harmonics}
\label{l:2}
%
\par We are now interested in whether an analog of the spheroidal harmonic differential equation $\Lop$~(\ceqn{Lop3}) may be constructed for the physical adjoint harmonics. 
Such an operator should be manifestly consistent with the bi-orthogonality between the spheroidal harmonics and their adjoints.
Recalling our discussion of the fixed oblateness spheroidals, it is clear that the adjoint spheroidal harmonics must be eigenfunctions of ${\mcl}$'s adjoint,
\begin{align}
	\label{aLop2}
	\adj{\mcl} \; &= \; \sum_{\lp} \, \ket{\tilde{S}_\j}\bra{S_\j}  \, \adj{\mcl}_\j \; 
	\\ 
	\label{aLop2b}
	&= \; \sum_{\lp} \, -A^*_\j \, \ket{\tilde{S}_\j}\bra{S_\j} \; .
\end{align}
\Eqns{aLop2}{aLop2b} generalize the same relationship for the fixed oblateness harmonics presented during our discussion of the adjoint eigenfunctions under fixed oblateness~(\ceqn{LSc}).
In \eqn{aLop2}, we have used that fact adjugating a product of operators reverses ordering~\cite{lax2002functional,Christensen2003}.
In \eqn{aLop2b}, we have simply adjugated the last statement of \eqn{Lop2}.
\par Interestingly, \eqn{aLop2} communicates that, if there exist operators $\tmcl_\k$ such that $\ket{\tilde{S}_\k}$ are eigenfunctions, then
\begin{align}
	\label{aLop3}
	\sum_{\lp} \, \tmcl_\j \, \ket{\tilde{S}_\j}\bra{S_\j} \; = \; \sum_{\lp} \, \ket{\tilde{S}_\j}\bra{S_\j}  \, \adj{\mcl}_\j \; ,
\end{align}
with 
\begin{align}
	\label{aLop4}
	\tmcl_\j \, \ket{\tilde{S}_\j} \; = \; -A_\j^* \, \ket{\tilde{S}_\j} \; .
\end{align}
Perhaps uninterestingly, \eqn{aLop4} is the generalization of the adjoint eigenvalue relation for the fixed oblateness harmonics~(\ceqn{LSe}).
\Eqn{aLop3} is perhaps more interesting: applying $\bra{S_\k}$ on the left and $\ket{\tilde{S}_\k}$ on the right allows the extraction of terms
\begin{widetext}
\begin{align}
	\label{tL1}
	\sum_{\lp}  \bra{S_\k} \tmcl_\j  \ket{\tilde{S}_\j}\brak{S_\j}{\tilde{S}_\k}  &=  \sum_{\lp}  \brak{S_\k}{\tilde{S}_\j}\bra{S_\j}   \adj{\mcl}_\j  \ket{\tilde{S}_\k}
	\\ \label{tL2}
	\sum_{\lp}  \brak{S_\k}{ \tmcl_\j   \tilde{S}_\j}\brak{S_\j}{\tilde{S}_\k}  &=  \sum_{\lp}  \brak{S_\k}{\tilde{S}_\j}\brak{ {\mcl}_\j   S_\j}{\tilde{S}_\k}
	\\ \label{tL3}
	\brak{S_\k}{ \tmcl_\k \tilde{S}_\k} \; &= \; \brak{ {\mcl}_\k   S_\k}{\tilde{S}_\k} \; .
\end{align}
\end{widetext}
In \eqns{tL1}{tL3} we have taken care to render the connection between $\mcl_\k$ and $\tmcl_\k$, as it may not be immediately clear from \eqns{aLop3}{aLop4}.
In going from \eqn{tL1} to \eqn{tL2} we have grouped operators with harmonics that have the same label. 
In \eqn{tL2}'s right-hand side, we have used the defining property of the adjoint operator~(\ceqn{adjoint-condition}).
In \eqn{tL3} we have applied bi-orthogonality~(\ceqn{ORTH4}).
\par Together, \eqns{LopAction}{tL3} illustrate the required relationships between $\mcl_\k$ and $\tmcl_\k$.
\Eqn{LopAction} and \eqn{aLop4} communicate that $\tmcl_\k$ has the same eigenvalues as $\mcl_\k^*$ , and by \eqn{ajdL} we recall that
\begin{align}
	\label{LadjIsConjL}
	\adj{\mcl}_\k=\mcl^*_\k \; .
\end{align}
In this sense, we say that $\tmcl_\k$ is \textit{isospectral} with $\adj{\mcl}_\k$~\cite{Mostafazadeh:2001jk,ROSASORTIZ201826}.
Finally, \eqn{tL3} communicates that in the case of fixed oblateness, $\tmcl_\k$ would simply be the adjoint of $\mcl_\k$.
\par The condition of isospectrality is most interesting.
Two operators are isospectral if there exists an operator, $\mcp$, with inverse $\mcq$, such that 
\begin{align}
	\label{wind1}
	\mcp \, \amcl_\k \; &= \; \tmcl_\k \, \mcp 
	\\
	\label{wind1b}
	\amcl_\k \, \mcq \; &= \; \mcq \, \tmcl_\k \; ,
\end{align}
or, equivalently, since $\adj{\mcl}_\k=\mcl^*_\k$,
\begin{align}
	\label{wind2}
	\tmcl_\k \; = \; \mcp \, \cmcl_\k \, \mcq \; .
\end{align}
\Eqn{wind1} presents what are called inter-winding relationships~\cite{Andrianov:1984er,ROSASORTIZ201826}.
\Eqn{wind2} communicates that, given $\mcp$ and $\mcq$, we may transform $\cmcl_\k$ into $\tmcl_\k$.
\par The use of \eqns{wind1}{wind1b} is that they relate eigenfunctions of $\cmcl_\k$ to those of $\tmcl_\k$. For example, applying $\ket{S_\k^*}$ on the left of \eqn{wind1} gives
\begin{align}
	\label{PaS}
	\tmcl_\k \, \mcp \, \ket{S^*_\k}  \; &= \; \mcp \, \cmcl_\k \, \ket{S^*_\k}
	\\ 
	\label{PaSb}
	\tmcl_\k \, \mcp \, \ket{S^*_\k}  \; &= \; -A^*_\k \, \mcp \, \ket{S^*_\k} \; .
\end{align}
In going from \eqn{PaS} to \eqn{PaSb}, we apply the eigenvalue relationship appropriate for the conjugate harmonics~(\ceqn{LSe}).
\Eqn{PaSb} communicates that $\mcp \, \ket{S^*_\k}$ is an eigenfunction of $\tmcl_\k$, and so $\mcp \, \ket{S^*_\k}$ must be an \as{} function.
Thus $\mcp$ maps conjugated spheroidal harmonics to physical adjoint harmonics, and $\mcq$ must have the opposite effect  
\begin{align}
	\label{pqAct}
	\ket{\tilde{S}_\k} \; = \; \mcp \, \ket{S^*_\k} \; ,
	\\
	\label{pqActb}
	\ket{S^*_\k} \; = \; \mcq \, \ket{\tilde{S}_\k} \; .
\end{align}
\par Like $\mct$, which uses bi-orthogonality to map spherical harmonics into spheroidals, we may write $\mcp$ and $\mcq$ as a sum over projectors
\begin{align}
	\label{PQ}
	\mcp \; = \; \sum_{\lp} \, \ket{\tilde{S}_\j}\bra{\tilde{S}^*_\j}\;,
	\\
	\label{PQb}
	\mcq \; = \; \sum_{\lp} \, \ket{S^*_\j}\bra{S_\j}\;.
\end{align}
In \eqn{PQ} we have noted and made use of the fact that the complex conjugates of the physical spheroidals and their adjoints are also bi-orthogonal,
\begin{align}
	\label{sassy1}
	\brak{\tilde{S}^*_\j}{S^*_\k} \; = \;  \brak{\tilde{S}_\j}{S_\k}^* \; = \;  \delta_{\lp\l} \; .
\end{align}
Using \eqn{sassy1} it may be easily verified that the $\mcp$ and $\mcq$ of \eqn{PQ} have the properties given in \eqns{pqAct}{pqActb}. 
Together, \eqn{wind2} and \eqn{PQ} allow $\tmcl_\k$ to be written as 
\begin{align}
	\label{wind4}
	\tmcl_\k \; = \; \sum_{\lp,\l''} \, \ket{\tilde{S}_\j}\bra{\tilde{S}^*_\j} \, \cmcl_\k\, \ket{S^*_{\l'' m n}}\bra{S_{\l'' m n}}\;.
\end{align}
\par \Eqn{wind4} presents a matrix representation for $\tmcl_\k$, the operator for which the adjoint-spherical harmonics are eigenfunctions.
It is fair to think of $\tmcl_\k$ as a ``{heterogeneous} adjoint'', as \eqn{wind4} communicates that it relies on multiple physical oblatenesses rather than one.
It may be of future interest to determine whether $\tmcl_\k$ has a linear differential form that does not require prior knowledge of its eigenfunctions.
%
\definecolor{UrlColor}{rgb}{0.75, 0, 0}
\bibliographystyle{apsrev4-1}
\bibliography{p1.bib}

\begin{thebibliography}{63}%
\makeatletter
\providecommand \@ifxundefined [1]{%
 \@ifx{#1\undefined}
}%
\providecommand \@ifnum [1]{%
 \ifnum #1\expandafter \@firstoftwo
 \else \expandafter \@secondoftwo
 \fi
}%
\providecommand \@ifx [1]{%
 \ifx #1\expandafter \@firstoftwo
 \else \expandafter \@secondoftwo
 \fi
}%
\providecommand \natexlab [1]{#1}%
\providecommand \enquote  [1]{``#1''}%
\providecommand \bibnamefont  [1]{#1}%
\providecommand \bibfnamefont [1]{#1}%
\providecommand \citenamefont [1]{#1}%
\providecommand \href@noop [0]{\@secondoftwo}%
\providecommand \href [0]{\begingroup \@sanitize@url \@href}%
\providecommand \@href[1]{\@@startlink{#1}\@@href}%
\providecommand \@@href[1]{\endgroup#1\@@endlink}%
\providecommand \@sanitize@url [0]{\catcode `\\12\catcode `\$12\catcode
  `\&12\catcode `\#12\catcode `\^12\catcode `\_12\catcode `\%12\relax}%
\providecommand \@@startlink[1]{}%
\providecommand \@@endlink[0]{}%
\providecommand \url  [0]{\begingroup\@sanitize@url \@url }%
\providecommand \@url [1]{\endgroup\@href {#1}{\urlprefix }}%
\providecommand \urlprefix  [0]{URL }%
\providecommand \Eprint [0]{\href }%
\providecommand \doibase [0]{http://dx.doi.org/}%
\providecommand \selectlanguage [0]{\@gobble}%
\providecommand \bibinfo  [0]{\@secondoftwo}%
\providecommand \bibfield  [0]{\@secondoftwo}%
\providecommand \translation [1]{[#1]}%
\providecommand \BibitemOpen [0]{}%
\providecommand \bibitemStop [0]{}%
\providecommand \bibitemNoStop [0]{.\EOS\space}%
\providecommand \EOS [0]{\spacefactor3000\relax}%
\providecommand \BibitemShut  [1]{\csname bibitem#1\endcsname}%
\let\auto@bib@innerbib\@empty
\bibitem [{\citenamefont {Abbott}\ \emph {et~al.}(2019)\citenamefont {Abbott}
  \emph {et~al.}}]{LIGOScientific:2018mvr}%
  \BibitemOpen
  \bibfield  {author} {\bibinfo {author} {\bibfnamefont {B.}~\bibnamefont
  {Abbott}} \emph {et~al.} (\bibinfo {collaboration} {LIGO Scientific,
  Virgo}),\ }\href {\doibase 10.1103/PhysRevX.9.031040} {\bibfield  {journal}
  {\bibinfo  {journal} {Phys. Rev. X}\ }\textbf {\bibinfo {volume} {9}},\
  \bibinfo {pages} {031040} (\bibinfo {year} {2019})},\ \Eprint
  {http://arxiv.org/abs/1811.12907} {arXiv:1811.12907 [astro-ph.HE]}
  \BibitemShut {NoStop}%
\bibitem [{\citenamefont {Abbott}\ \emph {et~al.}(2020)\citenamefont {Abbott}
  \emph {et~al.}}]{LIGOScientific:2020stg}%
  \BibitemOpen
  \bibfield  {author} {\bibinfo {author} {\bibfnamefont {R.}~\bibnamefont
  {Abbott}} \emph {et~al.} (\bibinfo {collaboration} {LIGO Scientific,
  Virgo}),\ }\href@noop {} {\  (\bibinfo {year} {2020})},\ \Eprint
  {http://arxiv.org/abs/2004.08342} {arXiv:2004.08342 [astro-ph.HE]}
  \BibitemShut {NoStop}%
\bibitem [{\citenamefont {Teukolsky}(1973)}]{Teukolsky:1973ha}%
  \BibitemOpen
  \bibfield  {author} {\bibinfo {author} {\bibfnamefont {S.~A.}\ \bibnamefont
  {Teukolsky}},\ }\href {\doibase 10.1086/152444} {\bibfield  {journal}
  {\bibinfo  {journal} {Astrophys. J.}\ }\textbf {\bibinfo {volume} {185}},\
  \bibinfo {pages} {635} (\bibinfo {year} {1973})}\BibitemShut {NoStop}%
\bibitem [{\citenamefont {Newman}\ and\ \citenamefont {Penrose}(1966)}]{NP66}%
  \BibitemOpen
  \bibfield  {author} {\bibinfo {author} {\bibfnamefont {E.~T.}\ \bibnamefont
  {Newman}}\ and\ \bibinfo {author} {\bibfnamefont {R.}~\bibnamefont
  {Penrose}},\ }\href {\doibase 10.1063/1.1931221} {\bibfield  {journal}
  {\bibinfo  {journal} {Journal of Mathematical Physics}\ }\textbf {\bibinfo
  {volume} {7}},\ \bibinfo {pages} {863} (\bibinfo {year} {1966})},\ \Eprint
  {http://arxiv.org/abs/https://doi.org/10.1063/1.1931221}
  {https://doi.org/10.1063/1.1931221} \BibitemShut {NoStop}%
\bibitem [{\citenamefont {Thorne}(1980)}]{Thorne:1980}%
  \BibitemOpen
  \bibfield  {author} {\bibinfo {author} {\bibfnamefont {K.~S.}\ \bibnamefont
  {Thorne}},\ }\href {\doibase 10.1103/RevModPhys.52.299} {\bibfield  {journal}
  {\bibinfo  {journal} {Rev. Mod. Phys.}\ }\textbf {\bibinfo {volume} {52}},\
  \bibinfo {pages} {299} (\bibinfo {year} {1980})}\BibitemShut {NoStop}%
\bibitem [{\citenamefont {Ruiz}\ \emph {et~al.}(2008)\citenamefont {Ruiz},
  \citenamefont {Takahashi}, \citenamefont {Alcubierre},\ and\ \citenamefont
  {Nunez}}]{Ruiz:2007yx}%
  \BibitemOpen
  \bibfield  {author} {\bibinfo {author} {\bibfnamefont {M.}~\bibnamefont
  {Ruiz}}, \bibinfo {author} {\bibfnamefont {R.}~\bibnamefont {Takahashi}},
  \bibinfo {author} {\bibfnamefont {M.}~\bibnamefont {Alcubierre}}, \ and\
  \bibinfo {author} {\bibfnamefont {D.}~\bibnamefont {Nunez}},\ }\href
  {\doibase 10.1007/s10714-007-0570-8} {\bibfield  {journal} {\bibinfo
  {journal} {Gen. Rel. Grav.}\ }\textbf {\bibinfo {volume} {40}},\ \bibinfo
  {pages} {2467} (\bibinfo {year} {2008})},\ \Eprint
  {http://arxiv.org/abs/0707.4654} {arXiv:0707.4654 [gr-qc]} \BibitemShut
  {NoStop}%
\bibitem [{\citenamefont {{Breuer}}\ \emph {et~al.}(1977)\citenamefont
  {{Breuer}}, \citenamefont {{Ryan}},\ and\ \citenamefont
  {{Waller}}}]{Breuer1977}%
  \BibitemOpen
  \bibfield  {author} {\bibinfo {author} {\bibfnamefont {R.~A.}\ \bibnamefont
  {{Breuer}}}, \bibinfo {author} {\bibfnamefont {M.~P.}\ \bibnamefont {{Ryan}},
  \bibfnamefont {Jr.}}, \ and\ \bibinfo {author} {\bibfnamefont
  {S.}~\bibnamefont {{Waller}}},\ }\href {\doibase 10.1098/rspa.1977.0187}
  {\bibfield  {journal} {\bibinfo  {journal} {Proceedings of the Royal Society
  of London Series A}\ }\textbf {\bibinfo {volume} {358}},\ \bibinfo {pages}
  {71} (\bibinfo {year} {1977})}\BibitemShut {NoStop}%
\bibitem [{\citenamefont {London}\ \emph {et~al.}(2014)\citenamefont {London},
  \citenamefont {Shoemaker},\ and\ \citenamefont {Healy}}]{London:2014cma}%
  \BibitemOpen
  \bibfield  {author} {\bibinfo {author} {\bibfnamefont {L.}~\bibnamefont
  {London}}, \bibinfo {author} {\bibfnamefont {D.}~\bibnamefont {Shoemaker}}, \
  and\ \bibinfo {author} {\bibfnamefont {J.}~\bibnamefont {Healy}},\ }\href
  {\doibase 10.1103/PhysRevD.90.124032} {\bibfield  {journal} {\bibinfo
  {journal} {Phys. Rev.}\ }\textbf {\bibinfo {volume} {D90}},\ \bibinfo {pages}
  {124032} (\bibinfo {year} {2014})},\ \Eprint {http://arxiv.org/abs/1404.3197}
  {arXiv:1404.3197 [gr-qc]} \BibitemShut {NoStop}%
\bibitem [{\citenamefont {Leaver}(1985)}]{leaver85}%
  \BibitemOpen
  \bibfield  {author} {\bibinfo {author} {\bibfnamefont {E.}~\bibnamefont
  {Leaver}},\ }\href {\doibase 10.1098/rspa.1985.0119} {\bibfield  {journal}
  {\bibinfo  {journal} {Proc. Roy. Soc. Lond. A}\ }\textbf {\bibinfo {volume}
  {A402}},\ \bibinfo {pages} {285} (\bibinfo {year} {1985})}\BibitemShut
  {NoStop}%
\bibitem [{\citenamefont {Hughes}(2000)}]{Hughes:1999bq}%
  \BibitemOpen
  \bibfield  {author} {\bibinfo {author} {\bibfnamefont {S.~A.}\ \bibnamefont
  {Hughes}},\ }\href {\doibase 10.1103/PhysRevD.65.069902} {\bibfield
  {journal} {\bibinfo  {journal} {Phys. Rev. D}\ }\textbf {\bibinfo {volume}
  {61}},\ \bibinfo {pages} {084004} (\bibinfo {year} {2000})},\ \bibinfo {note}
  {[Erratum: Phys.Rev.D 63, 049902 (2001), Erratum: Phys.Rev.D 65, 069902
  (2002), Erratum: Phys.Rev.D 67, 089901 (2003), Erratum: Phys.Rev.D 78, 109902
  (2008), Erratum: Phys.Rev.D 90, 109904 (2014)]},\ \Eprint
  {http://arxiv.org/abs/gr-qc/9910091} {arXiv:gr-qc/9910091} \BibitemShut
  {NoStop}%
\bibitem [{\citenamefont {London}\ \emph {et~al.}(2018)\citenamefont {London},
  \citenamefont {Khan}, \citenamefont {Fauchon-Jones}, \citenamefont {García},
  \citenamefont {Hannam}, \citenamefont {Husa}, \citenamefont
  {Jiménez-Forteza}, \citenamefont {Kalaghatgi}, \citenamefont {Ohme},\ and\
  \citenamefont {Pannarale}}]{London:2017bcn}%
  \BibitemOpen
  \bibfield  {author} {\bibinfo {author} {\bibfnamefont {L.}~\bibnamefont
  {London}}, \bibinfo {author} {\bibfnamefont {S.}~\bibnamefont {Khan}},
  \bibinfo {author} {\bibfnamefont {E.}~\bibnamefont {Fauchon-Jones}}, \bibinfo
  {author} {\bibfnamefont {C.}~\bibnamefont {García}}, \bibinfo {author}
  {\bibfnamefont {M.}~\bibnamefont {Hannam}}, \bibinfo {author} {\bibfnamefont
  {S.}~\bibnamefont {Husa}}, \bibinfo {author} {\bibfnamefont {X.}~\bibnamefont
  {Jiménez-Forteza}}, \bibinfo {author} {\bibfnamefont {C.}~\bibnamefont
  {Kalaghatgi}}, \bibinfo {author} {\bibfnamefont {F.}~\bibnamefont {Ohme}}, \
  and\ \bibinfo {author} {\bibfnamefont {F.}~\bibnamefont {Pannarale}},\ }\href
  {\doibase 10.1103/PhysRevLett.120.161102} {\bibfield  {journal} {\bibinfo
  {journal} {Phys. Rev. Lett.}\ }\textbf {\bibinfo {volume} {120}},\ \bibinfo
  {pages} {161102} (\bibinfo {year} {2018})},\ \Eprint
  {http://arxiv.org/abs/1708.00404} {arXiv:1708.00404 [gr-qc]} \BibitemShut
  {NoStop}%
\bibitem [{\citenamefont {Holzegel}\ and\ \citenamefont
  {Smulevici}(2013)}]{Holzegel:2013kna}%
  \BibitemOpen
  \bibfield  {author} {\bibinfo {author} {\bibfnamefont {G.}~\bibnamefont
  {Holzegel}}\ and\ \bibinfo {author} {\bibfnamefont {J.}~\bibnamefont
  {Smulevici}},\ }\href {\doibase 10.2140/apde.2014.7.1057} {\  (\bibinfo
  {year} {2013}),\ 10.2140/apde.2014.7.1057},\ \Eprint
  {http://arxiv.org/abs/1303.5944} {arXiv:1303.5944 [gr-qc]} \BibitemShut
  {NoStop}%
\bibitem [{\citenamefont {Berti}\ \emph
  {et~al.}(2006{\natexlab{a}})\citenamefont {Berti}, \citenamefont {Cardoso},\
  and\ \citenamefont {Casals}}]{Berti:2005gp}%
  \BibitemOpen
  \bibfield  {author} {\bibinfo {author} {\bibfnamefont {E.}~\bibnamefont
  {Berti}}, \bibinfo {author} {\bibfnamefont {V.}~\bibnamefont {Cardoso}}, \
  and\ \bibinfo {author} {\bibfnamefont {M.}~\bibnamefont {Casals}},\ }\href
  {\doibase 10.1103/PhysRevD.73.109902, 10.1103/PhysRevD.73.024013} {\bibfield
  {journal} {\bibinfo  {journal} {Phys. Rev.}\ }\textbf {\bibinfo {volume}
  {D73}},\ \bibinfo {pages} {024013} (\bibinfo {year} {2006}{\natexlab{a}})},\
  \bibinfo {note} {[Erratum: Phys. Rev.D73,109902(2006)]},\ \Eprint
  {http://arxiv.org/abs/gr-qc/0511111} {arXiv:gr-qc/0511111 [gr-qc]}
  \BibitemShut {NoStop}%
\bibitem [{\citenamefont {Berti}\ and\ \citenamefont
  {Klein}(2014)}]{Berti:2014fga}%
  \BibitemOpen
  \bibfield  {author} {\bibinfo {author} {\bibfnamefont {E.}~\bibnamefont
  {Berti}}\ and\ \bibinfo {author} {\bibfnamefont {A.}~\bibnamefont {Klein}},\
  }\href {\doibase 10.1103/PhysRevD.90.064012} {\bibfield  {journal} {\bibinfo
  {journal} {Phys. Rev. D}\ }\textbf {\bibinfo {volume} {90}},\ \bibinfo
  {pages} {064012} (\bibinfo {year} {2014})},\ \Eprint
  {http://arxiv.org/abs/1408.1860} {arXiv:1408.1860 [gr-qc]} \BibitemShut
  {NoStop}%
\bibitem [{\citenamefont {Kelly}\ and\ \citenamefont
  {Baker}(2013)}]{Kelly:2012nd}%
  \BibitemOpen
  \bibfield  {author} {\bibinfo {author} {\bibfnamefont {B.~J.}\ \bibnamefont
  {Kelly}}\ and\ \bibinfo {author} {\bibfnamefont {J.~G.}\ \bibnamefont
  {Baker}},\ }\href {\doibase 10.1103/PhysRevD.87.084004} {\bibfield  {journal}
  {\bibinfo  {journal} {Phys. Rev. D}\ }\textbf {\bibinfo {volume} {87}},\
  \bibinfo {pages} {084004} (\bibinfo {year} {2013})},\ \Eprint
  {http://arxiv.org/abs/1212.5553} {arXiv:1212.5553 [gr-qc]} \BibitemShut
  {NoStop}%
\bibitem [{\citenamefont {García-Quirós}\ \emph {et~al.}(2020)\citenamefont
  {García-Quirós}, \citenamefont {Colleoni}, \citenamefont {Husa},
  \citenamefont {Estellés}, \citenamefont {Pratten}, \citenamefont
  {Ramos-Buades}, \citenamefont {Mateu-Lucena},\ and\ \citenamefont
  {Jaume}}]{Garcia-Quiros:2020qpx}%
  \BibitemOpen
  \bibfield  {author} {\bibinfo {author} {\bibfnamefont {C.}~\bibnamefont
  {García-Quirós}}, \bibinfo {author} {\bibfnamefont {M.}~\bibnamefont
  {Colleoni}}, \bibinfo {author} {\bibfnamefont {S.}~\bibnamefont {Husa}},
  \bibinfo {author} {\bibfnamefont {H.}~\bibnamefont {Estellés}}, \bibinfo
  {author} {\bibfnamefont {G.}~\bibnamefont {Pratten}}, \bibinfo {author}
  {\bibfnamefont {A.}~\bibnamefont {Ramos-Buades}}, \bibinfo {author}
  {\bibfnamefont {M.}~\bibnamefont {Mateu-Lucena}}, \ and\ \bibinfo {author}
  {\bibfnamefont {R.}~\bibnamefont {Jaume}},\ }\href@noop {} {\  (\bibinfo
  {year} {2020})},\ \Eprint {http://arxiv.org/abs/2001.10914} {arXiv:2001.10914
  [gr-qc]} \BibitemShut {NoStop}%
\bibitem [{\citenamefont {Blackman}\ \emph {et~al.}(2017)\citenamefont
  {Blackman}, \citenamefont {Field}, \citenamefont {Scheel}, \citenamefont
  {Galley}, \citenamefont {Ott}, \citenamefont {Boyle}, \citenamefont {Kidder},
  \citenamefont {Pfeiffer},\ and\ \citenamefont
  {Szilágyi}}]{Blackman:2017pcm}%
  \BibitemOpen
  \bibfield  {author} {\bibinfo {author} {\bibfnamefont {J.}~\bibnamefont
  {Blackman}}, \bibinfo {author} {\bibfnamefont {S.~E.}\ \bibnamefont {Field}},
  \bibinfo {author} {\bibfnamefont {M.~A.}\ \bibnamefont {Scheel}}, \bibinfo
  {author} {\bibfnamefont {C.~R.}\ \bibnamefont {Galley}}, \bibinfo {author}
  {\bibfnamefont {C.~D.}\ \bibnamefont {Ott}}, \bibinfo {author} {\bibfnamefont
  {M.}~\bibnamefont {Boyle}}, \bibinfo {author} {\bibfnamefont {L.~E.}\
  \bibnamefont {Kidder}}, \bibinfo {author} {\bibfnamefont {H.~P.}\
  \bibnamefont {Pfeiffer}}, \ and\ \bibinfo {author} {\bibfnamefont
  {B.}~\bibnamefont {Szilágyi}},\ }\href {\doibase 10.1103/PhysRevD.96.024058}
  {\bibfield  {journal} {\bibinfo  {journal} {Phys. Rev. D}\ }\textbf {\bibinfo
  {volume} {96}},\ \bibinfo {pages} {024058} (\bibinfo {year} {2017})},\
  \Eprint {http://arxiv.org/abs/1705.07089} {arXiv:1705.07089 [gr-qc]}
  \BibitemShut {NoStop}%
\bibitem [{\citenamefont {Mehta}\ \emph {et~al.}(2019)\citenamefont {Mehta},
  \citenamefont {Tiwari}, \citenamefont {Johnson-McDaniel}, \citenamefont
  {Mishra}, \citenamefont {Varma},\ and\ \citenamefont
  {Ajith}}]{Mehta:2019wxm}%
  \BibitemOpen
  \bibfield  {author} {\bibinfo {author} {\bibfnamefont {A.~K.}\ \bibnamefont
  {Mehta}}, \bibinfo {author} {\bibfnamefont {P.}~\bibnamefont {Tiwari}},
  \bibinfo {author} {\bibfnamefont {N.~K.}\ \bibnamefont {Johnson-McDaniel}},
  \bibinfo {author} {\bibfnamefont {C.~K.}\ \bibnamefont {Mishra}}, \bibinfo
  {author} {\bibfnamefont {V.}~\bibnamefont {Varma}}, \ and\ \bibinfo {author}
  {\bibfnamefont {P.}~\bibnamefont {Ajith}},\ }\href {\doibase
  10.1103/PhysRevD.100.024032} {\bibfield  {journal} {\bibinfo  {journal}
  {Phys. Rev.}\ }\textbf {\bibinfo {volume} {D100}},\ \bibinfo {pages} {024032}
  (\bibinfo {year} {2019})},\ \Eprint {http://arxiv.org/abs/1902.02731}
  {arXiv:1902.02731 [gr-qc]} \BibitemShut {NoStop}%
\bibitem [{\citenamefont {Carullo}\ \emph {et~al.}(2018)\citenamefont {Carullo}
  \emph {et~al.}}]{Carullo:2018sfu}%
  \BibitemOpen
  \bibfield  {author} {\bibinfo {author} {\bibfnamefont {G.}~\bibnamefont
  {Carullo}} \emph {et~al.},\ }\href {\doibase 10.1103/PhysRevD.98.104020}
  {\bibfield  {journal} {\bibinfo  {journal} {Phys. Rev. D}\ }\textbf {\bibinfo
  {volume} {98}},\ \bibinfo {pages} {104020} (\bibinfo {year} {2018})},\
  \Eprint {http://arxiv.org/abs/1805.04760} {arXiv:1805.04760 [gr-qc]}
  \BibitemShut {NoStop}%
\bibitem [{\citenamefont {Ghosh}\ \emph {et~al.}(2016)\citenamefont {Ghosh}
  \emph {et~al.}}]{Ghosh:2016qgn}%
  \BibitemOpen
  \bibfield  {author} {\bibinfo {author} {\bibfnamefont {A.}~\bibnamefont
  {Ghosh}} \emph {et~al.},\ }\href {\doibase 10.1103/PhysRevD.94.021101}
  {\bibfield  {journal} {\bibinfo  {journal} {Phys. Rev.}\ }\textbf {\bibinfo
  {volume} {D94}},\ \bibinfo {pages} {021101} (\bibinfo {year} {2016})},\
  \Eprint {http://arxiv.org/abs/1602.02453} {arXiv:1602.02453 [gr-qc]}
  \BibitemShut {NoStop}%
\bibitem [{\citenamefont {Ota}\ and\ \citenamefont
  {Chirenti}(2020)}]{Ota:2019bzl}%
  \BibitemOpen
  \bibfield  {author} {\bibinfo {author} {\bibfnamefont {I.}~\bibnamefont
  {Ota}}\ and\ \bibinfo {author} {\bibfnamefont {C.}~\bibnamefont {Chirenti}},\
  }\href {\doibase 10.1103/PhysRevD.101.104005} {\bibfield  {journal} {\bibinfo
   {journal} {Phys. Rev. D}\ }\textbf {\bibinfo {volume} {101}},\ \bibinfo
  {pages} {104005} (\bibinfo {year} {2020})},\ \Eprint
  {http://arxiv.org/abs/1911.00440} {arXiv:1911.00440 [gr-qc]} \BibitemShut
  {NoStop}%
\bibitem [{\citenamefont {Bhagwat}\ \emph {et~al.}(2020)\citenamefont
  {Bhagwat}, \citenamefont {Forteza}, \citenamefont {Pani},\ and\ \citenamefont
  {Ferrari}}]{Bhagwat:2019dtm}%
  \BibitemOpen
  \bibfield  {author} {\bibinfo {author} {\bibfnamefont {S.}~\bibnamefont
  {Bhagwat}}, \bibinfo {author} {\bibfnamefont {X.~J.}\ \bibnamefont
  {Forteza}}, \bibinfo {author} {\bibfnamefont {P.}~\bibnamefont {Pani}}, \
  and\ \bibinfo {author} {\bibfnamefont {V.}~\bibnamefont {Ferrari}},\ }\href
  {\doibase 10.1103/PhysRevD.101.044033} {\bibfield  {journal} {\bibinfo
  {journal} {Phys. Rev. D}\ }\textbf {\bibinfo {volume} {101}},\ \bibinfo
  {pages} {044033} (\bibinfo {year} {2020})},\ \Eprint
  {http://arxiv.org/abs/1910.08708} {arXiv:1910.08708 [gr-qc]} \BibitemShut
  {NoStop}%
\bibitem [{\citenamefont {Mino}\ \emph {et~al.}(1997)\citenamefont {Mino},
  \citenamefont {Sasaki}, \citenamefont {Shibata}, \citenamefont {Tagoshi},\
  and\ \citenamefont {Tanaka}}]{Mino:1997bx}%
  \BibitemOpen
  \bibfield  {author} {\bibinfo {author} {\bibfnamefont {Y.}~\bibnamefont
  {Mino}}, \bibinfo {author} {\bibfnamefont {M.}~\bibnamefont {Sasaki}},
  \bibinfo {author} {\bibfnamefont {M.}~\bibnamefont {Shibata}}, \bibinfo
  {author} {\bibfnamefont {H.}~\bibnamefont {Tagoshi}}, \ and\ \bibinfo
  {author} {\bibfnamefont {T.}~\bibnamefont {Tanaka}},\ }\href {\doibase
  10.1143/PTPS.128.1} {\bibfield  {journal} {\bibinfo  {journal} {Progress of
  Theoretical Physics Supplement}\ }\textbf {\bibinfo {volume} {128}},\
  \bibinfo {pages} {1} (\bibinfo {year} {1997})},\ \Eprint
  {http://arxiv.org/abs/http://oup.prod.sis.lan/ptps/article-pdf/doi/10.1143/PTPS.128.1/5438984/128-1.pdf}
  {http://oup.prod.sis.lan/ptps/article-pdf/doi/10.1143/PTPS.128.1/5438984/128-1.pdf}
  \BibitemShut {NoStop}%
\bibitem [{\citenamefont {O'Sullivan}\ and\ \citenamefont
  {Hughes}(2014)}]{OSullivan:2014ywd}%
  \BibitemOpen
  \bibfield  {author} {\bibinfo {author} {\bibfnamefont {S.}~\bibnamefont
  {O'Sullivan}}\ and\ \bibinfo {author} {\bibfnamefont {S.~A.}\ \bibnamefont
  {Hughes}},\ }\href {\doibase 10.1103/PhysRevD.91.109901} {\bibfield
  {journal} {\bibinfo  {journal} {Phys. Rev. D}\ }\textbf {\bibinfo {volume}
  {90}},\ \bibinfo {pages} {124039} (\bibinfo {year} {2014})},\ \bibinfo {note}
  {[Erratum: Phys.Rev.D 91, 109901 (2015)]},\ \Eprint
  {http://arxiv.org/abs/1407.6983} {arXiv:1407.6983 [gr-qc]} \BibitemShut
  {NoStop}%
\bibitem [{\citenamefont {Blanchet}(2014)}]{Blanchet:2013haa}%
  \BibitemOpen
  \bibfield  {author} {\bibinfo {author} {\bibfnamefont {L.}~\bibnamefont
  {Blanchet}},\ }\href {\doibase 10.12942/lrr-2014-2} {\bibfield  {journal}
  {\bibinfo  {journal} {Living Rev. Rel.}\ }\textbf {\bibinfo {volume} {17}},\
  \bibinfo {pages} {2} (\bibinfo {year} {2014})},\ \Eprint
  {http://arxiv.org/abs/1310.1528} {arXiv:1310.1528 [gr-qc]} \BibitemShut
  {NoStop}%
\bibitem [{\citenamefont {London}(2021)}]{London:2021P2}%
  \BibitemOpen
  \bibfield  {author} {\bibinfo {author} {\bibfnamefont {L.}~\bibnamefont
  {London}},\ }\href@noop {} {\enquote {\bibinfo {title} {{Bi-orthogonal
  harmonics for the decomposition of gravitational radiation II: applications
  for extreme and comparable mass-ratio black hole binaries (Submittted to
  PRL)}},}\ } (\bibinfo {year} {2021})\BibitemShut {NoStop}%
\bibitem [{\citenamefont {Brauer}(1964)}]{brauer1964}%
  \BibitemOpen
  \bibfield  {author} {\bibinfo {author} {\bibfnamefont {F.}~\bibnamefont
  {Brauer}},\ }\href {\doibase 10.1307/mmj/1028999193} {\bibfield  {journal}
  {\bibinfo  {journal} {Michigan Math. J.}\ }\textbf {\bibinfo {volume} {11}},\
  \bibinfo {pages} {379} (\bibinfo {year} {1964})}\BibitemShut {NoStop}%
\bibitem [{\citenamefont {Christensen}(2003)}]{Christensen2003}%
  \BibitemOpen
  \bibfield  {author} {\bibinfo {author} {\bibfnamefont {O.}~\bibnamefont
  {Christensen}},\ }\href {\doibase 10.1007/978-0-8176-8224-8} {\emph {\bibinfo
  {title} {An Introduction to Frames and Riesz Bases}}}\ (\bibinfo  {publisher}
  {Birkh\"{a}user Boston},\ \bibinfo {year} {2003})\BibitemShut {NoStop}%
\bibitem [{\citenamefont {Green}\ \emph {et~al.}(2022)\citenamefont {Green}
  \emph {et~al.}}]{Green}%
  \BibitemOpen
  \bibfield  {author} {\bibinfo {author} {\bibfnamefont {S.}~\bibnamefont
  {Green}} \emph {et~al.},\ }\href@noop {} {\enquote {\bibinfo {title} {In
  preparation.}}\ } (\bibinfo {year} {2022})\BibitemShut {NoStop}%
\bibitem [{\citenamefont {Sberna}\ \emph {et~al.}(2022)\citenamefont {Sberna},
  \citenamefont {Bosch}, \citenamefont {East}, \citenamefont {Green},\ and\
  \citenamefont {Lehner}}]{Sberna:2021eui}%
  \BibitemOpen
  \bibfield  {author} {\bibinfo {author} {\bibfnamefont {L.}~\bibnamefont
  {Sberna}}, \bibinfo {author} {\bibfnamefont {P.}~\bibnamefont {Bosch}},
  \bibinfo {author} {\bibfnamefont {W.~E.}\ \bibnamefont {East}}, \bibinfo
  {author} {\bibfnamefont {S.~R.}\ \bibnamefont {Green}}, \ and\ \bibinfo
  {author} {\bibfnamefont {L.}~\bibnamefont {Lehner}},\ }\href {\doibase
  10.1103/PhysRevD.105.064046} {\bibfield  {journal} {\bibinfo  {journal}
  {Phys. Rev. D}\ }\textbf {\bibinfo {volume} {105}},\ \bibinfo {pages}
  {064046} (\bibinfo {year} {2022})},\ \Eprint
  {http://arxiv.org/abs/2112.11168} {arXiv:2112.11168 [gr-qc]} \BibitemShut
  {NoStop}%
\bibitem [{\citenamefont {London}\ \emph {et~al.}(2020)\citenamefont {London},
  \citenamefont {Fauchon},\ and\ \citenamefont {Hamilton}}]{positive:2020}%
  \BibitemOpen
  \bibfield  {author} {\bibinfo {author} {\bibfnamefont {L.}~\bibnamefont
  {London}}, \bibinfo {author} {\bibfnamefont {E.}~\bibnamefont {Fauchon}}, \
  and\ \bibinfo {author} {\bibfnamefont {E.}~\bibnamefont {Hamilton}},\ }\href
  {\doibase 10.5281/zenodo.3901856} {\enquote {\bibinfo {title}
  {llondon6/positive: map},}\ } (\bibinfo {year} {2020})\BibitemShut {NoStop}%
\bibitem [{\citenamefont {Courant}\ and\ \citenamefont
  {Hilbert.}(1954)}]{Courant1954}%
  \BibitemOpen
  \bibfield  {author} {\bibinfo {author} {\bibfnamefont {R.}~\bibnamefont
  {Courant}}\ and\ \bibinfo {author} {\bibfnamefont {D.}~\bibnamefont
  {Hilbert.}},\ }\href {\doibase 10.1002/qj.49708034534} {\bibfield  {journal}
  {\bibinfo  {journal} {Quarterly Journal of the Royal Meteorological Society}\
  }\textbf {\bibinfo {volume} {80}},\ \bibinfo {pages} {485} (\bibinfo {year}
  {1954})}\BibitemShut {NoStop}%
\bibitem [{\citenamefont {Berti}\ \emph
  {et~al.}(2006{\natexlab{b}})\citenamefont {Berti}, \citenamefont {Cardoso},\
  and\ \citenamefont {Will}}]{Berti:2005ys}%
  \BibitemOpen
  \bibfield  {author} {\bibinfo {author} {\bibfnamefont {E.}~\bibnamefont
  {Berti}}, \bibinfo {author} {\bibfnamefont {V.}~\bibnamefont {Cardoso}}, \
  and\ \bibinfo {author} {\bibfnamefont {C.~M.}\ \bibnamefont {Will}},\ }\href
  {\doibase 10.1103/PhysRevD.73.064030} {\bibfield  {journal} {\bibinfo
  {journal} {Phys. Rev.}\ }\textbf {\bibinfo {volume} {D73}},\ \bibinfo {pages}
  {064030} (\bibinfo {year} {2006}{\natexlab{b}})},\ \Eprint
  {http://arxiv.org/abs/gr-qc/0512160} {arXiv:gr-qc/0512160 [gr-qc]}
  \BibitemShut {NoStop}%
\bibitem [{\citenamefont {Nollert}(1999)}]{Nollert:1999ji}%
  \BibitemOpen
  \bibfield  {author} {\bibinfo {author} {\bibfnamefont {H.-P.}\ \bibnamefont
  {Nollert}},\ }\href {\doibase 10.1088/0264-9381/16/12/201} {\bibfield
  {journal} {\bibinfo  {journal} {Class. Quant. Grav.}\ }\textbf {\bibinfo
  {volume} {16}},\ \bibinfo {pages} {R159} (\bibinfo {year}
  {1999})}\BibitemShut {NoStop}%
\bibitem [{\citenamefont {Andersson}(1997)}]{Andersson:1996cm}%
  \BibitemOpen
  \bibfield  {author} {\bibinfo {author} {\bibfnamefont {N.}~\bibnamefont
  {Andersson}},\ }\href {\doibase 10.1103/PhysRevD.55.468} {\bibfield
  {journal} {\bibinfo  {journal} {Phys. Rev. D}\ }\textbf {\bibinfo {volume}
  {55}},\ \bibinfo {pages} {468} (\bibinfo {year} {1997})},\ \Eprint
  {http://arxiv.org/abs/gr-qc/9607064} {arXiv:gr-qc/9607064} \BibitemShut
  {NoStop}%
\bibitem [{\citenamefont {Bhagwat}\ \emph {et~al.}(2017)\citenamefont
  {Bhagwat}, \citenamefont {Okounkova}, \citenamefont {Ballmer}, \citenamefont
  {Brown}, \citenamefont {Giesler}, \citenamefont {Scheel},\ and\ \citenamefont
  {Teukolsky}}]{Bhagwat:2017tkm}%
  \BibitemOpen
  \bibfield  {author} {\bibinfo {author} {\bibfnamefont {S.}~\bibnamefont
  {Bhagwat}}, \bibinfo {author} {\bibfnamefont {M.}~\bibnamefont {Okounkova}},
  \bibinfo {author} {\bibfnamefont {S.~W.}\ \bibnamefont {Ballmer}}, \bibinfo
  {author} {\bibfnamefont {D.~A.}\ \bibnamefont {Brown}}, \bibinfo {author}
  {\bibfnamefont {M.}~\bibnamefont {Giesler}}, \bibinfo {author} {\bibfnamefont
  {M.~A.}\ \bibnamefont {Scheel}}, \ and\ \bibinfo {author} {\bibfnamefont
  {S.~A.}\ \bibnamefont {Teukolsky}},\ }\href@noop {} {\  (\bibinfo {year}
  {2017})},\ \Eprint {http://arxiv.org/abs/1711.00926} {arXiv:1711.00926
  [gr-qc]} \BibitemShut {NoStop}%
\bibitem [{\citenamefont {Mostafazadeh}(2002)}]{Mostafazadeh:2001jk}%
  \BibitemOpen
  \bibfield  {author} {\bibinfo {author} {\bibfnamefont {A.}~\bibnamefont
  {Mostafazadeh}},\ }\href {\doibase 10.1063/1.1418246} {\bibfield  {journal}
  {\bibinfo  {journal} {J. Math. Phys.}\ }\textbf {\bibinfo {volume} {43}},\
  \bibinfo {pages} {205} (\bibinfo {year} {2002})},\ \Eprint
  {http://arxiv.org/abs/math-ph/0107001} {arXiv:math-ph/0107001} \BibitemShut
  {NoStop}%
\bibitem [{\citenamefont {Shah}\ and\ \citenamefont
  {Whiting}(2016)}]{Shah:2015sva}%
  \BibitemOpen
  \bibfield  {author} {\bibinfo {author} {\bibfnamefont {A.~G.}\ \bibnamefont
  {Shah}}\ and\ \bibinfo {author} {\bibfnamefont {B.~F.}\ \bibnamefont
  {Whiting}},\ }\href {\doibase 10.1007/s10714-016-2064-z} {\bibfield
  {journal} {\bibinfo  {journal} {Gen. Rel. Grav.}\ }\textbf {\bibinfo {volume}
  {48}},\ \bibinfo {pages} {78} (\bibinfo {year} {2016})},\ \Eprint
  {http://arxiv.org/abs/1503.02618} {arXiv:1503.02618 [gr-qc]} \BibitemShut
  {NoStop}%
\bibitem [{\citenamefont {Cook}\ and\ \citenamefont
  {Zalutskiy}(2014)}]{Cook:2014cta}%
  \BibitemOpen
  \bibfield  {author} {\bibinfo {author} {\bibfnamefont {G.~B.}\ \bibnamefont
  {Cook}}\ and\ \bibinfo {author} {\bibfnamefont {M.}~\bibnamefont
  {Zalutskiy}},\ }\href {\doibase 10.1103/PhysRevD.90.124021} {\bibfield
  {journal} {\bibinfo  {journal} {Phys. Rev. D}\ }\textbf {\bibinfo {volume}
  {90}},\ \bibinfo {pages} {124021} (\bibinfo {year} {2014})},\ \Eprint
  {http://arxiv.org/abs/1410.7698} {arXiv:1410.7698 [gr-qc]} \BibitemShut
  {NoStop}%
\bibitem [{\citenamefont {London}\ and\ \citenamefont
  {Fauchon-Jones}(2019)}]{London:2018nxs}%
  \BibitemOpen
  \bibfield  {author} {\bibinfo {author} {\bibfnamefont {L.}~\bibnamefont
  {London}}\ and\ \bibinfo {author} {\bibfnamefont {E.}~\bibnamefont
  {Fauchon-Jones}},\ }\href {\doibase 10.1088/1361-6382/ab2f11} {\bibfield
  {journal} {\bibinfo  {journal} {Class. Quant. Grav.}\ }\textbf {\bibinfo
  {volume} {36}},\ \bibinfo {pages} {235015} (\bibinfo {year} {2019})},\
  \Eprint {http://arxiv.org/abs/1810.03550} {arXiv:1810.03550 [gr-qc]}
  \BibitemShut {NoStop}%
\bibitem [{\citenamefont {Maga\~na Zertuche}\ \emph {et~al.}(2021)\citenamefont
  {Maga\~na Zertuche} \emph {et~al.}}]{MaganaZertuche:2021syq}%
  \BibitemOpen
  \bibfield  {author} {\bibinfo {author} {\bibfnamefont {L.}~\bibnamefont
  {Maga\~na Zertuche}} \emph {et~al.},\ }\href@noop {} {\  (\bibinfo {year}
  {2021})},\ \Eprint {http://arxiv.org/abs/2110.15922} {arXiv:2110.15922
  [gr-qc]} \BibitemShut {NoStop}%
\bibitem [{\citenamefont {Kamaretsos}\ \emph
  {et~al.}(2012{\natexlab{a}})\citenamefont {Kamaretsos}, \citenamefont
  {Hannam}, \citenamefont {Husa},\ and\ \citenamefont
  {Sathyaprakash}}]{Kamaretsos:2011um}%
  \BibitemOpen
  \bibfield  {author} {\bibinfo {author} {\bibfnamefont {I.}~\bibnamefont
  {Kamaretsos}}, \bibinfo {author} {\bibfnamefont {M.}~\bibnamefont {Hannam}},
  \bibinfo {author} {\bibfnamefont {S.}~\bibnamefont {Husa}}, \ and\ \bibinfo
  {author} {\bibfnamefont {B.}~\bibnamefont {Sathyaprakash}},\ }\href {\doibase
  10.1103/PhysRevD.85.024018} {\bibfield  {journal} {\bibinfo  {journal}
  {Phys.Rev.}\ }\textbf {\bibinfo {volume} {D85}},\ \bibinfo {pages} {024018}
  (\bibinfo {year} {2012}{\natexlab{a}})},\ \Eprint
  {http://arxiv.org/abs/1107.0854} {arXiv:1107.0854 [gr-qc]} \BibitemShut
  {NoStop}%
\bibitem [{\citenamefont {Husa}\ \emph {et~al.}(2016)\citenamefont {Husa},
  \citenamefont {Khan}, \citenamefont {Hannam}, \citenamefont {Pürrer},
  \citenamefont {Ohme}, \citenamefont {Jiménez~Forteza},\ and\ \citenamefont
  {Bohé}}]{Husa:2015iqa}%
  \BibitemOpen
  \bibfield  {author} {\bibinfo {author} {\bibfnamefont {S.}~\bibnamefont
  {Husa}}, \bibinfo {author} {\bibfnamefont {S.}~\bibnamefont {Khan}}, \bibinfo
  {author} {\bibfnamefont {M.}~\bibnamefont {Hannam}}, \bibinfo {author}
  {\bibfnamefont {M.}~\bibnamefont {Pürrer}}, \bibinfo {author} {\bibfnamefont
  {F.}~\bibnamefont {Ohme}}, \bibinfo {author} {\bibfnamefont {X.}~\bibnamefont
  {Jiménez~Forteza}}, \ and\ \bibinfo {author} {\bibfnamefont
  {A.}~\bibnamefont {Bohé}},\ }\href {\doibase 10.1103/PhysRevD.93.044006}
  {\bibfield  {journal} {\bibinfo  {journal} {Phys. Rev.}\ }\textbf {\bibinfo
  {volume} {D93}},\ \bibinfo {pages} {044006} (\bibinfo {year} {2016})},\
  \Eprint {http://arxiv.org/abs/1508.07250} {arXiv:1508.07250 [gr-qc]}
  \BibitemShut {NoStop}%
\bibitem [{\citenamefont {Khan}\ \emph {et~al.}(2016)\citenamefont {Khan},
  \citenamefont {Husa}, \citenamefont {Hannam}, \citenamefont {Ohme},
  \citenamefont {Pürrer}, \citenamefont {Jiménez~Forteza},\ and\
  \citenamefont {Bohé}}]{Khan:2015jqa}%
  \BibitemOpen
  \bibfield  {author} {\bibinfo {author} {\bibfnamefont {S.}~\bibnamefont
  {Khan}}, \bibinfo {author} {\bibfnamefont {S.}~\bibnamefont {Husa}}, \bibinfo
  {author} {\bibfnamefont {M.}~\bibnamefont {Hannam}}, \bibinfo {author}
  {\bibfnamefont {F.}~\bibnamefont {Ohme}}, \bibinfo {author} {\bibfnamefont
  {M.}~\bibnamefont {Pürrer}}, \bibinfo {author} {\bibfnamefont
  {X.}~\bibnamefont {Jiménez~Forteza}}, \ and\ \bibinfo {author}
  {\bibfnamefont {A.}~\bibnamefont {Bohé}},\ }\href {\doibase
  10.1103/PhysRevD.93.044007} {\bibfield  {journal} {\bibinfo  {journal} {Phys.
  Rev.}\ }\textbf {\bibinfo {volume} {D93}},\ \bibinfo {pages} {044007}
  (\bibinfo {year} {2016})},\ \Eprint {http://arxiv.org/abs/1508.07253}
  {arXiv:1508.07253 [gr-qc]} \BibitemShut {NoStop}%
\bibitem [{\citenamefont {Hamilton}\ \emph {et~al.}(2021)\citenamefont
  {Hamilton}, \citenamefont {London}, \citenamefont {Thompson}, \citenamefont
  {Fauchon-Jones}, \citenamefont {Hannam}, \citenamefont {Kalaghatgi},
  \citenamefont {Khan}, \citenamefont {Pannarale},\ and\ \citenamefont
  {Vano-Vinuales}}]{Hamilton:2021pkf}%
  \BibitemOpen
  \bibfield  {author} {\bibinfo {author} {\bibfnamefont {E.}~\bibnamefont
  {Hamilton}}, \bibinfo {author} {\bibfnamefont {L.}~\bibnamefont {London}},
  \bibinfo {author} {\bibfnamefont {J.~E.}\ \bibnamefont {Thompson}}, \bibinfo
  {author} {\bibfnamefont {E.}~\bibnamefont {Fauchon-Jones}}, \bibinfo {author}
  {\bibfnamefont {M.}~\bibnamefont {Hannam}}, \bibinfo {author} {\bibfnamefont
  {C.}~\bibnamefont {Kalaghatgi}}, \bibinfo {author} {\bibfnamefont
  {S.}~\bibnamefont {Khan}}, \bibinfo {author} {\bibfnamefont {F.}~\bibnamefont
  {Pannarale}}, \ and\ \bibinfo {author} {\bibfnamefont {A.}~\bibnamefont
  {Vano-Vinuales}},\ }\href@noop {} {\  (\bibinfo {year} {2021})},\ \Eprint
  {http://arxiv.org/abs/2107.08876} {arXiv:2107.08876 [gr-qc]} \BibitemShut
  {NoStop}%
\bibitem [{\citenamefont {Berti}\ \emph {et~al.}(2016)\citenamefont {Berti},
  \citenamefont {Sesana}, \citenamefont {Barausse}, \citenamefont {Cardoso},\
  and\ \citenamefont {Belczynski}}]{Berti:2016lat}%
  \BibitemOpen
  \bibfield  {author} {\bibinfo {author} {\bibfnamefont {E.}~\bibnamefont
  {Berti}}, \bibinfo {author} {\bibfnamefont {A.}~\bibnamefont {Sesana}},
  \bibinfo {author} {\bibfnamefont {E.}~\bibnamefont {Barausse}}, \bibinfo
  {author} {\bibfnamefont {V.}~\bibnamefont {Cardoso}}, \ and\ \bibinfo
  {author} {\bibfnamefont {K.}~\bibnamefont {Belczynski}},\ }\href {\doibase
  10.1103/PhysRevLett.117.101102} {\bibfield  {journal} {\bibinfo  {journal}
  {Phys. Rev. Lett.}\ }\textbf {\bibinfo {volume} {117}},\ \bibinfo {pages}
  {101102} (\bibinfo {year} {2016})},\ \Eprint
  {http://arxiv.org/abs/1605.09286} {arXiv:1605.09286 [gr-qc]} \BibitemShut
  {NoStop}%
\bibitem [{\citenamefont {Kamaretsos}\ \emph
  {et~al.}(2012{\natexlab{b}})\citenamefont {Kamaretsos}, \citenamefont
  {Hannam},\ and\ \citenamefont {Sathyaprakash}}]{Kamaretsos:2012bs}%
  \BibitemOpen
  \bibfield  {author} {\bibinfo {author} {\bibfnamefont {I.}~\bibnamefont
  {Kamaretsos}}, \bibinfo {author} {\bibfnamefont {M.}~\bibnamefont {Hannam}},
  \ and\ \bibinfo {author} {\bibfnamefont {B.}~\bibnamefont {Sathyaprakash}},\
  }\href {\doibase 10.1103/PhysRevLett.109.141102} {\bibfield  {journal}
  {\bibinfo  {journal} {Phys.Rev.Lett.}\ }\textbf {\bibinfo {volume} {109}},\
  \bibinfo {pages} {141102} (\bibinfo {year} {2012}{\natexlab{b}})},\ \Eprint
  {http://arxiv.org/abs/1207.0399} {arXiv:1207.0399 [gr-qc]} \BibitemShut
  {NoStop}%
\bibitem [{\citenamefont {Keitel}\ \emph {et~al.}(2017)\citenamefont {Keitel}
  \emph {et~al.}}]{Keitel:2016krm}%
  \BibitemOpen
  \bibfield  {author} {\bibinfo {author} {\bibfnamefont {D.}~\bibnamefont
  {Keitel}} \emph {et~al.},\ }\href {\doibase 10.1103/PhysRevD.96.024006}
  {\bibfield  {journal} {\bibinfo  {journal} {Phys. Rev.}\ }\textbf {\bibinfo
  {volume} {D96}},\ \bibinfo {pages} {024006} (\bibinfo {year} {2017})},\
  \Eprint {http://arxiv.org/abs/1612.09566} {arXiv:1612.09566 [gr-qc]}
  \BibitemShut {NoStop}%
\bibitem [{\citenamefont {Jaramillo}\ \emph {et~al.}(2020)\citenamefont
  {Jaramillo}, \citenamefont {Panosso~Macedo},\ and\ \citenamefont
  {Al~Sheikh}}]{Jaramillo:2020tuu}%
  \BibitemOpen
  \bibfield  {author} {\bibinfo {author} {\bibfnamefont {J.~L.}\ \bibnamefont
  {Jaramillo}}, \bibinfo {author} {\bibfnamefont {R.}~\bibnamefont
  {Panosso~Macedo}}, \ and\ \bibinfo {author} {\bibfnamefont {L.}~\bibnamefont
  {Al~Sheikh}},\ }\href@noop {} {\  (\bibinfo {year} {2020})},\ \Eprint
  {http://arxiv.org/abs/2004.06434} {arXiv:2004.06434 [gr-qc]} \BibitemShut
  {NoStop}%
\bibitem [{\citenamefont {Axler}(2015)}]{axler2015linear}%
  \BibitemOpen
  \bibfield  {author} {\bibinfo {author} {\bibfnamefont {S.}~\bibnamefont
  {Axler}},\ }\href@noop {} {\emph {\bibinfo {title} {Linear algebra done
  right}}}\ (\bibinfo  {publisher} {Springer},\ \bibinfo {address} {Cham},\
  \bibinfo {year} {2015})\BibitemShut {NoStop}%
\bibitem [{\citenamefont {Seidel}(1989)}]{Seidel:1988ue}%
  \BibitemOpen
  \bibfield  {author} {\bibinfo {author} {\bibfnamefont {E.}~\bibnamefont
  {Seidel}},\ }\href {\doibase 10.1088/0264-9381/6/7/012} {\bibfield  {journal}
  {\bibinfo  {journal} {Class. Quant. Grav.}\ }\textbf {\bibinfo {volume}
  {6}},\ \bibinfo {pages} {1057} (\bibinfo {year} {1989})}\BibitemShut
  {NoStop}%
\bibitem [{\citenamefont {Press}\ and\ \citenamefont
  {Teukolsky}(1973)}]{Press:1973zz}%
  \BibitemOpen
  \bibfield  {author} {\bibinfo {author} {\bibfnamefont {W.~H.}\ \bibnamefont
  {Press}}\ and\ \bibinfo {author} {\bibfnamefont {S.~A.}\ \bibnamefont
  {Teukolsky}},\ }\href {\doibase 10.1086/152445} {\bibfield  {journal}
  {\bibinfo  {journal} {Astrophys. J.}\ }\textbf {\bibinfo {volume} {185}},\
  \bibinfo {pages} {649} (\bibinfo {year} {1973})}\BibitemShut {NoStop}%
\bibitem [{\citenamefont {Yang}\ \emph {et~al.}(2012)\citenamefont {Yang},
  \citenamefont {Nichols}, \citenamefont {Zhang}, \citenamefont {Zimmerman},
  \citenamefont {Zhang},\ and\ \citenamefont {Chen}}]{Yang:2012he}%
  \BibitemOpen
  \bibfield  {author} {\bibinfo {author} {\bibfnamefont {H.}~\bibnamefont
  {Yang}}, \bibinfo {author} {\bibfnamefont {D.~A.}\ \bibnamefont {Nichols}},
  \bibinfo {author} {\bibfnamefont {F.}~\bibnamefont {Zhang}}, \bibinfo
  {author} {\bibfnamefont {A.}~\bibnamefont {Zimmerman}}, \bibinfo {author}
  {\bibfnamefont {Z.}~\bibnamefont {Zhang}}, \ and\ \bibinfo {author}
  {\bibfnamefont {Y.}~\bibnamefont {Chen}},\ }\href {\doibase
  10.1103/PhysRevD.86.104006} {\bibfield  {journal} {\bibinfo  {journal} {Phys.
  Rev. D}\ }\textbf {\bibinfo {volume} {86}},\ \bibinfo {pages} {104006}
  (\bibinfo {year} {2012})},\ \Eprint {http://arxiv.org/abs/1207.4253}
  {arXiv:1207.4253 [gr-qc]} \BibitemShut {NoStop}%
\bibitem [{\citenamefont {Cotesta}\ \emph {et~al.}(2018)\citenamefont
  {Cotesta}, \citenamefont {Buonanno}, \citenamefont {Bohé}, \citenamefont
  {Taracchini}, \citenamefont {Hinder},\ and\ \citenamefont
  {Ossokine}}]{Cotesta:2018fcv}%
  \BibitemOpen
  \bibfield  {author} {\bibinfo {author} {\bibfnamefont {R.}~\bibnamefont
  {Cotesta}}, \bibinfo {author} {\bibfnamefont {A.}~\bibnamefont {Buonanno}},
  \bibinfo {author} {\bibfnamefont {A.}~\bibnamefont {Bohé}}, \bibinfo
  {author} {\bibfnamefont {A.}~\bibnamefont {Taracchini}}, \bibinfo {author}
  {\bibfnamefont {I.}~\bibnamefont {Hinder}}, \ and\ \bibinfo {author}
  {\bibfnamefont {S.}~\bibnamefont {Ossokine}},\ }\href {\doibase
  10.1103/PhysRevD.98.084028} {\bibfield  {journal} {\bibinfo  {journal} {Phys.
  Rev. D}\ }\textbf {\bibinfo {volume} {98}},\ \bibinfo {pages} {084028}
  (\bibinfo {year} {2018})},\ \Eprint {http://arxiv.org/abs/1803.10701}
  {arXiv:1803.10701 [gr-qc]} \BibitemShut {NoStop}%
\bibitem [{\citenamefont {Hughes}\ \emph {et~al.}(2019)\citenamefont {Hughes},
  \citenamefont {Apte}, \citenamefont {Khanna},\ and\ \citenamefont
  {Lim}}]{Hughes:2019zmt}%
  \BibitemOpen
  \bibfield  {author} {\bibinfo {author} {\bibfnamefont {S.~A.}\ \bibnamefont
  {Hughes}}, \bibinfo {author} {\bibfnamefont {A.}~\bibnamefont {Apte}},
  \bibinfo {author} {\bibfnamefont {G.}~\bibnamefont {Khanna}}, \ and\ \bibinfo
  {author} {\bibfnamefont {H.}~\bibnamefont {Lim}},\ }\href {\doibase
  10.1103/PhysRevLett.123.161101} {\bibfield  {journal} {\bibinfo  {journal}
  {Phys. Rev. Lett.}\ }\textbf {\bibinfo {volume} {123}},\ \bibinfo {pages}
  {161101} (\bibinfo {year} {2019})},\ \Eprint
  {http://arxiv.org/abs/1901.05900} {arXiv:1901.05900 [gr-qc]} \BibitemShut
  {NoStop}%
\bibitem [{\citenamefont {London}(2020)}]{London:2018gaq}%
  \BibitemOpen
  \bibfield  {author} {\bibinfo {author} {\bibfnamefont {L.}~\bibnamefont
  {London}},\ }\href {\doibase 10.1103/PhysRevD.102.084052} {\enquote {\bibinfo
  {title} {{Modeling ringdown. II. Aligned-spin binary black holes,
  implications for data analysis and fundamental theory}},}\ } (\bibinfo {year}
  {2020}),\ \Eprint {http://arxiv.org/abs/1801.08208} {arXiv:1801.08208
  [gr-qc]} \BibitemShut {NoStop}%
\bibitem [{\citenamefont {Le~Tiec}\ \emph {et~al.}(2010)\citenamefont
  {Le~Tiec}, \citenamefont {Blanchet},\ and\ \citenamefont
  {Will}}]{LeTiec:2009yg}%
  \BibitemOpen
  \bibfield  {author} {\bibinfo {author} {\bibfnamefont {A.}~\bibnamefont
  {Le~Tiec}}, \bibinfo {author} {\bibfnamefont {L.}~\bibnamefont {Blanchet}}, \
  and\ \bibinfo {author} {\bibfnamefont {C.~M.}\ \bibnamefont {Will}},\ }\href
  {\doibase 10.1088/0264-9381/27/1/012001} {\bibfield  {journal} {\bibinfo
  {journal} {Class. Quant. Grav.}\ }\textbf {\bibinfo {volume} {27}},\ \bibinfo
  {pages} {012001} (\bibinfo {year} {2010})},\ \Eprint
  {http://arxiv.org/abs/0910.4594} {arXiv:0910.4594 [gr-qc]} \BibitemShut
  {NoStop}%
\bibitem [{\citenamefont {Ossokine}\ \emph {et~al.}(2020)\citenamefont
  {Ossokine} \emph {et~al.}}]{Ossokine:2020kjp}%
  \BibitemOpen
  \bibfield  {author} {\bibinfo {author} {\bibfnamefont {S.}~\bibnamefont
  {Ossokine}} \emph {et~al.},\ }\href {\doibase 10.1103/PhysRevD.102.044055}
  {\bibfield  {journal} {\bibinfo  {journal} {Phys. Rev. D}\ }\textbf {\bibinfo
  {volume} {102}},\ \bibinfo {pages} {044055} (\bibinfo {year} {2020})},\
  \Eprint {http://arxiv.org/abs/2004.09442} {arXiv:2004.09442 [gr-qc]}
  \BibitemShut {NoStop}%
\bibitem [{\citenamefont {Giesler}\ \emph {et~al.}(2019)\citenamefont
  {Giesler}, \citenamefont {Isi}, \citenamefont {Scheel},\ and\ \citenamefont
  {Teukolsky}}]{Giesler:2019uxc}%
  \BibitemOpen
  \bibfield  {author} {\bibinfo {author} {\bibfnamefont {M.}~\bibnamefont
  {Giesler}}, \bibinfo {author} {\bibfnamefont {M.}~\bibnamefont {Isi}},
  \bibinfo {author} {\bibfnamefont {M.~A.}\ \bibnamefont {Scheel}}, \ and\
  \bibinfo {author} {\bibfnamefont {S.}~\bibnamefont {Teukolsky}},\ }\href
  {\doibase 10.1103/PhysRevX.9.041060} {\bibfield  {journal} {\bibinfo
  {journal} {Phys. Rev. X}\ }\textbf {\bibinfo {volume} {9}},\ \bibinfo {pages}
  {041060} (\bibinfo {year} {2019})},\ \Eprint
  {http://arxiv.org/abs/1903.08284} {arXiv:1903.08284 [gr-qc]} \BibitemShut
  {NoStop}%
\bibitem [{\citenamefont {Abbott}\ \emph {et~al.}(2021)\citenamefont {Abbott}
  \emph {et~al.}}]{LIGOScientific:2020ibl}%
  \BibitemOpen
  \bibfield  {author} {\bibinfo {author} {\bibfnamefont {R.}~\bibnamefont
  {Abbott}} \emph {et~al.} (\bibinfo {collaboration} {LIGO Scientific,
  Virgo}),\ }\href {\doibase 10.1103/PhysRevX.11.021053} {\bibfield  {journal}
  {\bibinfo  {journal} {Phys. Rev. X}\ }\textbf {\bibinfo {volume} {11}},\
  \bibinfo {pages} {021053} (\bibinfo {year} {2021})},\ \Eprint
  {http://arxiv.org/abs/2010.14527} {arXiv:2010.14527 [gr-qc]} \BibitemShut
  {NoStop}%
\bibitem [{\citenamefont {Rosas-Ortiz}\ and\ \citenamefont
  {Zelaya}(2018)}]{ROSASORTIZ201826}%
  \BibitemOpen
  \bibfield  {author} {\bibinfo {author} {\bibfnamefont {O.}~\bibnamefont
  {Rosas-Ortiz}}\ and\ \bibinfo {author} {\bibfnamefont {K.}~\bibnamefont
  {Zelaya}},\ }\href {\doibase https://doi.org/10.1016/j.aop.2017.10.020}
  {\bibfield  {journal} {\bibinfo  {journal} {Annals of Physics}\ }\textbf
  {\bibinfo {volume} {388}},\ \bibinfo {pages} {26 } (\bibinfo {year}
  {2018})}\BibitemShut {NoStop}%
\bibitem [{\citenamefont {Andrianov}\ \emph {et~al.}(1984)\citenamefont
  {Andrianov}, \citenamefont {Borisov},\ and\ \citenamefont
  {Ioffe}}]{Andrianov:1984er}%
  \BibitemOpen
  \bibfield  {author} {\bibinfo {author} {\bibfnamefont {A.}~\bibnamefont
  {Andrianov}}, \bibinfo {author} {\bibfnamefont {N.}~\bibnamefont {Borisov}},
  \ and\ \bibinfo {author} {\bibfnamefont {M.~V.}\ \bibnamefont {Ioffe}},\
  }\href {\doibase 10.1007/BF01029109} {\bibfield  {journal} {\bibinfo
  {journal} {Theor. Math. Phys.}\ }\textbf {\bibinfo {volume} {61}},\ \bibinfo
  {pages} {1078} (\bibinfo {year} {1984})}\BibitemShut {NoStop}%
\bibitem [{\citenamefont {Lax}(2002)}]{lax2002functional}%
  \BibitemOpen
  \bibfield  {author} {\bibinfo {author} {\bibfnamefont {P.}~\bibnamefont
  {Lax}},\ }\href@noop {} {\emph {\bibinfo {title} {Functional analysis}}}\
  (\bibinfo  {publisher} {Wiley},\ \bibinfo {address} {New York},\ \bibinfo
  {year} {2002})\BibitemShut {NoStop}%
\end{thebibliography}%
\end{document}